\documentclass[a4paper, 12pt]{report}
\usepackage{lgrind}
\usepackage{comment}
\usepackage{graphicx}
\usepackage{tabularx}
\usepackage{amssymb}
\usepackage{amsmath}
\usepackage[T2A]{fontenc}
\usepackage[english]{babel}
\usepackage{color}
\usepackage{ulem}
\usepackage{xcolor}
\usepackage{comment}

\usepackage[export]{adjustbox}
\usepackage{subcaption}

\usepackage[intoc, english]{nomencl}
\makenomenclature
\usepackage{etoolbox}

\usepackage{indentfirst}

\usepackage{authblk}

\usepackage{hyperref}

\graphicspath{ {images/} }

\begin{document}


\title{Fourier modal method for the description of nanoparticle lattices in the dipole approximation}

\renewcommand{\thefootnote}{\fnsymbol{footnote}}

\author[1,2]{Ilia M. Fradkin\thanks{Ilia.Fradkin@skoltech.ru}}
\author[1]{Sergey A. Dyakov}
\author[1]{Nikolay A. Gippius}
\affil[1]{Skolkovo Institute of Science and Technology, Nobel Street 3, Moscow 143025, Russia}
\affil[2]{Moscow Institute of Physics and Technology, Institutskiy pereulok 9, Moscow Region 141701, Russia}
\date{\today}


\maketitle

\begin{abstract}
Rigorous coupled-wave analysis (RCWA) is a very effective tool for the studying optical properties of multilayered vertically invariant periodic structures. However, it fails to deal with arrays of small particles because of high gradients in a local field. In this thesis, we implement discrete dipole approximation (DDA) for the construction of scattering matrices of arrays of resonant nanoparticles.
This strongly speeds up the calculations and therefore provides an opportunity for thorough consideration of various layered structures with small periodic inclusions in terms of the RCWA.
We study in detail three main stages of the method: calculation of polarizability tensor of a single nanoparticle, effective polarizability of this particle in a lattice and corresponding scattering matrix of the layer for further integration in the conventional RCWA approach.
We demonstrate the performance of the proposed method by considering plasmonic lattices embedded in a homogeneous ambiance and placed inside and onto optical waveguides and compare our results with experimental papers. Such phenomena as localized surface plasmon resonances (LSPRs) and lattice plasmon resonances (LPRs) are observed as well as their hybridization with photonic guided modes. High accuracy and fast convergence of our approach are shown by a comparison with other computational approaches. Typical limits of applicability of our approximate method are determined by an exploration of the dependence of its error on the parameters of the structure.

This paper is an extended version of our article \cite{fradkin2019fourier}.
\end{abstract}




\tableofcontents



\renewcommand\nomgroup[1]{%
  \item[\bfseries
  \ifstrequal{#1}{P}{Physical quantities:}{%
  \ifstrequal{#1}{A}{Technical abbreviations:}{%
  \ifstrequal{#1}{O}{Other Symbols}{}}}%
]}
 
 
\nomenclature[A]{DDA}{Discrete Dipole Approximation}
\nomenclature[A]{RCWA}{Rigorous Coupled Wave Analysis}
\nomenclature[A]{FMM}{Fourier Modal Method}
\nomenclature[A]{ASR}{Adaptive Spatial Resolution}
\nomenclature[A]{FEM}{Finite Element Method}
\nomenclature[A]{FDTD}{Finite-Difference Time-Domain method}
\nomenclature[A]{LSPR}{Localized Surface Plasmon Resonance}
\nomenclature[A]{LPR}{Lattice Plasmon Resonance}
\nomenclature[A]{MLWA}{Modified Long-Wavelength Approximation}

\nomenclature[P]{$c$}{speed of light in a vacuum}
\nomenclature[P]{$\hat{C}(\mathbf{k}_\parallel)$}{dynamic interaction constant}
\nomenclature[P]{$k_0$}{wavevector in a vacuum}
\nomenclature[P]{$k$}{wavevector in a medium}
\nomenclature[P]{$\mathbf{k}_\parallel$}{in-plain component of a wavevector}
\nomenclature[P]{$h$}{distance between an interface and center of a particle}
\nomenclature[P]{$\mathbf{P}$}{dipole moment of a particle}
\nomenclature[P]{$\hat{G}$}{dyadic Green's function in real space}
\nomenclature[P]{$\hat{M}$}{dyadic Green's function in reciprocal space}
\nomenclature[P]{$\hat{\alpha}$}{polarizability tensor of a particle}
\nomenclature[P]{$\hat{\alpha}^{\mathrm{eff}}(\mathbf{k}_\parallel)$}{effective polarizability tensor of a particle in a certain lattice}
\nomenclature[P]{$\hat{I}$}{identity matrix}
\nomenclature[P]{$\mathbb{S}$}{scattering matrix}
\nomenclature[P]{$\vec{d}$}{amplitude of the wave propagating downwards}
\nomenclature[P]{$\vec{u}$}{amplitude of the wave propagating upwards}
\nomenclature[P]{$\mathbf{g}$}{translation vector of reciprocal lattice}
\nomenclature[P]{$\mathbf{t}$}{translation vector of a lattice}

\printnomenclature[7em]

\chapter{Introduction}

Periodic photonic structures are one of the most important low-level components in modern photonics since they form the basis for plenty of optical elements and devices. They include bandgap materials \cite{bykov1972spontaneous, yablonovitch1993photonic, sakoda2004optical}, diffraction gratings \cite{wood1902xlii, strutt1907dynamical, fano1941theory, fano1961effects}, frequency selective surfaces \cite{munk2005frequency}, antennas \cite{adato2009ultra, dregely20113d}, waveguides \cite{christ2003waveguide, benisty2000radiation, tikhodeev2002}, metasurfaces \cite{yu2014flat, wang2016optically, kildishev2013planar, li2017nonlinear, dyakov2018magnetic}, biosensors\cite{Shen2013,baba2015biosensing} etc. Spatial periodicity naturally suggests that the electromagnetic field in periodic structures can be expanded into the Fourier series, that appears to be an effective tool for calculating of optical properties of such structures. Indeed, this is a basic idea of the rigorous coupled-wave analysis (RCWA) \cite{moharam1995,tikhodeev2002} proved itself to be an extremely fast and efficient computational method.
However, inclusions much smaller than a wavelength give rise to high gradients in the near field, which forces to take lots of harmonics into account and can significantly slow down this approach. This fundamental drawback of all Fourier-modal methods, which originates from the Gibbs phenomenon \cite{GIBBS1898,Hewitt1979}, is most pronounced for high-contrast inclusions, such as plasmonic nanoparticles. Moreover, occurring of localized surface plasmon resonances makes the contribution of these inclusions determinative for the optical properties of the whole structure.

To overcome this problem, several approaches have been developed. One of them is the Li factorization rules \cite{li1997new}, which solve the problem of poor convergence at concurrent jump discontinuities. Yet another approach is the use of an adaptive spatial resolution by choosing specially designed coordinate transformation which increases the resolution around the material boundaries \cite{granet2002parametric,weiss2009matched}. These methods significantly improve the convergence of the RCWA numerical scheme. However, if the size of metal inclusions is much smaller than the structure spatial period, practically unrealizable number of Fourier harmonics is required for the solution of Maxwell's equations to be converged. As a result, the RCWA fails to describe optical properties of a periodic array of metallic particles in a dielectric.

At the same time, the light scattering by relatively small particles in most cases might be approximately described by substituting them with ideal electric dipoles, which brings us to discrete dipole approximation \cite{chaumet2003} (DDA).

\begin{figure}[h]
    \centering
    \includegraphics[width=0.6\columnwidth]{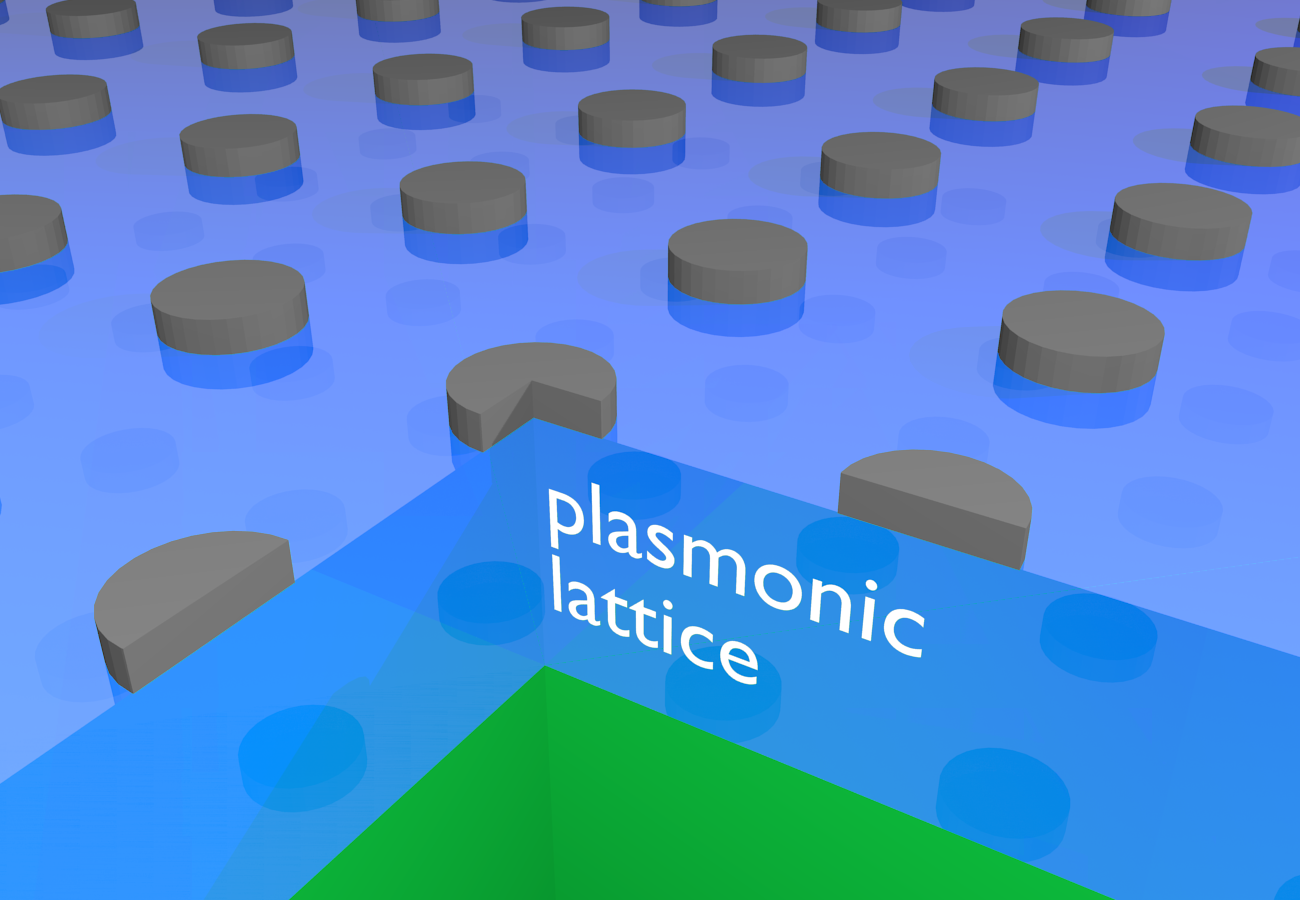}
    \caption{Sketch of a plasmonic lattice embedded in a homogeneous layer of a typical layered structure.}
    \label{fig:lattice}
\end{figure}

In this thesis, we report the approach for calculation of a scattering matrix of a plasmonic lattice, which is based on a determination of an effective polarizability tensor of nanoparticles arranged in periodic arrays. We combine three widespread numerical methods: finite element method (FEM), DDA, and RCWA in order to implement each of them on a specific stage of computation and obtain precise results in a fast way. It helps to study any layered structures with inclusions of plasmonic lattices for any angles of incident light. Such an approach makes it possible to observe dispersion of lattice waves and phenomena of out-of-plane polarization of plasmonic particles \cite{meinzer2014,zhou2011}. To illustrate the feasibility of the proposed method, we consider the same plasmonic lattice in a homogeneous ambience, on a photonic waveguide and inside it, the formation of photonic band structure, strong coupling of photonic guided modes with both LSPRs and LPRs. Also we reproduce several experiments and compare them with our results.
High accuracy and fast convergence of our approach are shown by a comparison with conventional RCWA, RCWA enhanced by adaptive spatial resolution and FEM. Typical limits of applicability of our approximate method are determined by an exploration of the dependence of its error on the parameters of the structure.

\clearpage
\newpage

\chapter{Effective polarizability}
\label{eff_pol}
Dipole approximation makes it possible to split the problem of determination of particles' polarization into two parts, which deal with a problem in different scales. Indeed, when a particle is incorporated in a certain structure of permittivity $\varepsilon^{\mathrm{bg}}(\mathbf{r})$ so that this permittivity is constant over the volume of a particle then it can be conveniently substituted with a solitary dipole (see Appendix \ref{App alpha}). The dipole moment of the $i$-th particle, $\mathbf{P}_i$, is determined as a polarizability tensor, $\hat{\alpha}$, acting on a background low-gradient electric field, $\mathbf{E}^{\mathrm{bg}}_i$, which excites electron oscillations in a plasmonic nanoparticle:
\begin{equation}
    \mathbf{P}_i = \hat{\alpha} \mathbf{E}^{\mathrm{bg}}_i.
    \label{eq:1}
\end{equation}
In this paper, we define dipole moment $\mathbf{P}_i$ as a dipole moment of free charges, which generates the same far-field as a real particle. To give an example, such a definition makes $\hat{\alpha}$ tensor $\varepsilon^{\mathrm{bg}}$ times larger in comparison with a tensor conventionally defined for a particle in a homogeneous medium \cite{bohren2008absorption}. Computation of this polarizability tensor is the first subproblem, which can be easily solved numerically by any near-field computational method for a particle of any shape (see Appendix \ref{App alpha}) or even analytically for particles of trivial shape \cite{landau2013electrodynamics, draine1988}. The second subproblem is a determination of the background field itself, $\mathbf{E}_i^{\mathrm{bg}}$, which is a sum of an incident waves' field, $\mathbf{E}^{0}_i$, which would have been in the absence of the lattice and electric field scattered by all the neighboring particles of a lattice at the coordinate of the considered particle, $\mathbf{r}_i$:

\begin{equation}
    \mathbf{E}^{\mathrm{bg}}_i=\mathbf{E}^{0}_i+
    \sum_{j\neq i}
    \hat{G}(\mathbf{r}_i,\mathbf{r}_j)\mathbf{P}_j,
    \label{eq:2}
\end{equation}
where $\hat{G}(\mathbf{r}_i,\mathbf{r}_j)$ is the dyadic Green's function showing electric field induced at the point $\mathbf{r}_i$ by a dipole at the coordinate $\mathbf{r}_j$ in a considered, not necessarily homogeneous environment. It should be emphasized that in this expression Green's function acts on a dipole moment of free charges and is defined accordingly. 

Background electric field can be found immediately by solving this linear algebraic system consisting of $3N$ equations ($N$ is a number of particles) via any specialized method \cite{Evlyukhin2010}, which is a general approach for DDA method. However, occurrence of periodicity makes it possible to express the solution of algebraic system in a simple form.

Indeed, let us consider an infinite lattice consisting of the same particles. In this case according to Bloch theorem the Floquet periodicity occur, which means that $\mathbf{P}_j=\mathbf{P}_i e^{i\mathbf{k}_\parallel(\mathbf{r}_j-\mathbf{r}_i)}$. Substituting this expression into Eqn. \ref{eq:2} gives us:

\begin{equation}
    \mathbf{E}^{\mathrm{bg}}_i=\mathbf{E}^{0}_i+
    \sum_{j\neq i}
    \hat{G}(\mathbf{r}_i,\mathbf{r}_j) e^{i\mathbf{k}_\parallel(\mathbf{r}_j-\mathbf{r}_i)}\mathbf{P}_i.
    \label{eq:2.5}
\end{equation}

Application of Eqn. \ref{eq:1}, finally results in a following expression, which variations has been observed in literature \cite{Evlyukhin2010,Humphrey2014,schatz2001,Zhao2003,Augie2008,rodriguez2012,Belov2005,Garcia2007}:

\begin{equation}
    \mathbf{E}^{\mathrm{bg}}_i=
    (\hat{I}-\hat{C}(\mathbf{k}_\parallel)\hat{\alpha})^{-1}\mathbf{E}^{0}_i,
    \label{eq:3}
\end{equation}
where, $\hat{I}$, is the identity matrix, $\mathbf{k}_\parallel$ is the in-plain component of photon wavevector and $\hat{C}$ tensor is a so-called dynamic interaction constant \cite{Belov2005}, which is a sum of dyadic Green's function over the lattice (see Appendix \ref{App sum} for details of practical calculation):

\begin{equation}
    \hat{C}(\mathbf{k}_\parallel)=
    \sum_{j\neq i}
    \hat{G}(\mathbf{r}_i,\mathbf{r}_j) e^{-i\mathbf{k}_\parallel(\mathbf{r}_i-\mathbf{r}_j)}.
    \label{eq:4}
\end{equation}

Relation \ref{eq:3} allows to introduce effective polarizability tensor, $\hat{\alpha}^{\mathrm{eff}}$, which connects dipole moment with an incident electric field $\mathbf{P}_i=\hat{\alpha}^{\mathrm{eff}}\mathbf{E}^{0}_i$:
\begin{equation}
    \hat{\alpha}^{\mathrm{eff}}=
    \hat{\alpha}(\hat{I}-\hat{C}(\mathbf{k}_\parallel)\hat{\alpha})^{-1}.
    \label{eq:5}
\end{equation}

As it can be seen from this expression, effective polarizability, $\hat{\alpha}^{\mathrm{eff}}$, has resonances of two types. The first one occurs, when the ordinary individual polarizability $\hat{\alpha}$ experiences wide LSPR. Another resonance with non-trivial dispersion (LPR) is associated with collective oscillations of the lattice and occurs when the condition $\hat{C}(\mathbf{k}_\parallel)=\hat{\alpha}^{-1}$ is fulfilled.

This approach allows to describe the structure in a very simple way but naturally has inherent limitations. Indeed, dipole approximation works until background field does not change on the dimensions of a particle, which results in a requirement for a particle to be much less than a wavelength (far-field limit) and a period of a structure (near-field limit). Also, an electric field of an ideal dipole and a real particle match each other for distances larger than several particle sizes. This means that all the in the vicinity of particles should be accounted in $\hat{\alpha}^{\mathrm{eff}}$ tensor by an appropriate choice of Green's function {and polarizability of a single particle $\hat{\alpha}$}.

Dipole approximation allows considering a wide range of structures. However, the proposed approach can be naturally extended by taking into account higher order multipole moments. Even accounting for quadrupole moment makes it possible to consider larger particles, place them closer to each other and observe quadrupole resonances. For instance, this can be potentially profitable for the description of dense plasmonic metasurfaces or periodic structures of large dielectric particles.

\chapter{Scattering matrix calculation}\label{rcwa_corrections}

Calculation of the effective polarizability, $\hat{\alpha}^{\mathrm{eff}}$, of a particle in a lattice is an important stage. However, our final goal is to obtain the total scattering matrix of the entire structure containing resonant particles. For this hereafter, we will use the formalism of RCWA which was conveniently developed for calculating of optical properties of vertically invariant periodic structures (e.g., multilayered gratings). The total scattering matrix of a multilayered structure is calculated iteratively \cite{ko88}; at each iteration step the scattering matrix of the $i$-th layer is connected with the total scattering matrix of layers 1 to $i-1$ found at previous iteration step. Usually, the scattering matrix of each layer is found by decomposing the electromagnetic field into spatial Fourier harmonics.

\begin{figure*}[h]
    \centering
    \includegraphics[width=1\linewidth]{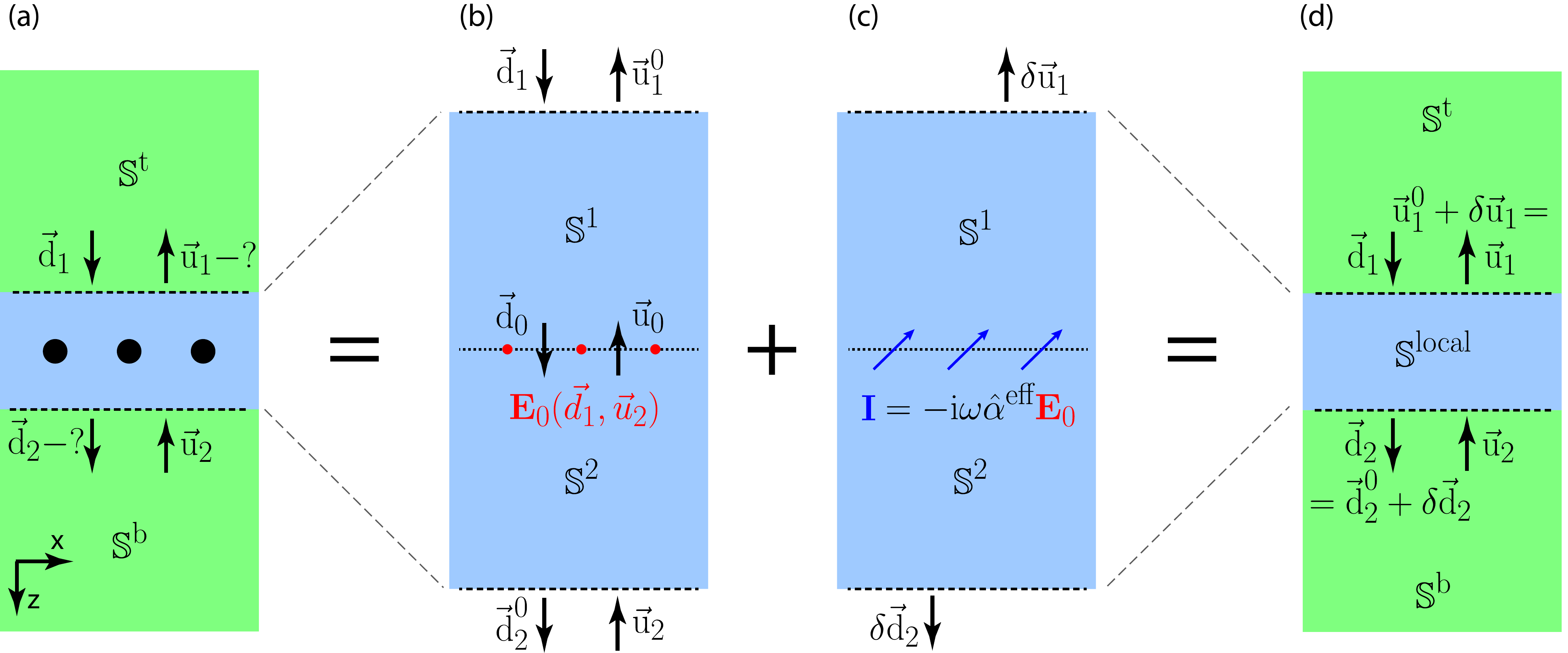}
    \caption{(Color online) Calculation of the scattering matrix of a plasmonic lattice in dipole approximation. (a) Nanoparticles lattice in a layered medium. (b) Calculation of external field at the position of nanoparticles (red points) in the layered medium without nanoparticles. (c) Calculation of current densities (blue arrows) at the position of nanoparticles. (d) Calculation of the local scattering matrix. In panels (a)-(d) dashed lines separate the local dielectric environment of the nanoparticles (blue color) from the outer dielectric environment (green color). Both local and outer dielectric environments might include any number of vertically homogeneous layers and interfaces between them.}
    \label{fig:sketch}
\end{figure*}

As mentioned above, such structures as small metallic particles cause high-gradient fields, which requires taking into account a very large number of spatial harmonics to resolve them in RCWA. However, strongly evanescent behavior of high-$\mathbf{k}_\parallel$ harmonics leads to their confinement inside a layer of several dozens nanometers thickness, which contains plasmonic nanoparticles (blue region in Fig.\,\ref{fig:sketch}). This means, that construction of the scattering matrix of this layer as a whole, $\mathbb{S}^{\mathrm{local}}$, via DDA gives us a possibility to work only with low-$\mathbf{k}_\parallel$ harmonics (see Fig. \ref{fig:sketch} (a)). In such an approach, all high-$\mathbf{k}_\parallel$ effects are described by the effective $\hat{\alpha}^{\mathrm{eff}}$, while low-$\mathbf{k}_\parallel$ effects are treated conventionally by RCWA. Once the local scattering matrix $\mathbb{S}^{\mathrm{local}}$ is calculated, the initial RCWA problem of calculation of the total scattering matrix takes on the task of coordination of adjacent layers, which is consistent with the original spirit of RCWA. 

By definition, the scattering matrix of the considered layer, $\mathbb{S}^{\mathrm{local}}$, connects amplitudes of the incoming and outgoing waves on the boundaries of the considered layer (see Fig. \ref{fig:sketch} (a)) \cite{tikhodeev2002}:
\begin{equation}
    \begin{bmatrix}\vec{\mathrm{d}}_2\\\vec{\mathrm{u}}_1\end{bmatrix}
    =\mathbb{S}^{\mathrm{local}}\begin{bmatrix}\vec{\mathrm{d}}_1\\\vec{\mathrm{u}}_2\end{bmatrix}.
    \label{eq:n:d-1}
\end{equation}
{Hereinafter, we use the symbols $\vec{\mathrm{d}}$ and $\vec{\mathrm{u}}$ for the amplitudes of positively and negatively propagating waves taken at positions specified by the subscripts as shown in Fig. \ref{fig:sketch}. Please note that $z$ axis is directed from top to bottom.}

Elaborating the idea discussed in the section \ref{eff_pol} we represent this matrix as a sum of two terms $\mathbb{S}^{\mathrm{local}}=\mathbb{S}^{\mathrm{local}}_0+\delta \mathbb{S}^{\mathrm{local}}$. The first term corresponds to a matrix calculated in the assumption of the absence of a lattice inclusion (see Fig. \ref{fig:sketch} (b)), whereas the second one (see Fig. \ref{fig:sketch} (c)) accounts for the radiation of the currents induced in the particles. Finally, we substitute the complex layer, which contains nanoparticles lattice by a black box, which is fully described by the matrix, $\mathbb{S}^{\mathrm{local}}$, which has a small number of non-zero elements, describing low-$\mathbf{k}_\parallel$ harmonics.

We start with the consideration of an empty layer without plasmonic inclusions. Scattering matrix $\mathbb{S}_0^{\mathrm{local}}$ connects amplitudes on the boundaries:
\begin{equation}
    \begin{bmatrix}\vec{\mathrm{d}}_2^0\\\vec{\mathrm{u}}_1^0\end{bmatrix}
    =\mathbb{S}^{\mathrm{local}}_0\begin{bmatrix}\vec{\mathrm{d}}_1\\\vec{\mathrm{u}}_2\end{bmatrix}.
    \label{eq:n:d0}
\end{equation}
However, for our purposes, it is very important to know the vector of amplitudes $[\vec{\mathrm{d}}_0, \vec{\mathrm{u}}_0]^T$ at the plane, which will further contain the dipole lattice. Therefore, we introduce scattering matrices $\mathbb{S}^1$ and $\mathbb{S}^2$ of upper and lower parts of the layer respectively. These matrices act as follows:

\begin{gather}
    \begin{bmatrix}\vec{\mathrm{d}}_0\\\vec{\mathrm{u}}_1^0\end{bmatrix}
    =\mathbb{S}^{1}\begin{bmatrix}\vec{\mathrm{d}}_1\\\vec{\mathrm{u}}_0\end{bmatrix},
    \hspace{20pt}
    \begin{bmatrix}\vec{\mathrm{d}}_2^0\\\vec{\mathrm{u}}_0\end{bmatrix}
    =\mathbb{S}^{2}\begin{bmatrix}\vec{\mathrm{d}}_0\\\vec{\mathrm{u}}_2\end{bmatrix},\label{eq:n:d1}
\end{gather}
Moreover, they are obviously connected as:
\begin{equation}
    \mathbb{S}^{\mathrm{local}}_0=\mathbb{S}^1\otimes\mathbb{S}^{2},
    \label{eq:n:d2}
\end{equation}
where the operand $\otimes$ denotes the combination of two adjacent scattering matrices \cite{ko88} (see Appendix \ref{App rcwa} for details of its calculations).

It is convenient to introduce an auxiliary matrix $\mathbb{B}^{\mathrm{in}}$, which allows to determine $[\vec{\mathrm{d}}_0, \vec{\mathrm{u}}_0]^T$ vectors directly from the incoming waves amplitudes. Equations \ref{eq:n:d1} allows us to represent this matrix via $\mathbb{S}^1$ and $\mathbb{S}^2$ components:

\begin{equation}
    \begin{bmatrix}\vec{\mathrm{d}}_0\\\vec{\mathrm{u}}_0\end{bmatrix}=\mathbb{B}^{\mathrm{in}}\begin{bmatrix}\vec{\mathrm{d}}_1\\\vec{\mathrm{u}}_2\end{bmatrix},
    \hspace{10pt}
    \mathbb{B}^{\mathrm{in}}=
    \begin{bmatrix}
    \mathbb{D}_1\mathbb{S}^1_{11} & \mathbb{D}_1\mathbb{S}^1_{12}\mathbb{S}^2_{22} \\
    \mathbb{D}_2\mathbb{S}^2_{21}\mathbb{S}^1_{11} & \mathbb{D}_2\mathbb{S}^2_{22}, 
    \end{bmatrix},
    \label{eq:n:d3}
\end{equation}
where
\begin{align}
    \mathbb{D}_1 &= (1-\mathbb{S}^1_{12}\mathbb{S}^2_{21})^{-1}\\
    \mathbb{D}_2 &= (1-\mathbb{S}^2_{21}\mathbb{S}^1_{12})^{-1}.
\end{align}

Vector $[\vec{\mathrm{d}}_0, \vec{\mathrm{u}}_0]^T$ help us to determine subsequently the field, $\mathbf{E}^0$, induced at the position of the particle and corresponding current $\mathbf{I}=-i\omega\hat{\alpha}^{\mathrm{eff}}\mathbf{E}^0$. According to Maxwell's equations, the presence of currents leads to an appearance of discontinuities of the horizontal components of electric and magnetic fields \cite{lobanov2012emission,taniyama2008s}. As a result, the vectors of amplitudes taken at the coordinates infinitesimally above and below the dipole's plane (see Fig. \ref{fig:sketch} (c)) are connected as follows: 
\begin{equation}
\begin{bmatrix}\vec{\mathrm{d}}\\\vec{\mathrm{u}}\end{bmatrix}_{z_p{+}0}
-\begin{bmatrix}\vec{\mathrm{d}}\\\vec{\mathrm{u}}\end{bmatrix}_{z_p{-}0}=
\begin{bmatrix}\vec{\mathrm{j}}_d\\\vec{\mathrm{j}}_u\end{bmatrix} = \mathbb{A}[\hat{\alpha}^{\mathrm{eff}}]\begin{bmatrix}\vec{\mathrm{d}}_0\\\vec{\mathrm{u}}_0\end{bmatrix},
\label{eq:n:d4}
\end{equation}
where $[\vec{\mathrm{j}}_d, \vec{\mathrm{j}}_u]^T$ is the amplitude discontinuity vector determined by the induced current $\mathbf{I}$, which results in a connection with incoming waves' amplitudes via a special tensor, $\mathbb{A}[\hat{\alpha}^{\mathrm{eff}}]$, (see Appendix \ref{App rcwa}).

The last thing, which is left to do is the calculation of outgoing from the layer waves' amplitudes, determined by this emission. Similarly to $\mathbb{B}^{\mathrm{in}}$, it is very convenient to introduce another matrix $\mathbb{B}^{\mathrm{out}}$ connecting these discontinuities with outgoing waves' amplitudes associated with them. This matrix can be expressed through scattering matrices $\mathbb{S}^1$ and $\mathbb{S}^2$ components from the condition of the absence of incoming waves (see Fig. \ref{fig:sketch} (c)):
\begin{equation}
    \begin{bmatrix}\delta\vec{\mathrm{d}}_2\\\delta\vec{\mathrm{u}}_1\end{bmatrix}=\mathbb{B}^{\mathrm{out}}\begin{bmatrix}\vec{\mathrm{j}}_d\\\vec{\mathrm{j}}_u\end{bmatrix},
    \hspace{10pt}
    \mathbb{\mathbb{B}}^{\mathrm{out}}=
    \begin{bmatrix}
    \mathbb{S}^2_{11}\mathbb{D}_1 & -\mathbb{S}^2_{11}\mathbb{D}_1\mathbb{S}^1_{12}\\
    \mathbb{S}^1_{22}\mathbb{D}_2\mathbb{S}^2_{21} & -\mathbb{S}^1_{22}\mathbb{D}_2
    \end{bmatrix}.
    \label{eq:n:d5}
\end{equation}

It should be especially emphasized, that although there are no incoming waves in the subproblem of dipoles' radiation, their amplitudes depend on the incoming waves, which allows finding the correction to the scattering matrix, $\delta \mathbb{S}^{\mathrm{local}}$ defined as
\begin{equation}
    \begin{bmatrix}\delta\vec{\mathrm{d}}_2\\\delta\vec{\mathrm{u}}_1\end{bmatrix}
    =\delta\mathbb{S}^{\mathrm{local}}\begin{bmatrix}\vec{\mathrm{d}}_1\\\vec{\mathrm{u}}_2\end{bmatrix}.
    \label{eq:n:d6}
\end{equation}
Summing up, in order to do this, we just have to (I) calculate the incoming waves' amplitudes at a plane of the particles, (II) apply $\mathbb{A}[\hat{\alpha}^{\mathrm{eff}}]$ operator in order to find the amplitude discontinuity vector $[\vec{\mathrm{j}}_d, \vec{\mathrm{j}}_u]^T$ and finally (III) determine the correction to the outgoing from the layer waves. 

Combining equations (\ref{eq:n:d3} -- \ref{eq:n:d6}), we obtain the following expression for the correction to the scattering matrix of the layer:
\begin{equation}
\delta\mathbb{S}^{\mathrm{local}}=
\mathbb{B}^{\mathrm{out}}
\mathbb{A}
\mathbb{B}^{\mathrm{in}}.
\label{eq:local}
\end{equation}

When the local scattering matrix of a plasmonic layer is known, it can be easily inserted in any structure. Its total scattering matrix is then found as 
\begin{equation}
    \mathbb{S}^{\mathrm{tot}}=\mathbb{S}^t\otimes\mathbb{S}^{\mathrm{local}}\otimes\mathbb{S}^b.
    \label{eq:d6}
\end{equation}

\begin{figure}[h]
\centering
\includegraphics[width=0.6\columnwidth]{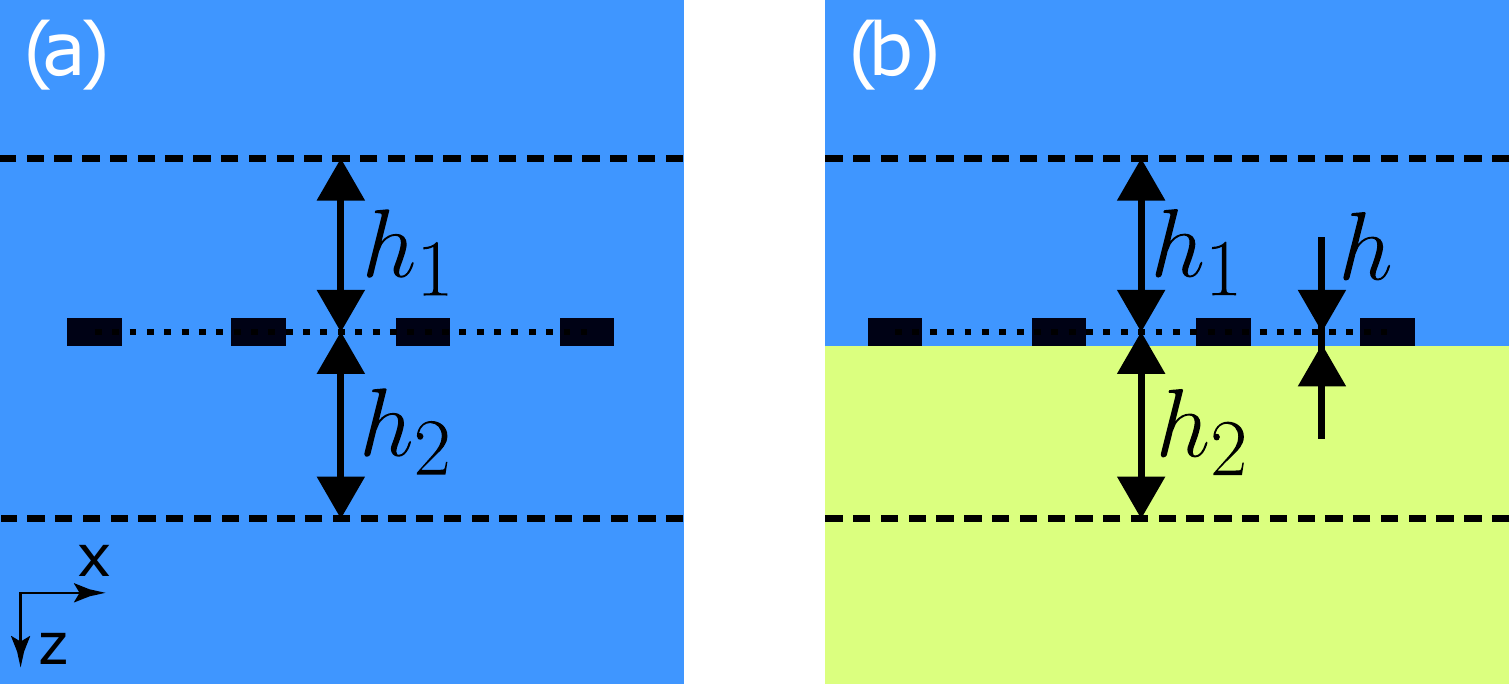}
\caption{(Color online) Sketch of the two most practical structues: a) nanoparticles in homogeneous medium and b) nanoparticles on an interface. Dashed lines bound the local dielectric environment of the nanoparticles. Dotted lines connects the nanoparticles centers.}
\label{fig:2cases}
\end{figure}

These rather general expressions can be potentially used for a description of rather complex structures, for example, plasmonic lattice inside a photonic crystal or a lattice on a metamaterial. However, in this paper, we implement our approach only for two most practical cases and it makes sense to specify $\mathbb{B}$ matrices for them. When the lattice is embedded in a homogeneous medium, $\mathbb{B}$ matrices take a very simple form:

\begin{equation}
    \mathbb{B}^{\mathrm{in}} =\begin{bmatrix}    e^{i k_0\hat{K}_z h_1} & \hat{0}\\  \hat{0}   & e^{i k_0\hat{K}_z h_2} \end{bmatrix},
    \hspace{5pt}
    \mathbb{B}^{\mathrm{out}} =\begin{bmatrix}    e^{i k_0\hat{K}_z h_2} & \hat{0}\\  \hat{0}   & -e^{i k_0\hat{K}_z h_1} \end{bmatrix},
    \label{eq:B}
\end{equation}
where $h_1$ and $h_2$ are the thicknesses of higher and lower layers included into the local environment (see Fig. \ref{fig:2cases}) and $\hat{K}_z$ satisfies the relation
\begin{equation}
    \hat{K}_x^2+\hat{K}_y^2+\hat{K}_z^2=\varepsilon \hat{I}.
    \label{eq:Bz}
\end{equation}
In Eqns.\,(\ref{eq:B}) and (\ref{eq:Bz}) $\hat{K}_x$, $\hat{K}_y$ and $\hat{K}_z$ are the dimensionless diagonal operators \cite{tikhodeev2002} and  
\begin{equation}
    \hat{K}_x = \frac{1}{k_0}\mathrm{diag}(k_x+\vec{g}_x),
    \hspace{20pt}
    \hat{K}_y = \frac{1}{k_0}\mathrm{diag}(k_y+\vec{g}_y),
    \label{eq:kxky}
    \end{equation}
where $\vec{g}_{x}$ and $\vec{g}_{y}$ are $1 \times N_g$ hypervectors of $x$- and $y$- projections of reciprocal lattice vectors. Matrix $\mathbb{A}[\hat{\alpha}^{\mathrm{eff}}]$ also takes a simple form discussed in details in Appendix \ref{App rcwa}.

The second most practical case is a lattice placed in close proximity of an interface between two homogeneous media. It should be noted, that even if the real particles lay right on the boundary, effective dipole lattice is placed at a level of the centers of these particles. Thus, for a dipole lattice situated at a distance of $h$ above an interface $\mathbb{B}$ matrices have the following form:

\begin{align}
    &\mathbb{B}^{\mathrm{in}} &=\begin{bmatrix}    e^{i k_0\hat{K}_z^{(1)} h_1} & \hat{0}\\ \mathbb{S}^{\mathrm{int}}_{21} e^{i k_0\hat{K}_z^{(1)} (h_1+2h)}   & \mathbb{S}^{\mathrm{int}}_{22} e^{i (k_0\hat{K}_z^{(2)} (h_2-h)+k_0\hat{K}_z^{(1)} h)} \end{bmatrix},\notag
    \\
    &\mathbb{B}^{\mathrm{out}} &=\begin{bmatrix}    \mathbb{S}^{\mathrm{int}}_{11} e^{i (k_0\hat{K}_z^{(2)} (h_2-h)+k_0\hat{K}_z^{(1)} h)} & \hat{0}\\  \mathbb{S}^{\mathrm{int}}_{21}e^{i k_0\hat{K}_z^{(1)} (h_1+2h)}    & -e^{i k_0\hat{K}_z^{(1)} h_1} \end{bmatrix},
    \label{eq:B_interface}
\end{align}
where $\mathbb{S}^{\mathrm{int}}$ is the scattering matrix of the interface and operators $\hat{K}_z^{(1)}$ and $\hat{K}_z^{(2)}$ are calculated above and below the interface correspondingly, $h_1$ and $h_2$ are defined according to the Fig. \ref{fig:2cases} b.

The matrix $\mathbb{A}[\hat{\alpha}^{\mathrm{eff}}]$ has the same form as for homogeneous environment (whereas $\hat{\alpha}^{\mathrm{eff}}$ used for its calculation of course differs) since the plane of particles' centers is fully inside a homogeneous medium.

It should be emphasized once again that the dyadic Green's function $\hat{G}$, which is used for a calculation of effective polarizability tensor $\hat{\alpha}^{\mathrm{eff}}$, accounts for the structure of the layer (for instance, presence of an interface) in the near field of metallic particles (see Appendix \ref{App sum}). Indeed, this fact allows us not to account for the self-influence of dipoles by means of RCWA and escape from operating with all the high-$\mathbf{k}_\parallel$ harmonics. Also, it should be noted, that, although the matrices $\mathbb{A}[\hat{\alpha}^{\mathrm{eff}}]$ of the same form are implemented for both considered cases they differ in tensors $\hat{\alpha}^{\mathrm{eff}}$ applied for their calculation, which, in turn, strongly depend on the presence of an interface and a distance $h$ to it.

There is an interesting fact that such splitting parameters as $h^1$ and $h^2$ can be chosen arbitrarily and even set to zero. However, the method performance is determined by the distance to the first boundary not accounted by the local scattering matrix. Indeed, the closer the boundary, the larger is the number of required harmonics and the less is our advantage over the conventional RCWA. Moreover, for distances of the order of particle's size our approach becomes inapplicable and an additional boundary should be included in a local environment to be accounted by the polarizability tensor.

\clearpage
\newpage

\chapter{Example: hybrid resonances}

Plasmonic metals play an important role in modern photonics, because of unique ability to enhance light-matter interaction via strong confinement of light in subwavelength dimensions. Very well-known localized surface plasmon resonances (LSPR) and surface plasmon resonances (SPR) have low Q-factors but can compete with conventional resonances because of extremely small mode volume. In some cases, an intermediate regime of middle Q-factor and mode volume is needed, which suggests the use of hybrid plasmon-photonic resonators \cite{lalanne2018,DeAngelis2009}. An elegant way of implementation of this idea is a construction a regular lattice of resonant plasmonic nanoparticles. Such a structure allows employing both advantages and features of plasmonic and periodic structures. Lattice plasmon resonances that occur in them are actively implemented in biosensors \cite{lodewijks2012,rodriguez2011,Shen2013}, sources of light \cite{guo2015,rodriguez2012,vecchi2009}, stretchable devices \cite{Yang2016}. They might be used for lasing \cite{zhou2013,wang2017}, strong coupling with emitting systems \cite{vakevainen2013,todisco2016} and other purposes \cite{Rajeeva2018, kravets2018plasmonic,wang2017rich}.

Plasmonic lattices have already been thoroughly studied theoretically. Such universal near-field methods as FEM or finite-difference time-domain method \cite{taflove2005} (FDTD) can be implemented for a description of any periodic structure. However, they are too computationally expensive for observation of angle-dependent spectra with a good resolution, which requires at least thousands of computations. RCWA \cite{tikhodeev2002,moharam1995} specialized for periodic structures as it was mentioned before also fails in this case. Plasmonic lattices not only consist from small particles but generate high contrast in permittivities of adjacent materials as well. These factors together with exciting physical phenomena make them the most promising candidate for the first application of our approach.

\begin{figure*}[h]
\centering
\includegraphics[width=1\linewidth]{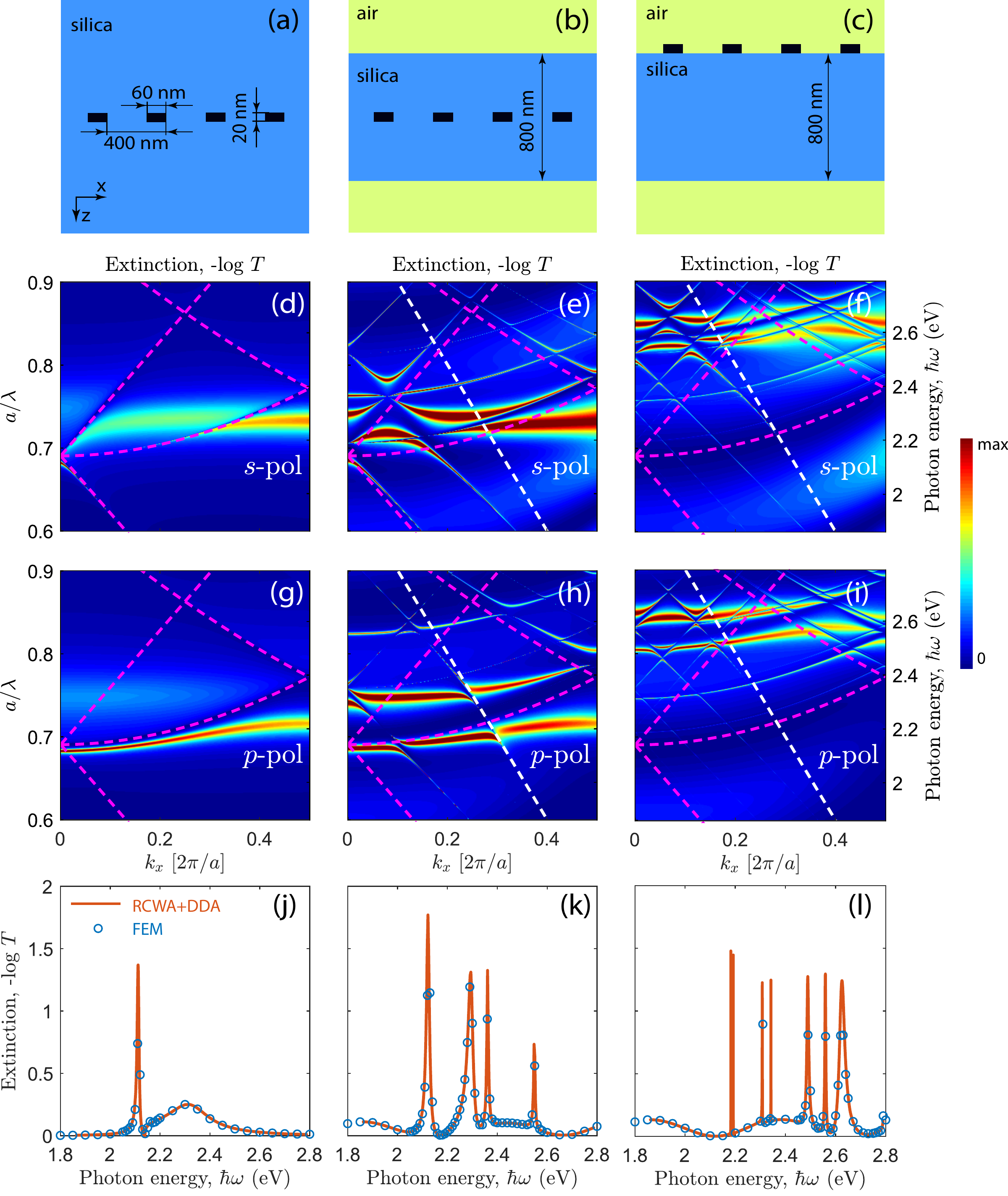}
\caption{(Color online) Schematics of the plasmonic lattice in different environments (a--c). In-plane wavevector and energy dependencies of extinction {($-\log T$, where $T$ is a transmission in the main diffraction order)} in $s$- and $p$-polarizations (d--i) for the case shown in panels (a--c) correspondingly. Color scale of panels (d--i) is explained on the right. Panels (j--l) show extinction spectra for the normal incidence of light, which corresponds to a $k_x=0$ section of angle-dependent spectra. They include the comparison of computations conducted via our approach with conventional FEM calculations.}
\label{sample}
\end{figure*}
\clearpage
\newpage

For this reason, in order to show the performance of our method and illustrate the physical phenomena, which can be investigated, we consider plasmonic lattice in $3$ different environments (see Fig. \ref{sample} (a--c)). We have chosen a square lattice of silver nanodisks with a period of $a=400$\,nm. Disks have the radius of $30$\,nm, the height of $20$\,nm and are described by Johnson-Christy optical constants \cite{JohnsonChristy1972}. In the first case, this lattice is embedded in an infinite surrounding of silica ($\varepsilon = 2.1$). In the second variant of the structure, the lattice is incorporated directly in the middle of this waveguide. And in the latter one, we deposit nanoparticles on an $800$\,nm membrane silica waveguide.

\section{Homogeneous environment}

To start with we have calculated extinction spectra {($-\log T$, where $T$ is a transmission in the main diffraction order)} of the plasmonic lattice in a bulk silica for both polarizations of light incident on a lattice along its crystallographic axis (see. Fig \ref{sample} d,g). Several specific phenomena are observed in this structure.

So-called Rayleigh anomalies \cite{maystre2012} (magenta dashed lines in Fig. \ref{sample} (d,g)) correspond to openings of different diffraction channels. In other words, for this relation of $\omega$ and $\mathbf{k}_\parallel$ all the particles radiate light coherently along the plane of a lattice, which results in in-phase contribution into the $\hat{C}$ tensor, its divergence and subsequent vanishing of the effective polarizability $\hat{\alpha}^{\mathrm{eff}}\rightarrow 0$. Hence, in dipole approximation lattice becomes effectively transparent under Rayleigh condition, which is clearly seen in Fig.\,\ref{sample} (d,g,j). At the same time, when we go to the energies lower than Rayleigh anomalies, interaction constant $\hat{C}$ slightly decreases, which leads to a fulfillment of the resonant condition for LPRs, observed in Fig.\,\ref{sample} (d,g,j) as well. These resonances are so-called Fano-Wood anomalies \cite{maystre2012}, which occur when one of the diffraction orders matches both frequency and an in-plane component of wavevector of the guided mode. LPRs in a homogeneous or almost homogeneous ambience have been thoroughly studied both theoretically and experimentally \cite{guo2017,Rajeeva2018,Augie2008,chu2008,kravets2008,rodriguez2011}. Their dispersion spectra have been observed experimentally and explained in \cite{guo2017}.

\begin{figure}[h!]
\centering
\includegraphics[width=0.6\columnwidth]{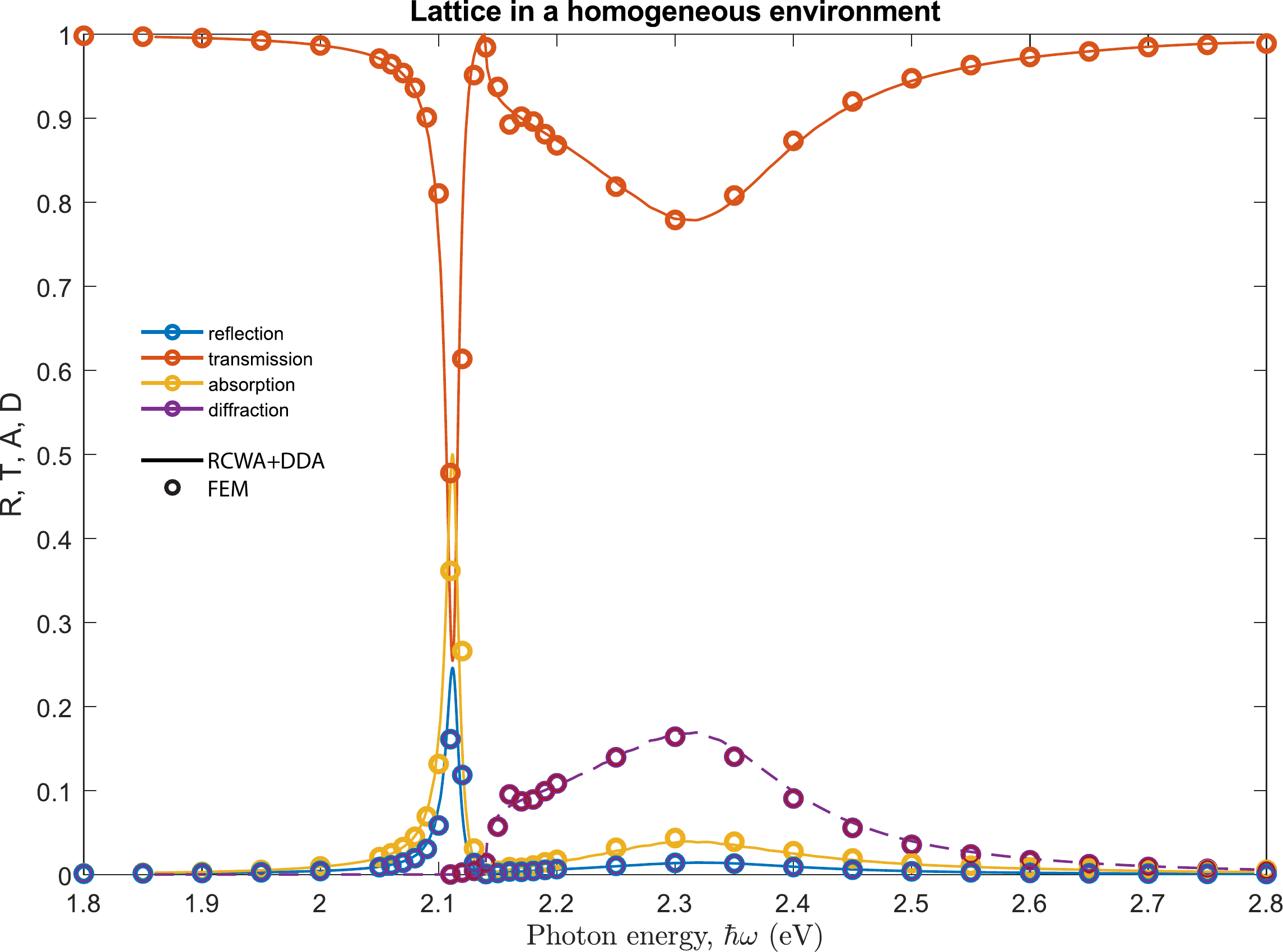}
\caption{(Color online) Reflection, transmission, absorption and diffraction spectra for the plasmonic lattice in a homogeneous environment. Solid lines represent spectra calculated by RCWA in DDA, while points represent spectra calculated by the FEM.}
\label{fig:suppl_homo}
\end{figure}

Extinction spectrum of the plasmonic lattice in a homogeneous environment under the normal incidence is shown in Fig.\,\ref{sample}(j), whereas  the reflection, transmission, absorption and diffraction spectra are depicted in Fig.\,\ref{fig:suppl_homo}.

\section{Lattice in a waveguide}

When the lattice is embedded in the middle of the waveguide of finite thickness (see Fig. \ref{sample} (b)) photonic guided modes come into play, while the positions of LSPR and LPR almost do not change. An interaction between plasmonic resonances with guided modes leads to an appearance of hybridized resonances (see Fig. \ref{sample} (e,h,k)), which represents the main difference with the lattice in the bulk space.
Since the lattice is located strictly in the middle of the symmetric membrane waveguide the hybridization is determined by the parity of the guided modes. Therefore, half of the modes, which have zero electric fields in the center of the waveguide are optically active.

The charachteristic example is observed for $p$-polarized wave at $\hbar\omega\approx2.15eV$ and $k_x\approx 0.11 \cdot 2\pi/a$ (see Fig. \ref{sample} (h)). In this point almost horizontal dispersion curve of lattice plasmon resonance is crossed by a second TM guided mode, which results in Rabi splitting.

\begin{figure}[h!]
\centering
\includegraphics[width=0.6\columnwidth]{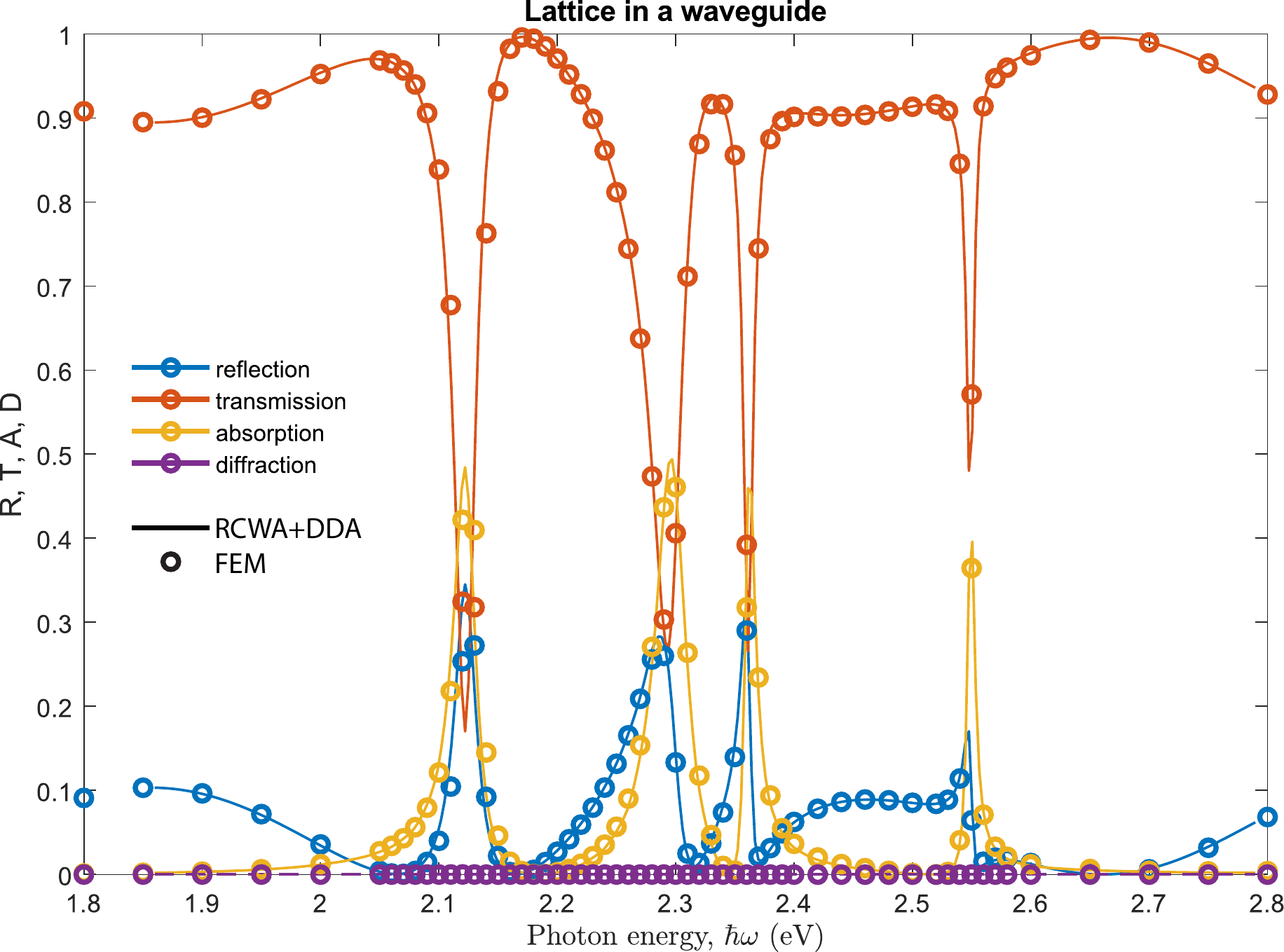}
\caption{(Color online) Reflection, transmission, absorption and diffraction spectra for the plasmonic lattice in a waveguide. Solid lines represent spectra calculated by RCWA in DDA, while points represent spectra calculated by the FEM.}
\label{fig:suppl_in}
\end{figure}

Reflection, transmission, absorption and diffraction spectra for the normal incidence of light are depicted in Fig.\,\ref{fig:suppl_in}.

\section{Lattice on a waveguide}

Despite the apparent similarity of the structures, the appearance of an interface in proximity of a lattice (see Fig. \ref{sample} (c)) strongly changes optical properties.
The difference comes from the fact that dipole located near an interface between two media almost does not radiate light along the boundary. It can be explained by the destructive interference of waves with their counterparts totally reflected in an opposite phase. Therefore, in this case, LSPRs are able to couple to each other only via near-field, which strongly suppresses any collective phenomena. However, the feedback required for observation of Fano-Wood anomalies can emerge again from the coupling of LSPRs to an external photonic guided mode.

In this way, we observe coupling between strongly blue-shifted LSPRs (see appendix \ref{App alpha}) and guided photonic modes of a membrane silica waveguide (see Fig. \ref{sample} (c)). Since the lattice is located on the surface of the waveguide, guided modes of both parities are excited in the structure (see Fig. \ref{sample} (f,i,l)), unlike the previous case.

\begin{figure}[h!]
\centering
\includegraphics[width=0.6\columnwidth]{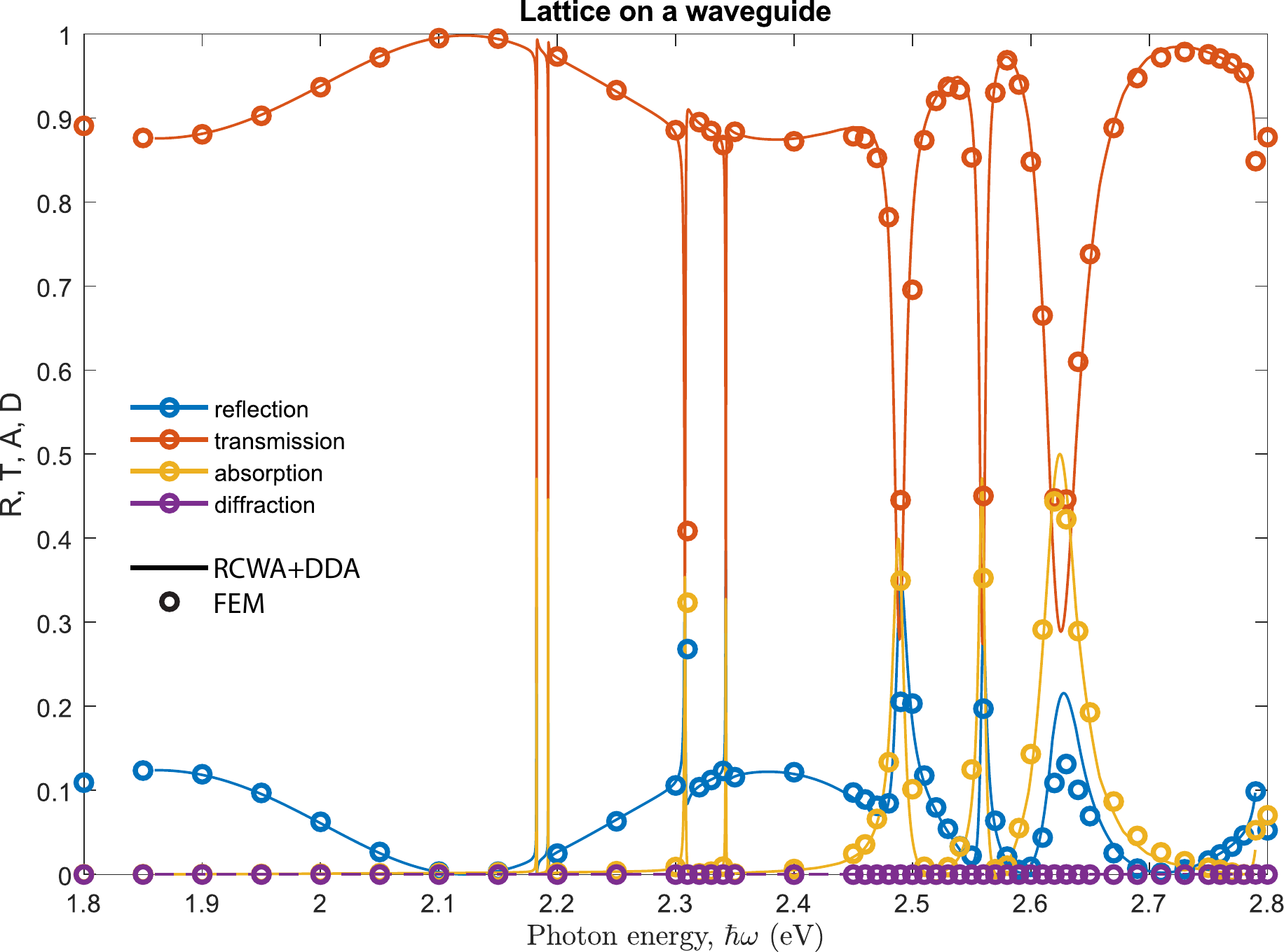}
\caption{(Color online) Reflection, transmission, absorption and diffraction spectra for the plasmonic lattice on a waveguide. Solid lines represent spectra calculated by RCWA in DDA, while points represent spectra calculated by the FEM.}
\label{fig:suppl_on}
\end{figure}

Reflection, transmission, absorption and diffraction spectra for the normal incidence of light are depicted in Fig.\,\ref{fig:suppl_on}.

\section{Optical modes in waveguides of various thickness}

Although reflection, transmission and extinction spectra are convenient for the comparison with experimental measurements, absorption spectra are the most convenient for the theoretical analysis of the band structure. First of all, absorption directly indicates excitation of the plasmonic lattice, since it is the only source of dissipation. Secondly, guided modes of the waveguide lead to a resonant suppression of the effective polarizability, $\hat{\alpha}$, and subsequent disappearance of absorption. Thus, absorption spectra allow us to observe resonant modes and see their plasmonic nature.

\begin{figure}[h!]
    \centering
    \begin{subfigure}{0.32\linewidth}
    \centering
    \includegraphics[width=\linewidth]{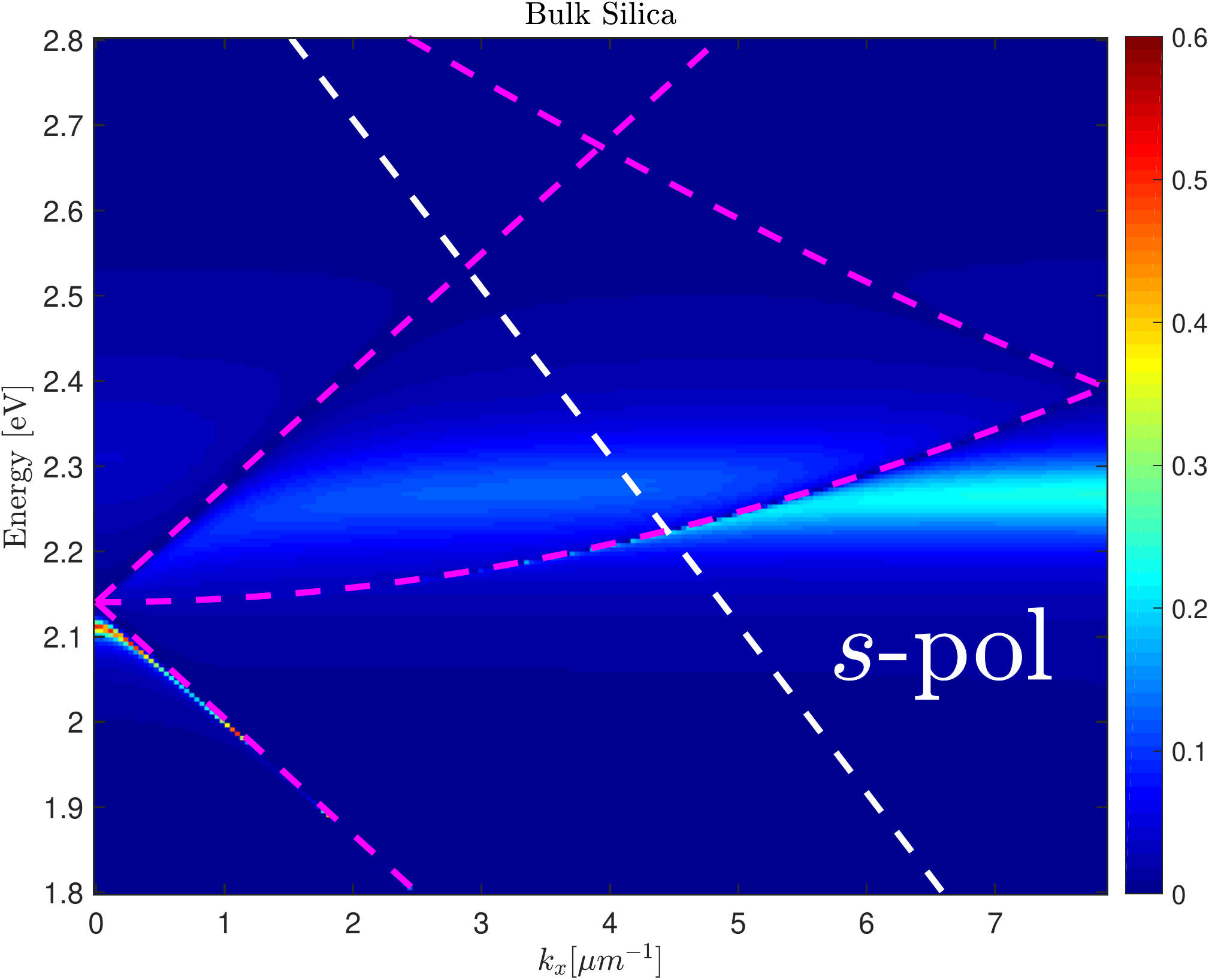}
    \end{subfigure}
    \hfill
    \begin{subfigure}{0.32\linewidth}
    \centering
    \includegraphics[width=\linewidth]{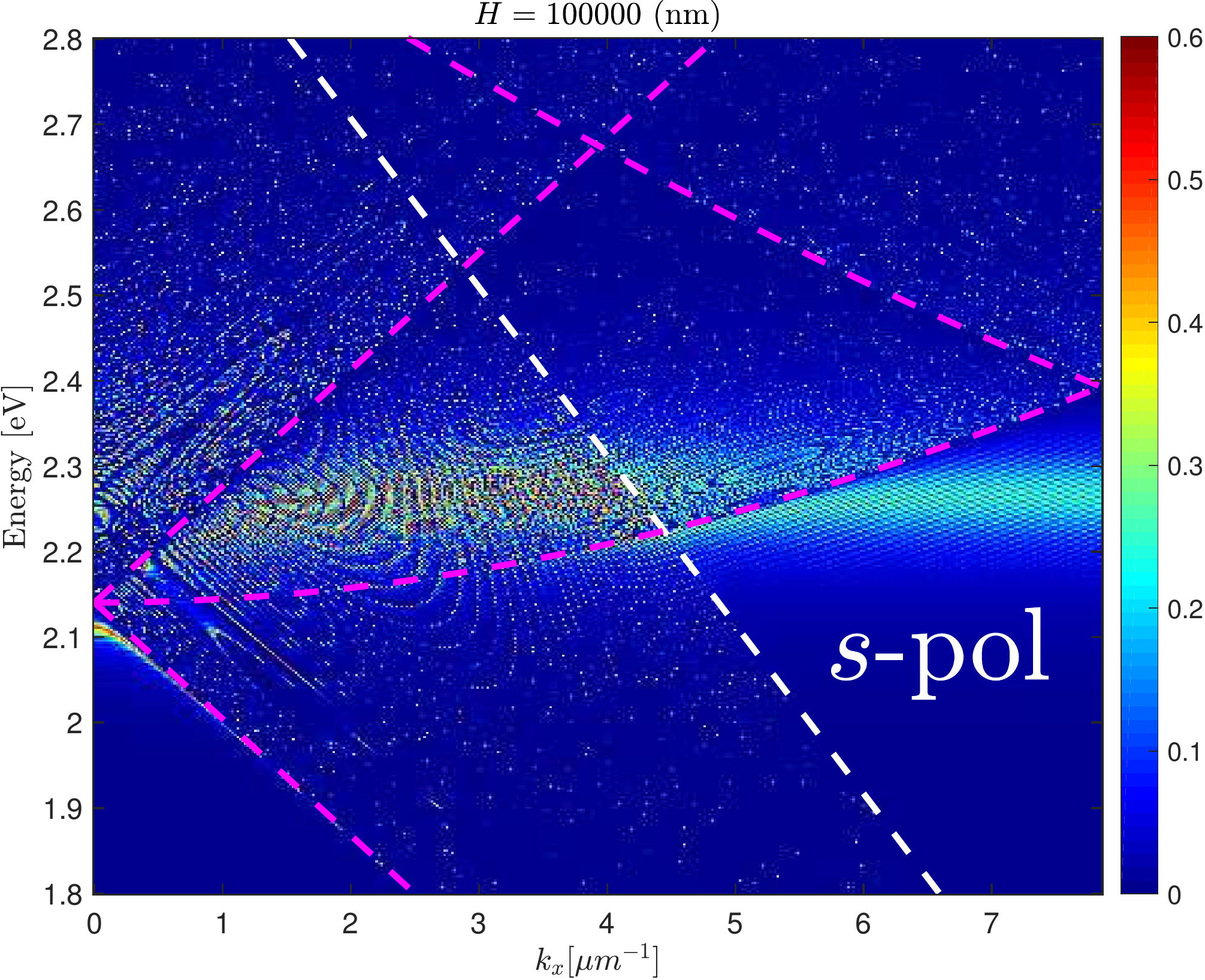}
    \end{subfigure}
    \hfill
    \begin{subfigure}{0.32\linewidth}
    \centering
    \includegraphics[width=\linewidth]{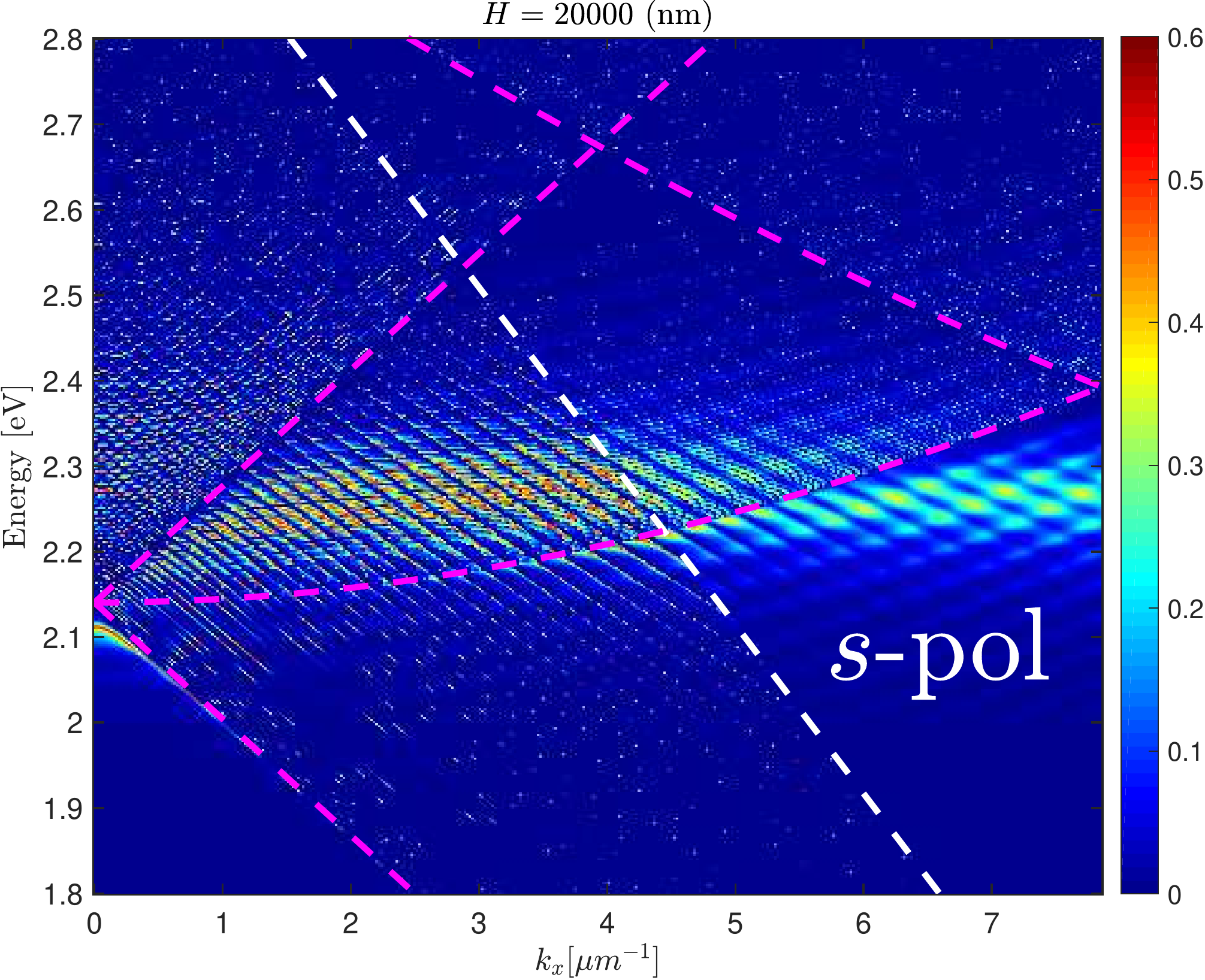}
    \end{subfigure}
    \centering
    \begin{subfigure}{0.32\linewidth}
    \centering
    \includegraphics[width=\linewidth]{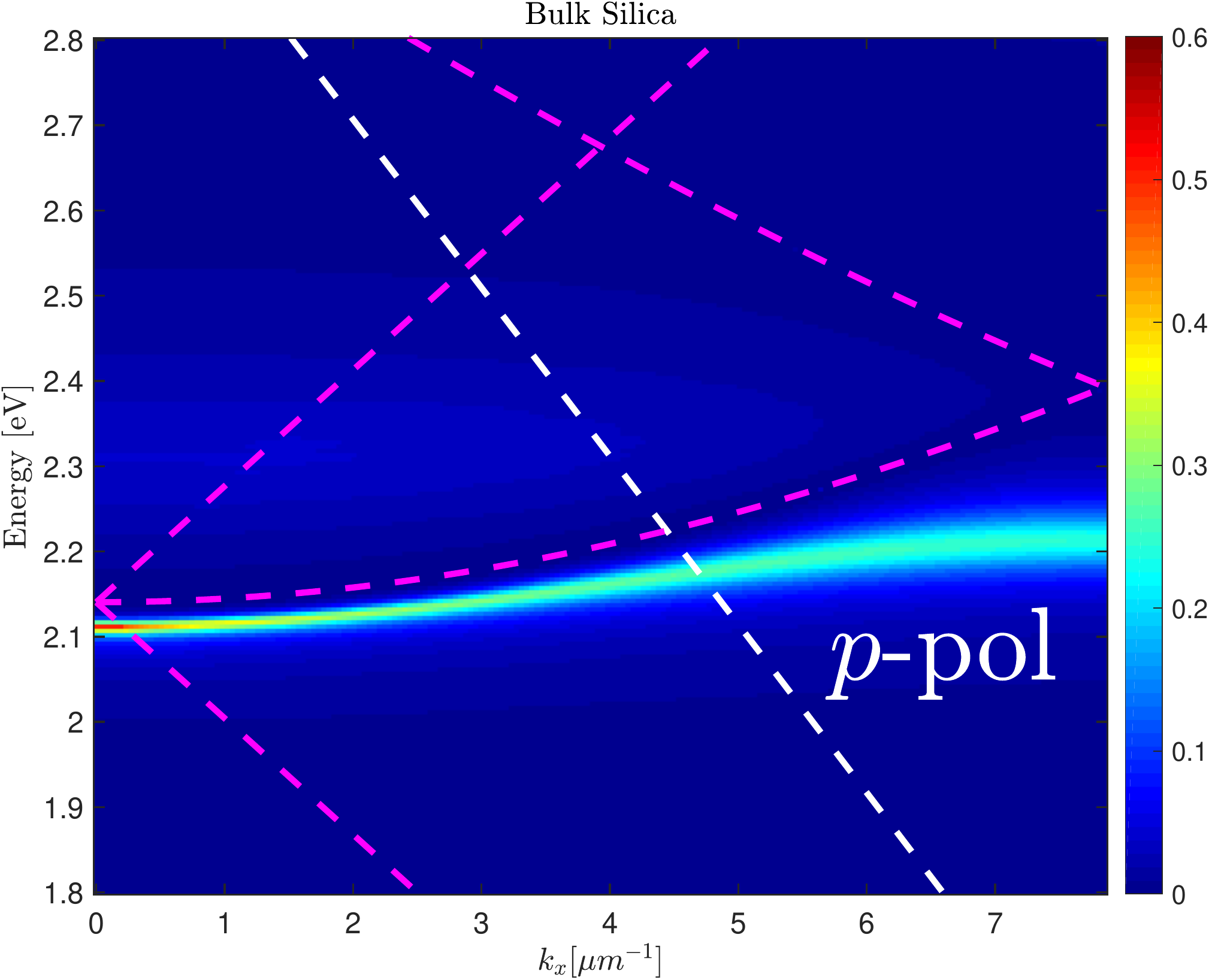}
    \end{subfigure}
    \hfill
    \begin{subfigure}{0.32\linewidth}
    \centering
    \includegraphics[width=\linewidth]{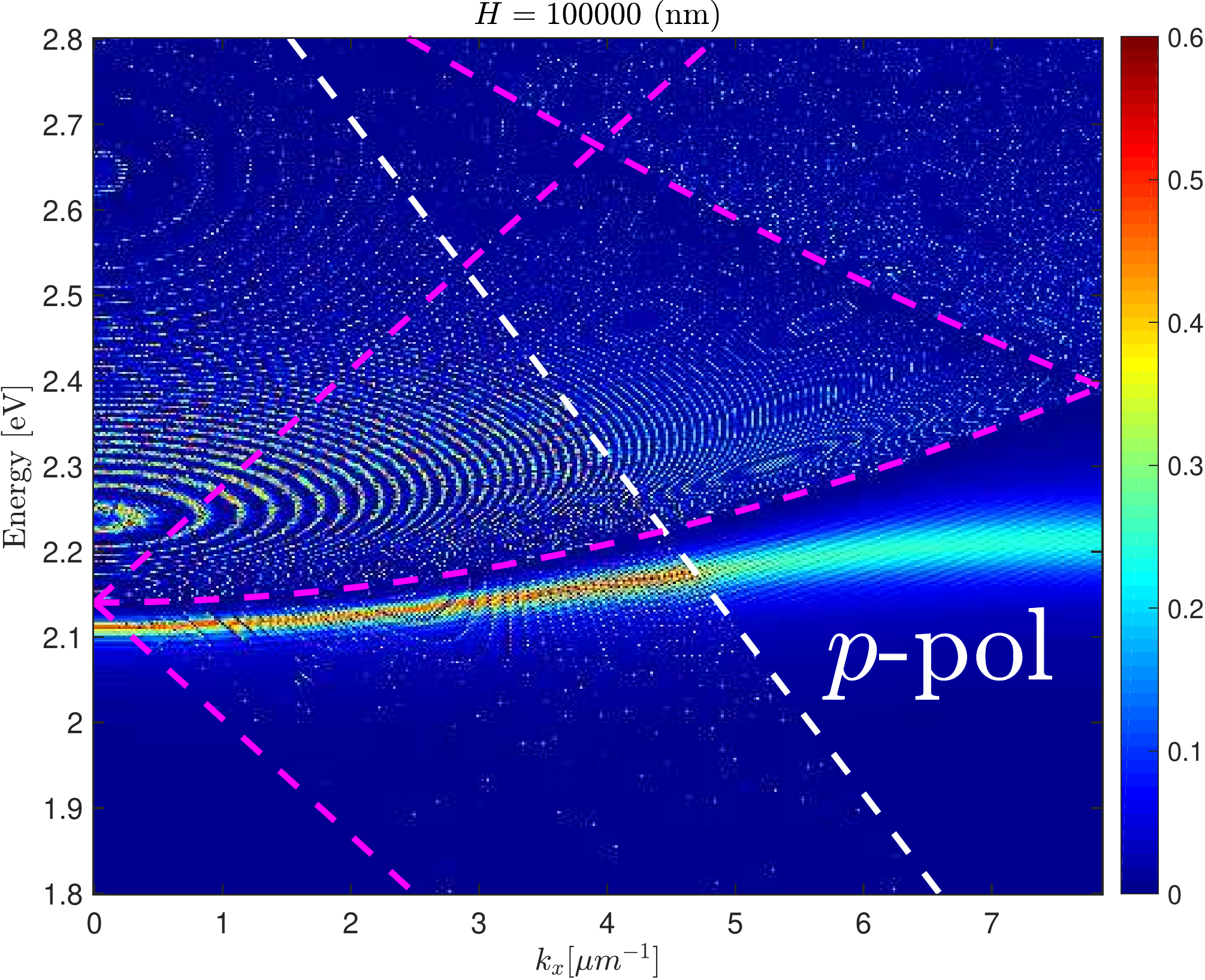}
    \end{subfigure}
    \hfill
    \begin{subfigure}{0.32\linewidth}
    \centering
    \includegraphics[width=\linewidth]{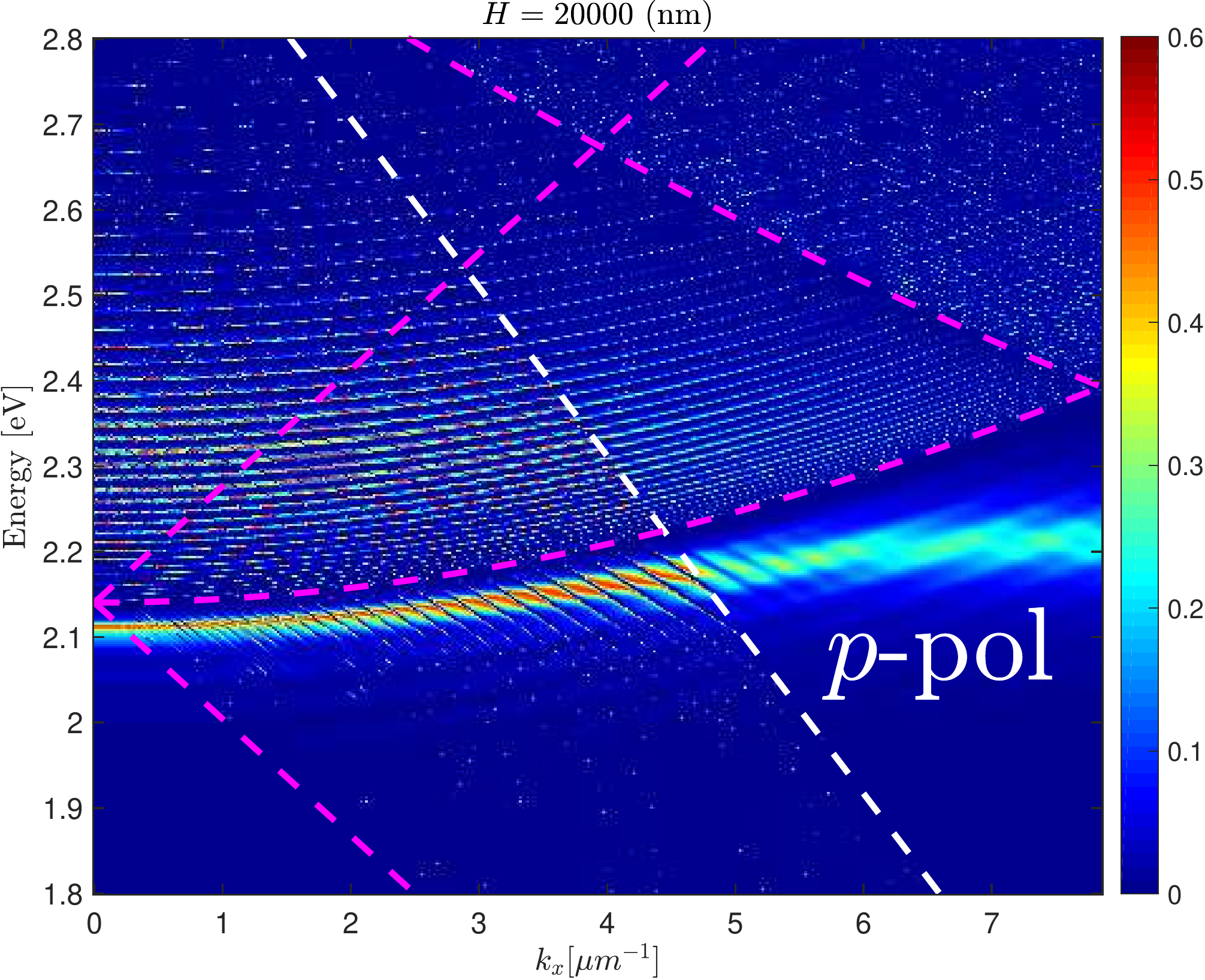}
    \end{subfigure}
    \caption{Absorption spectra in $s$ and $p$ polarizations of the plasmonic lattice of silver nanodisks in bulk silica, the center of $100\mu m$ and $20 \mu m$ membrane silica waveguides.}
    \label{fig:A_inf_100_20}
\end{figure}

In this section, we consider the same lattice of silver nanodisks inside a silica membrane waveguide but vary its thickness. First of all, we compare spectra of a lattice in a bulk silica (infinite thickness) with a case of very wide waveguides ($100 \mu m$ and $20 \mu m$). As can be seen from Fig. \ref{fig:A_inf_100_20} spectra in both $s$ and $p$ polarizations are very similar and does not have significant discrepancies. In a $100 \mu m$ waveguide finite thickness results in very dense lines of the waveguide and Fabry-Perot modes. Complicated patterns observed in the spectra, including characteristic concentric ellipses, are a result of Moire fringes originating from the crossing of very high-$Q$ resonances and computational grid. After the blurring, which is peculiar to any experiment, all these fringes will disappear and we will see the spectrum very similar to the case of bulk silica.

When the waveguide has the thickness of $20 \mu m$ Fabry-Perot and waveguide resonances become much more sparse and pronounced. They 'cut' both localized and lattice plasmon resonances, but the general picture does not change.

\begin{figure}[h]
    \centering
    \begin{subfigure}{0.32\linewidth}
    \centering
    \includegraphics[width=\linewidth]{inf_s.pdf}
    \end{subfigure}
    \hfill
    \begin{subfigure}{0.32\linewidth}
    \centering
    \includegraphics[width=\linewidth]{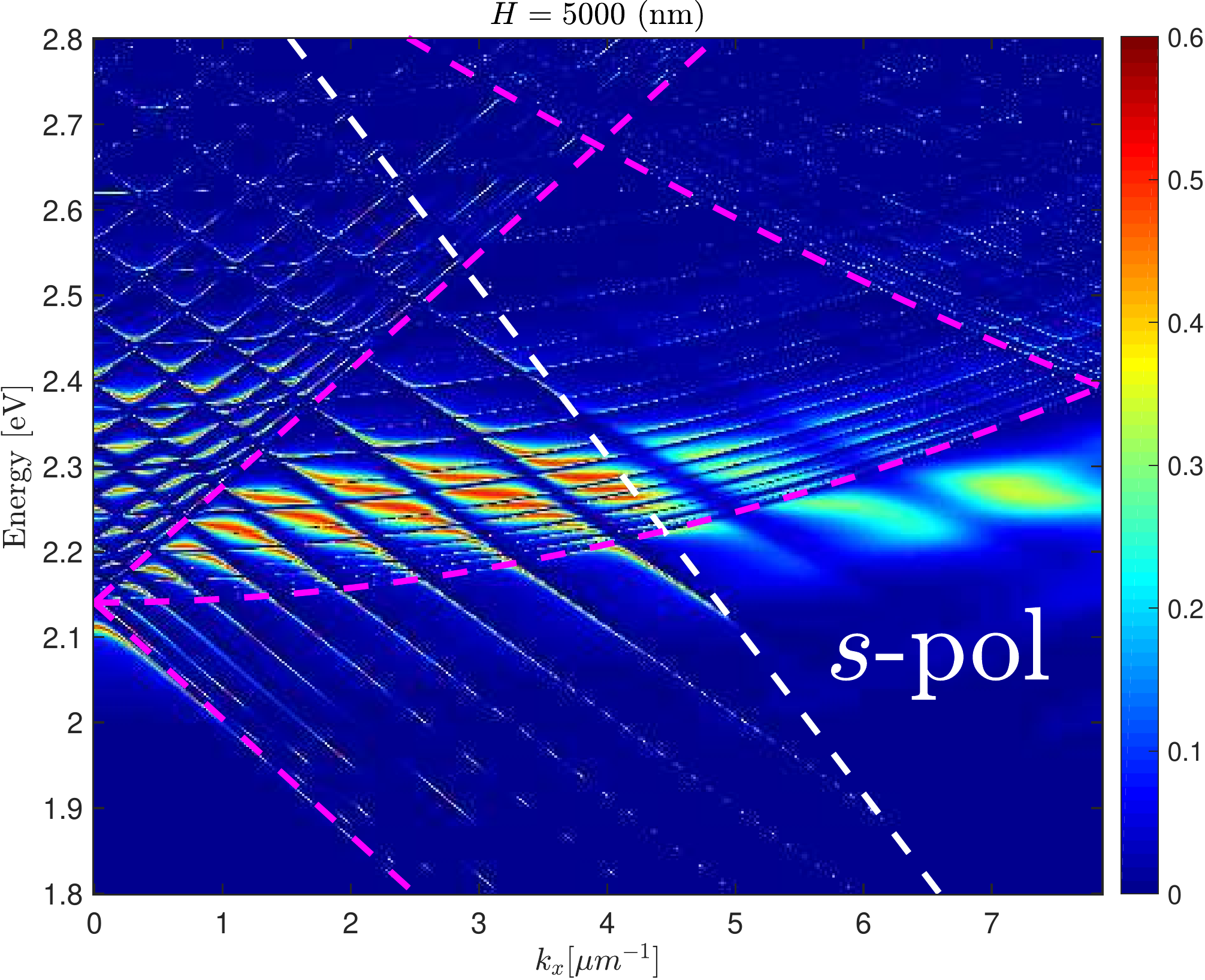}
    \end{subfigure}
    \hfill
    \begin{subfigure}{0.32\linewidth}
    \centering
    \includegraphics[width=\linewidth]{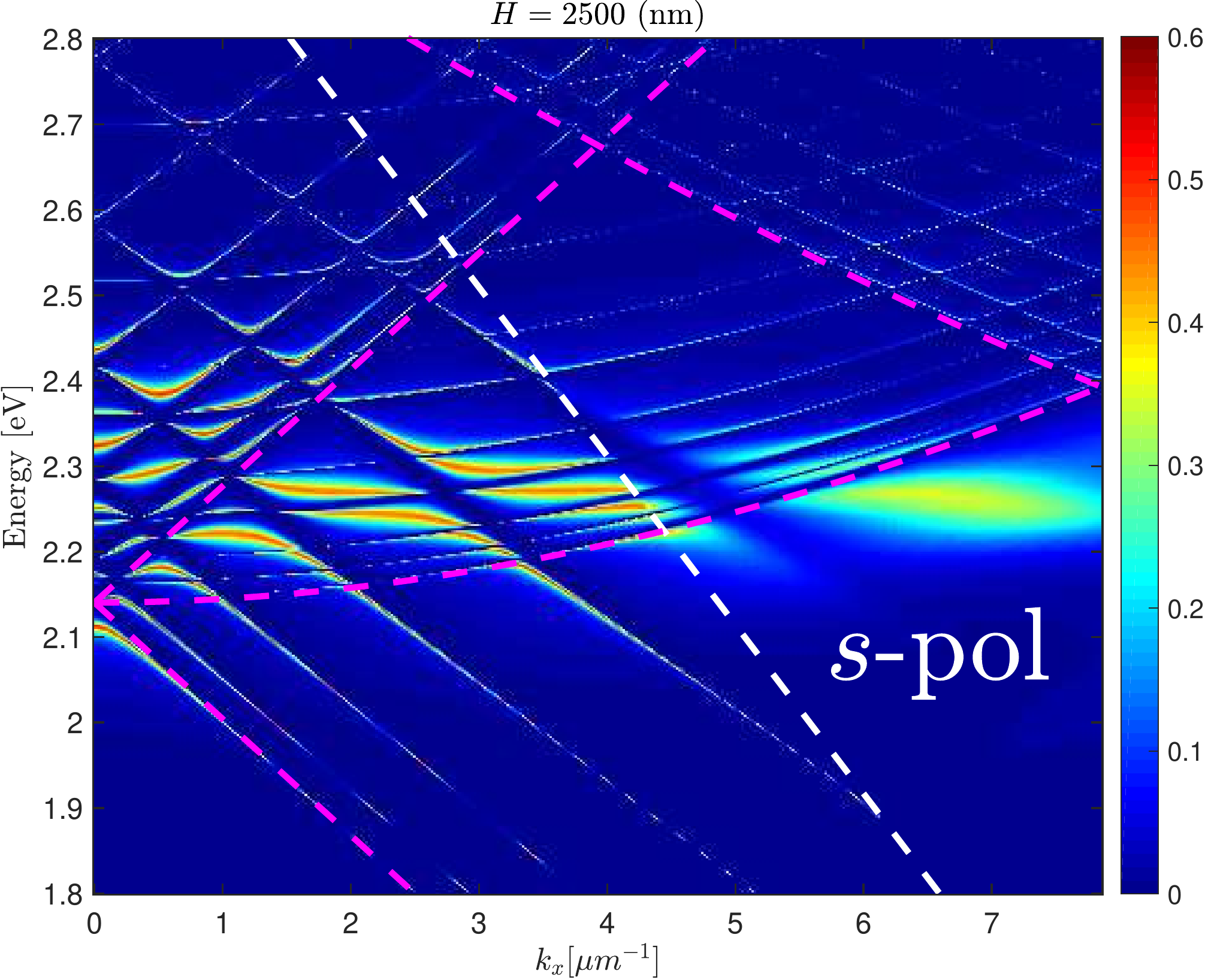}
    \end{subfigure}
    \centering
    \begin{subfigure}{0.32\linewidth}
    \centering
    \includegraphics[width=\linewidth]{inf_p.pdf}
    \end{subfigure}
    \hfill
    \begin{subfigure}{0.32\linewidth}
    \centering
    \includegraphics[width=\linewidth]{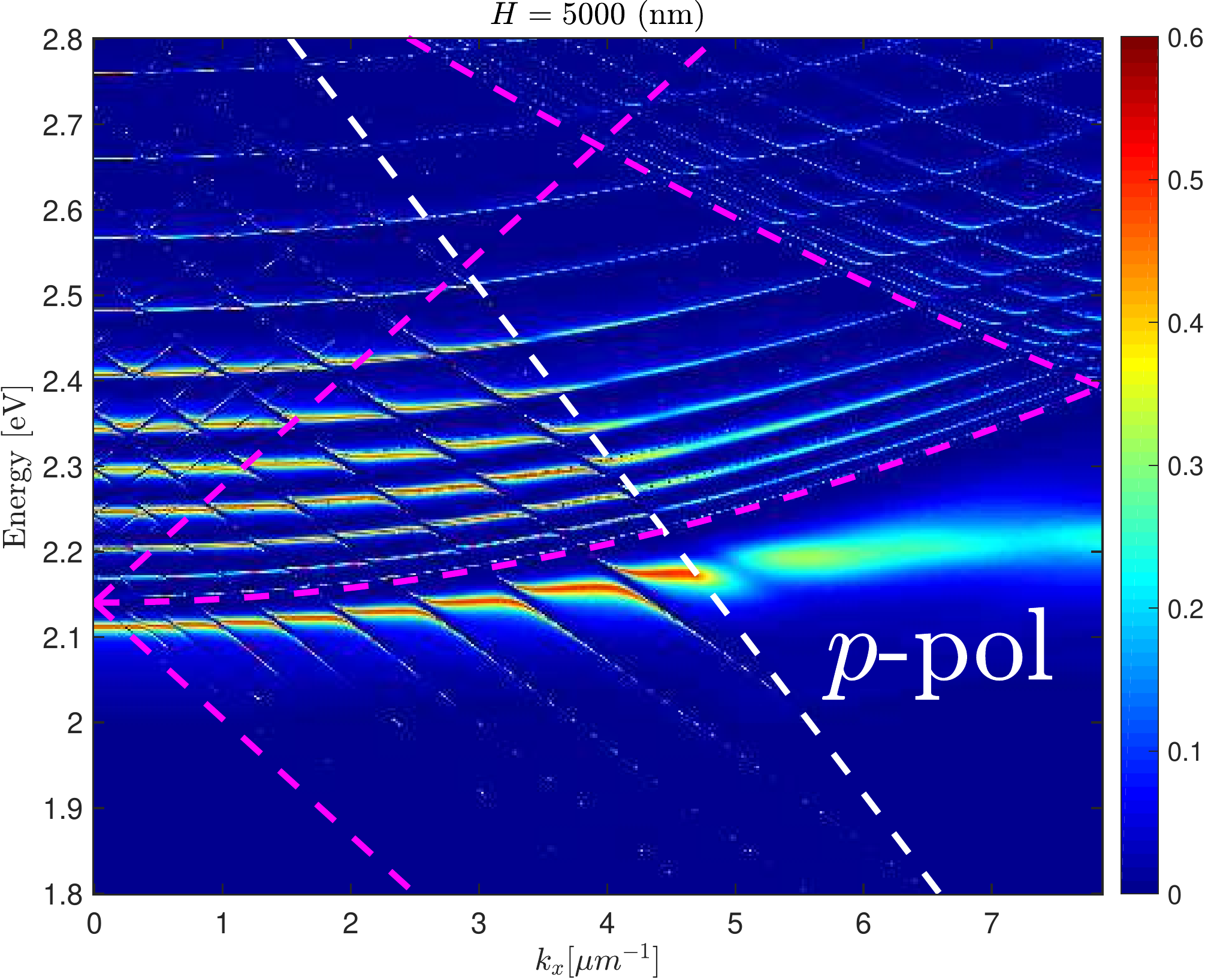}
    \end{subfigure}
    \hfill
    \begin{subfigure}{0.32\linewidth}
    \centering
    \includegraphics[width=\linewidth]{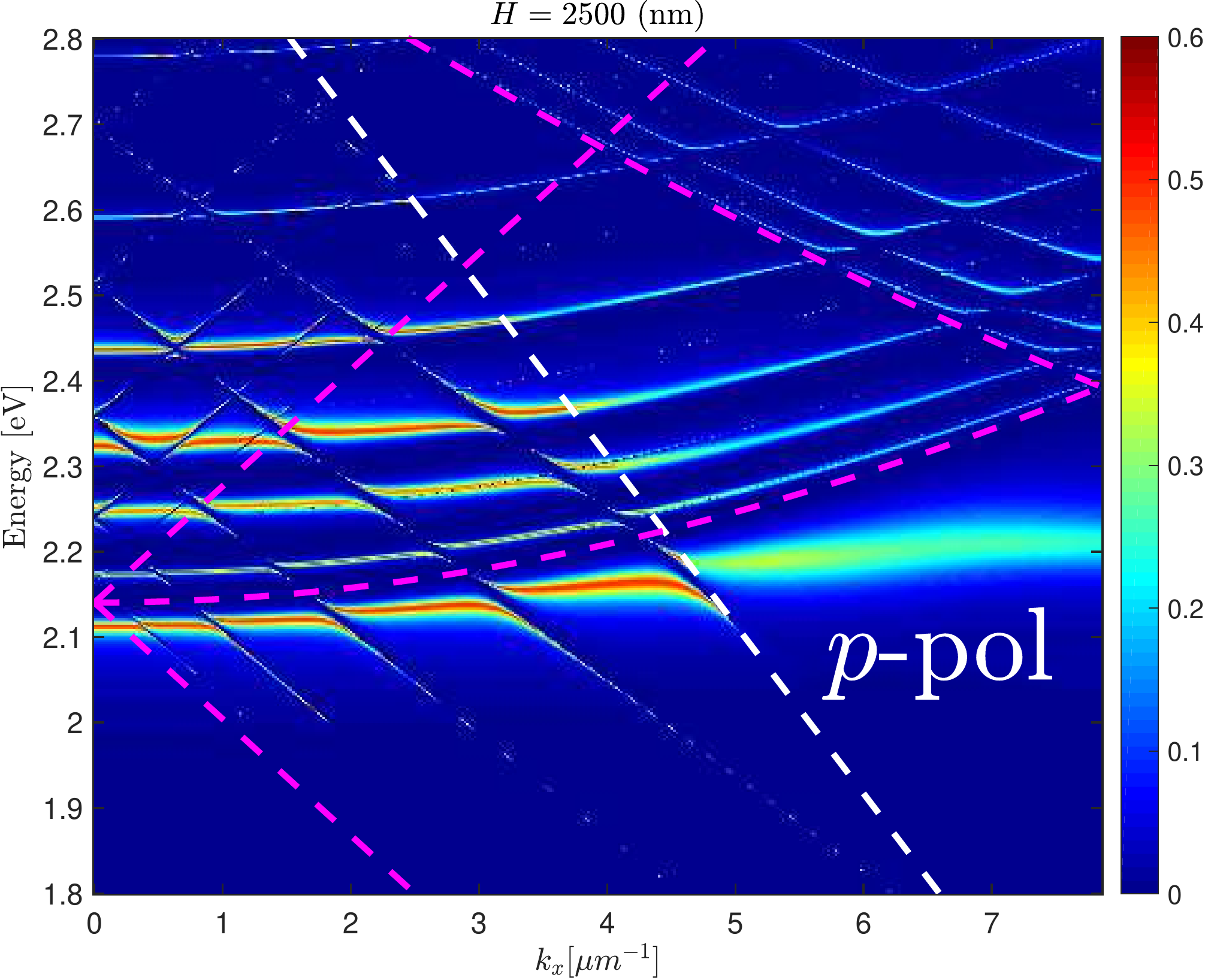}
    \end{subfigure}
    \caption{Absorption spectra in $s$ and $p$ polarizations of the plasmonic lattice of silver nanodisks in bulk silica, the center of $5\mu m$ and $2.5 \mu m$ membrane silica waveguides.}
    \label{fig:A_inf_5_2.5}
\end{figure}

Next, we consider and transition between ultra-wide waveguides and conventional ones. In Fig. \ref{fig:A_inf_5_2.5} extinctions spectra of $5 \mu m$ and $2.5 \mu m$ waveguides are presented. Now, there is a countable, but still a large number of resonances and their hybridizations. Lattice plasmon resonance is still distinguishable, whereas localized resonance is strongly coupled to plenty of guided modes.

\begin{figure}[h]
    \centering
    \begin{subfigure}{0.32\linewidth}
    \centering
    \includegraphics[width=\linewidth]{inf_s.pdf}
    \end{subfigure}
    \hfill
    \begin{subfigure}{0.32\linewidth}
    \centering
    \includegraphics[width=\linewidth]{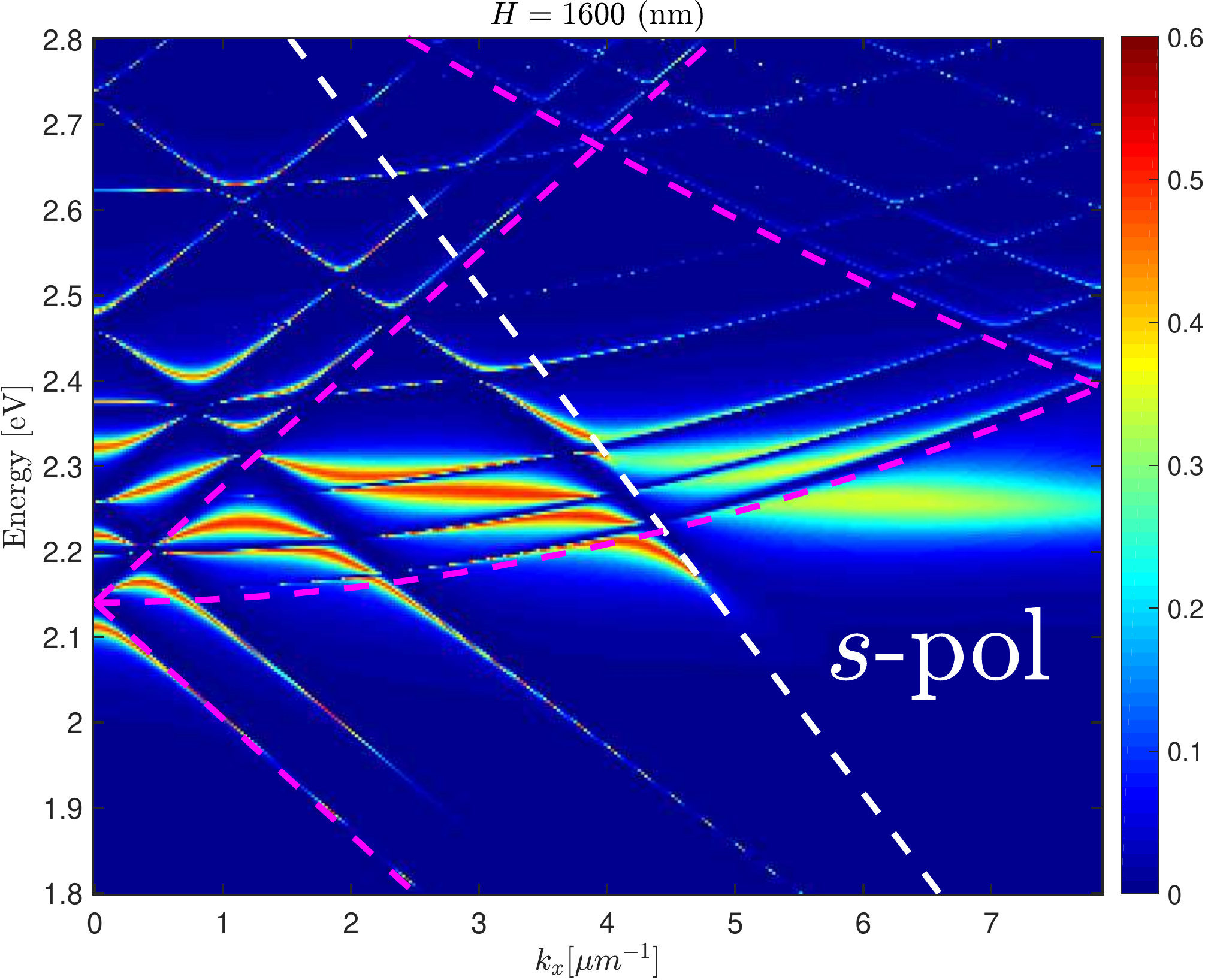}
    \end{subfigure}
    \hfill
    \begin{subfigure}{0.32\linewidth}
    \centering
    \includegraphics[width=\linewidth]{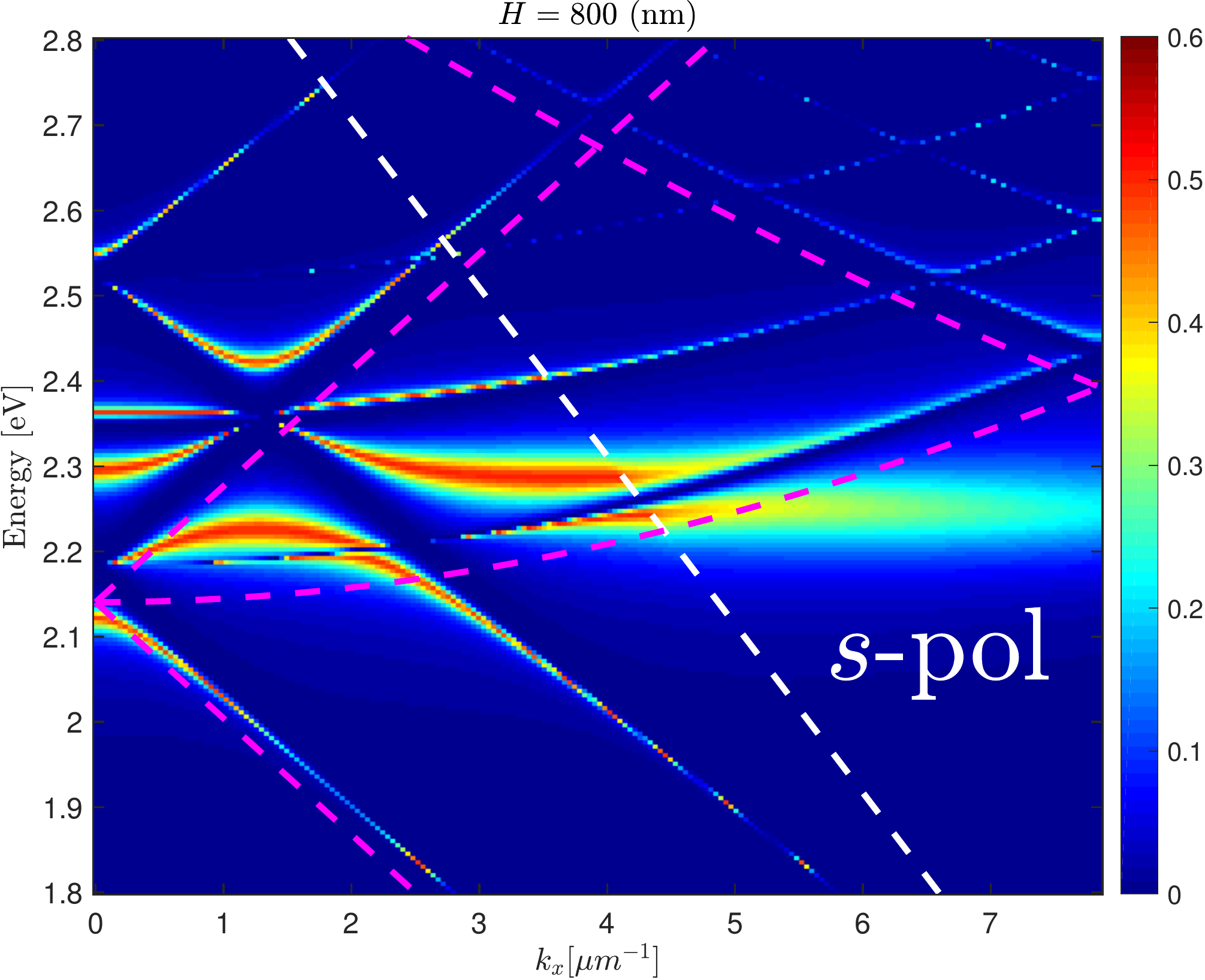}
    \end{subfigure}
    \centering
    \begin{subfigure}{0.32\linewidth}
    \centering
    \includegraphics[width=\linewidth]{inf_p.pdf}
    \end{subfigure}
    \hfill
    \begin{subfigure}{0.32\linewidth}
    \centering
    \includegraphics[width=\linewidth]{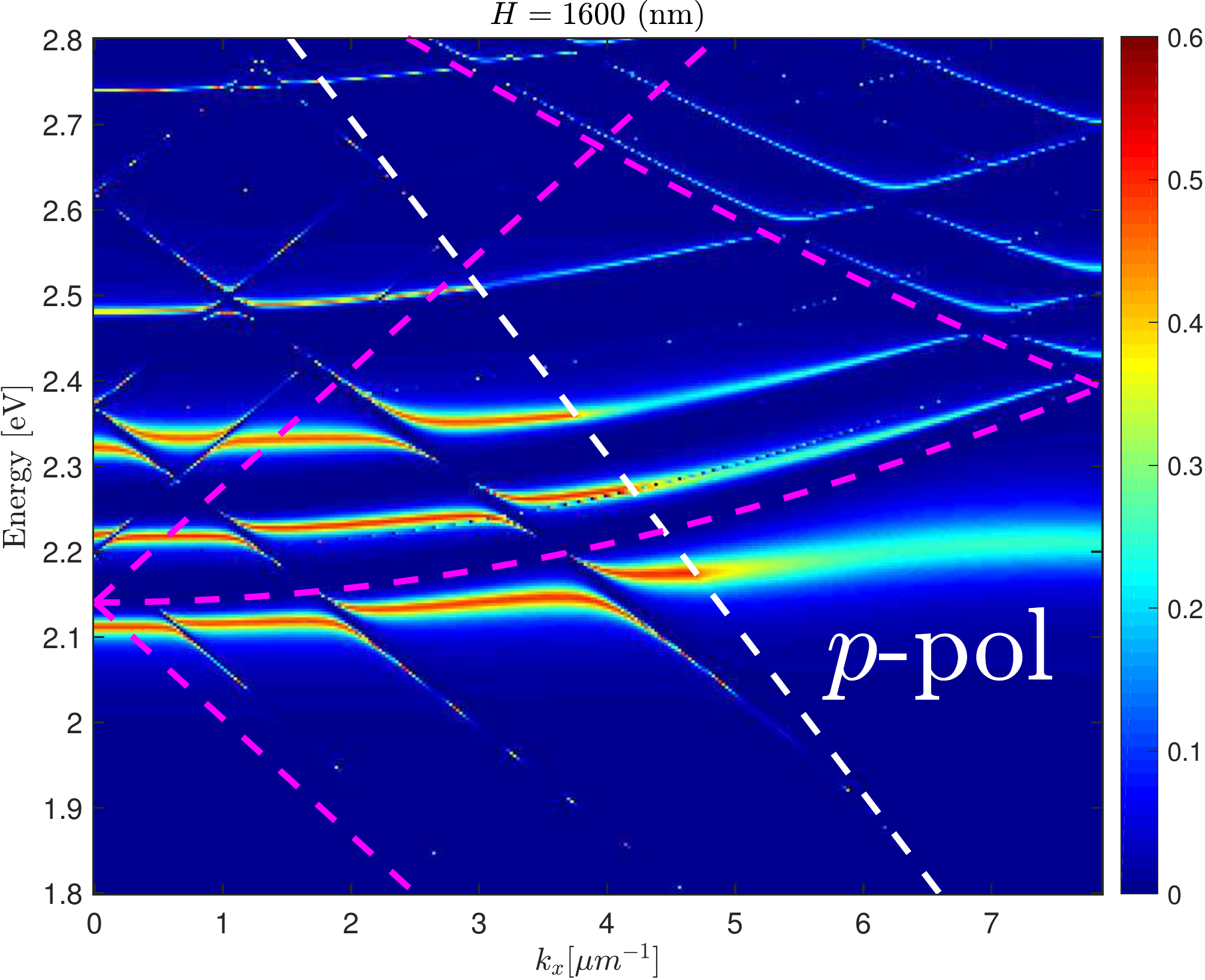}
    \end{subfigure}
    \hfill
    \begin{subfigure}{0.32\linewidth}
    \centering
    \includegraphics[width=\linewidth]{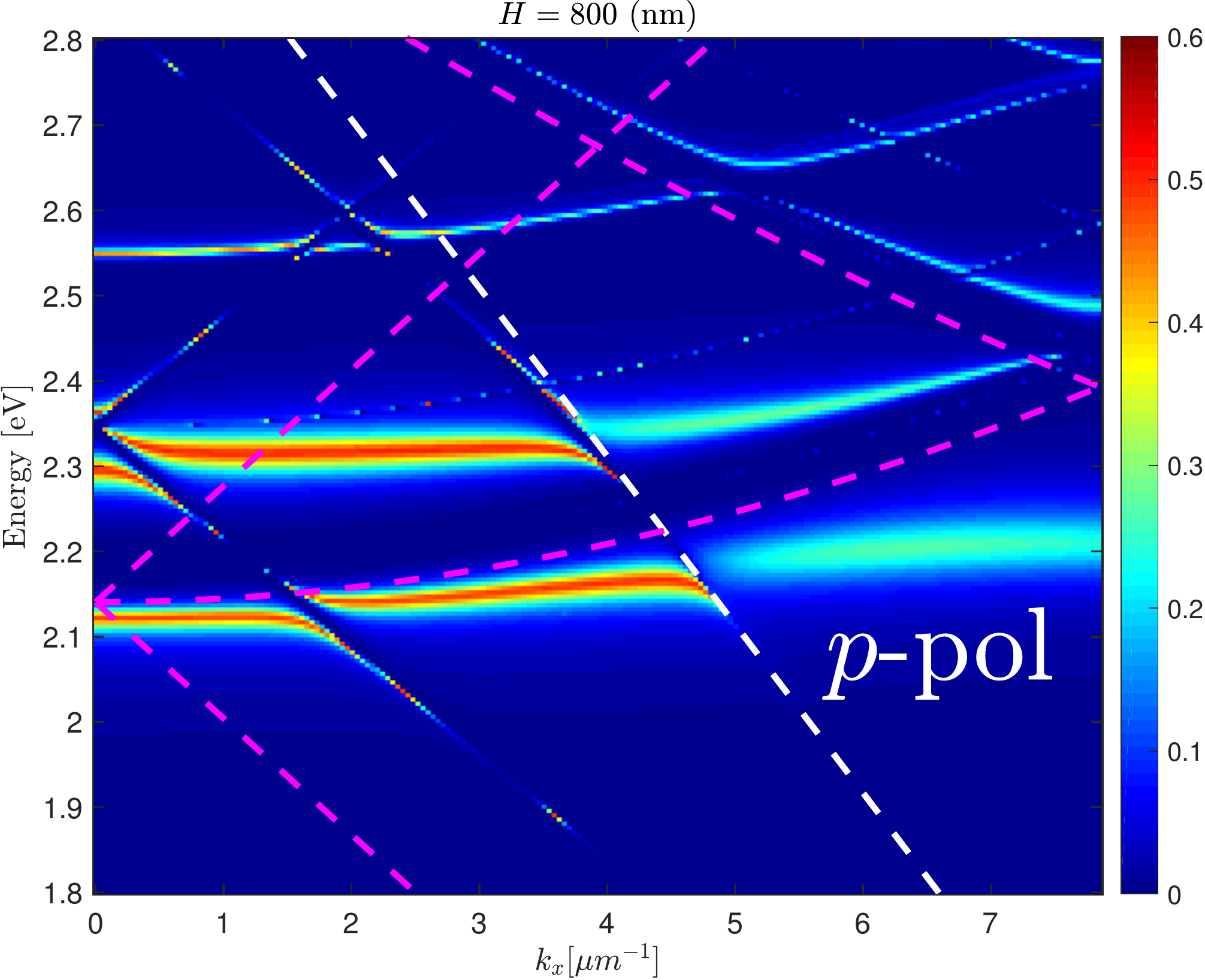}
    \end{subfigure}
    \caption{Absorption spectra in $s$ and $p$ polarizations of the plasmonic lattice of silver nanodisks in bulk silica, the center of $1.6\mu m$ and $0.8 \mu m$ membrane silica waveguides.}
    \label{fig:A_inf_1.6_0.8}
\end{figure}

For relatively thin waveguides ($1.6 \mu m$ and $0.8 \mu m$, see Fig. \ref{fig:A_inf_1.6_0.8}) several other features emerge. First of all, it is seen that Rayleigh anomalies and resonances associated with them slightly move upwards because of a slight increase of an average permittivity. Another feature, which becomes observable is that hybridized plasmon-photonic resonances are located below the eigenmodes of the waveguide, when $\mathrm{Re}\alpha(\omega)>0$ (red zone), whereas in the blue zone of the LSPR ($\mathrm{Re}\alpha(\omega)<0$) they swap places. It is another demonstration of well known resonances' splitting.

\begin{figure}[h]
    \centering
    \begin{subfigure}{0.32\linewidth}
    \centering
    \includegraphics[width=\linewidth]{inf_s.pdf}
    \end{subfigure}
    \hfill
    \begin{subfigure}{0.32\linewidth}
    \centering
    \includegraphics[width=\linewidth]{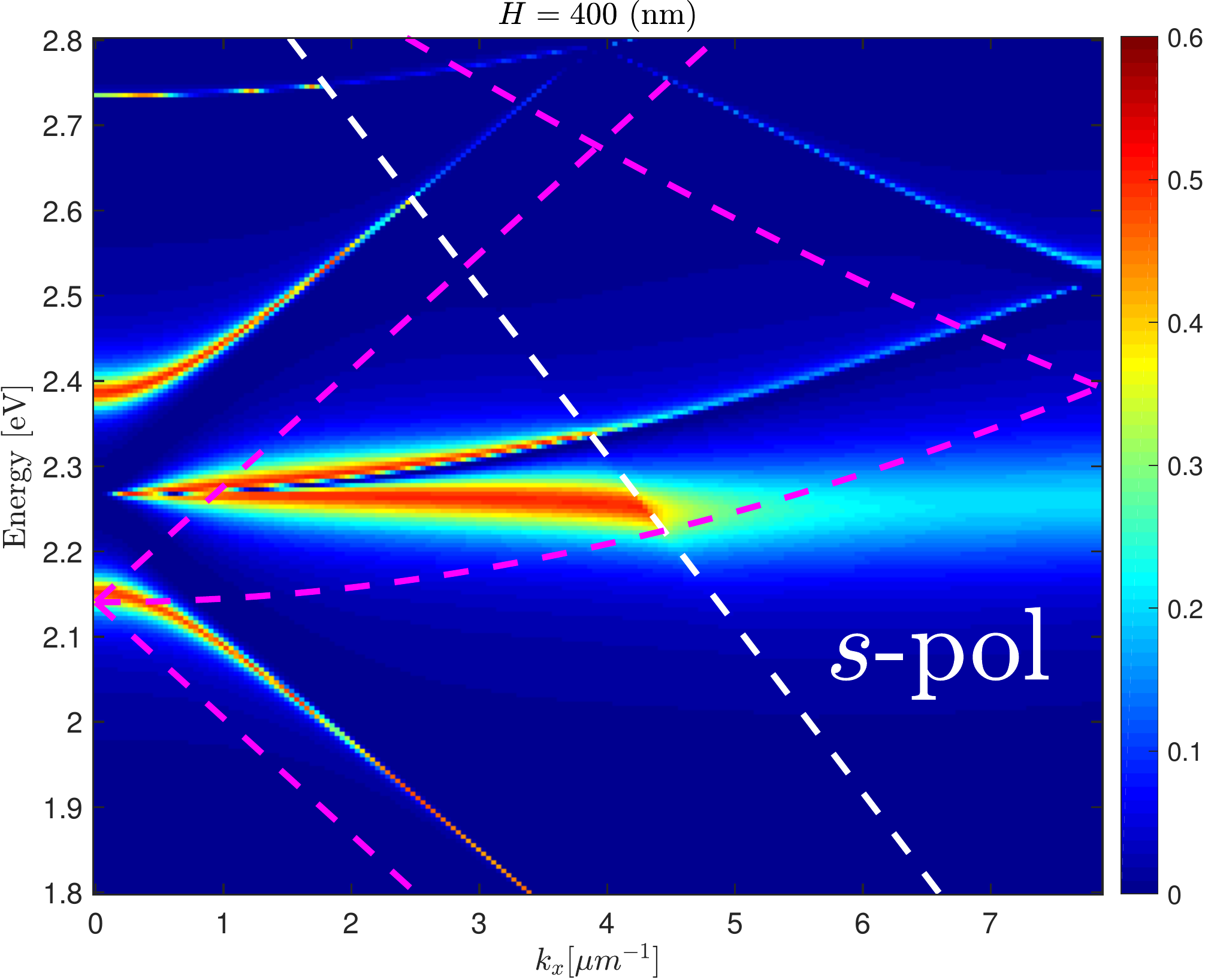}
    \end{subfigure}
    \hfill
    \begin{subfigure}{0.32\linewidth}
    \centering
    \includegraphics[width=\linewidth]{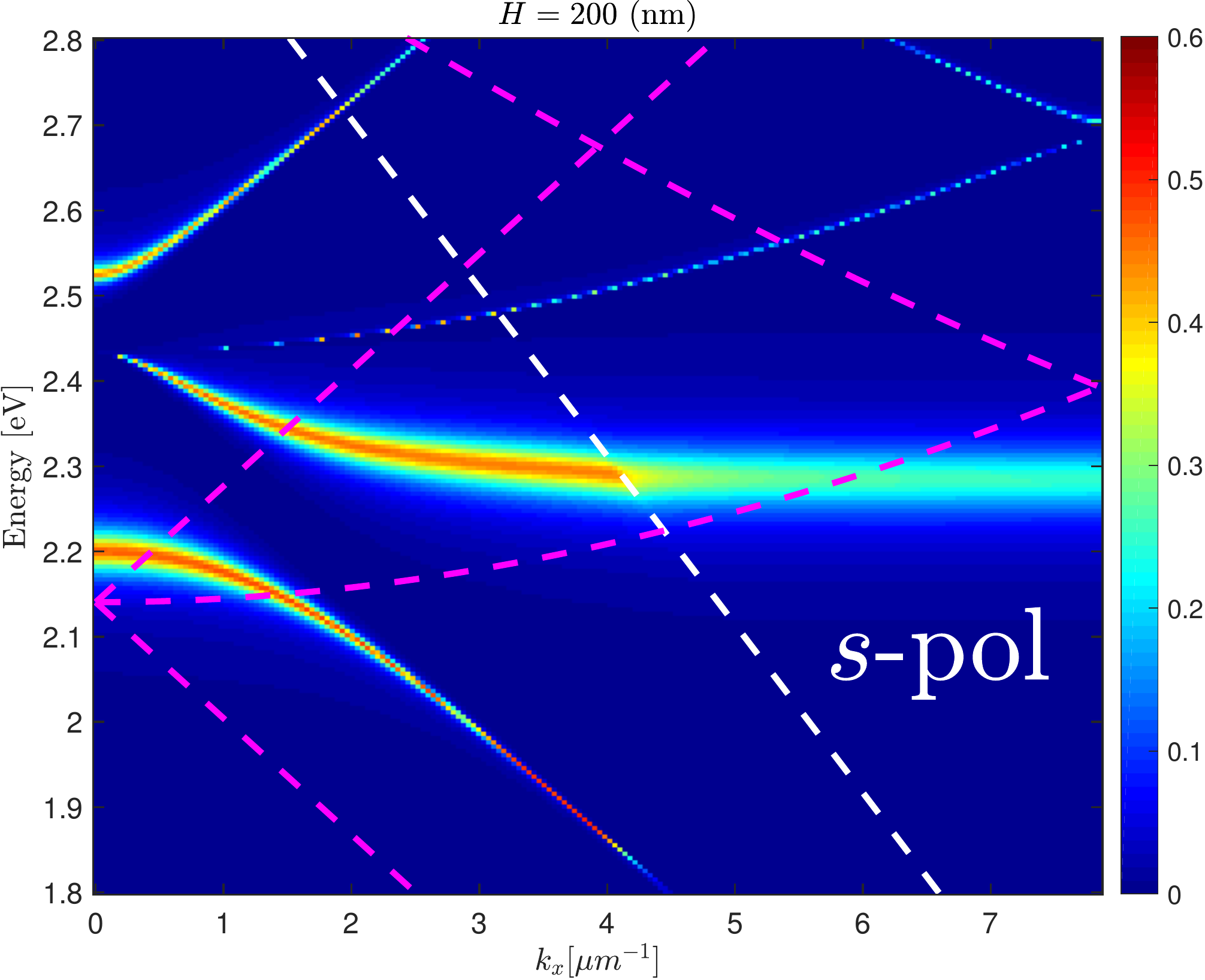}
    \end{subfigure}
    \centering
    \begin{subfigure}{0.32\linewidth}
    \centering
    \includegraphics[width=\linewidth]{inf_p.pdf}
    \end{subfigure}
    \hfill
    \begin{subfigure}{0.32\linewidth}
    \centering
    \includegraphics[width=\linewidth]{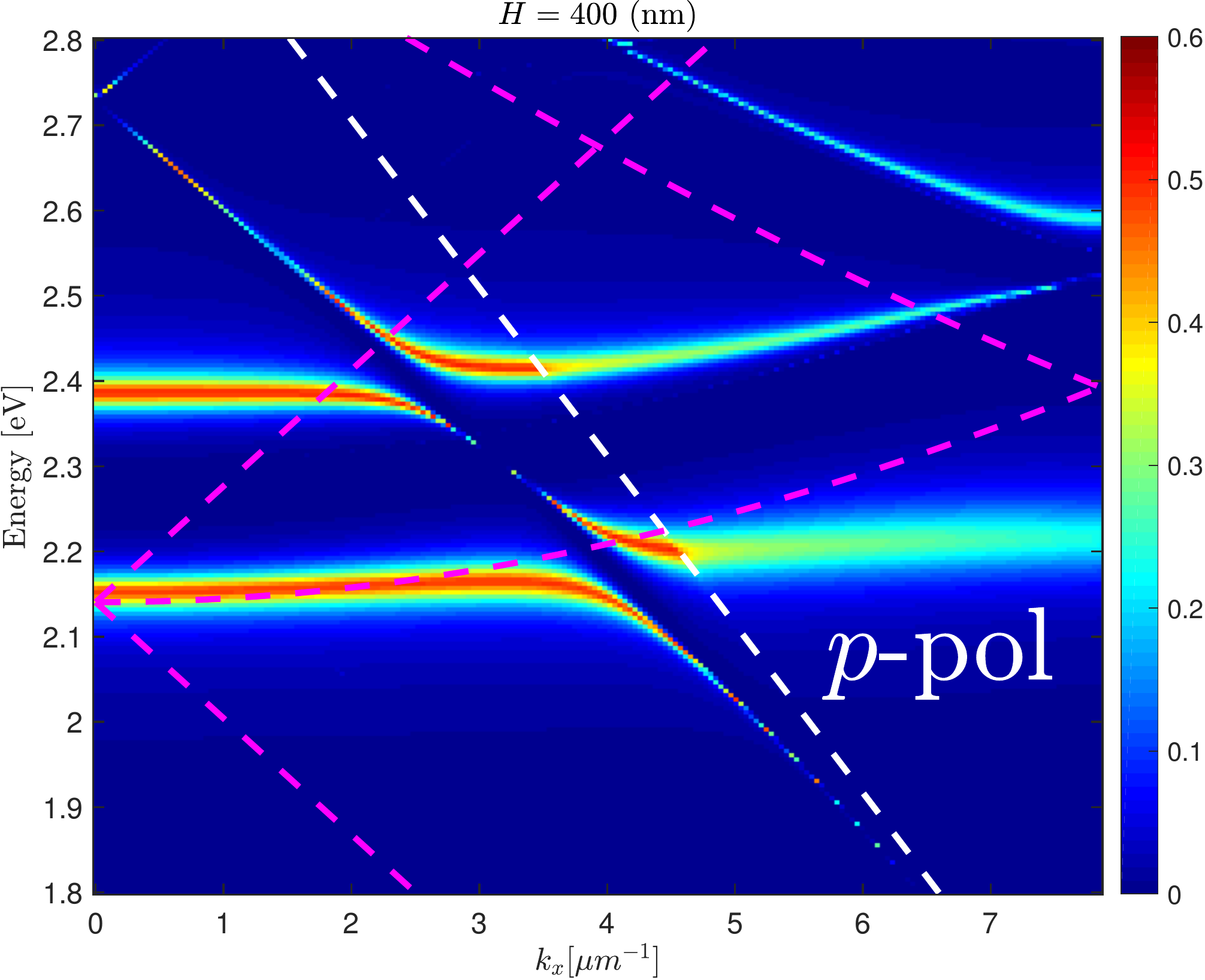}
    \end{subfigure}
    \hfill
    \begin{subfigure}{0.32\linewidth}
    \centering
    \includegraphics[width=\linewidth]{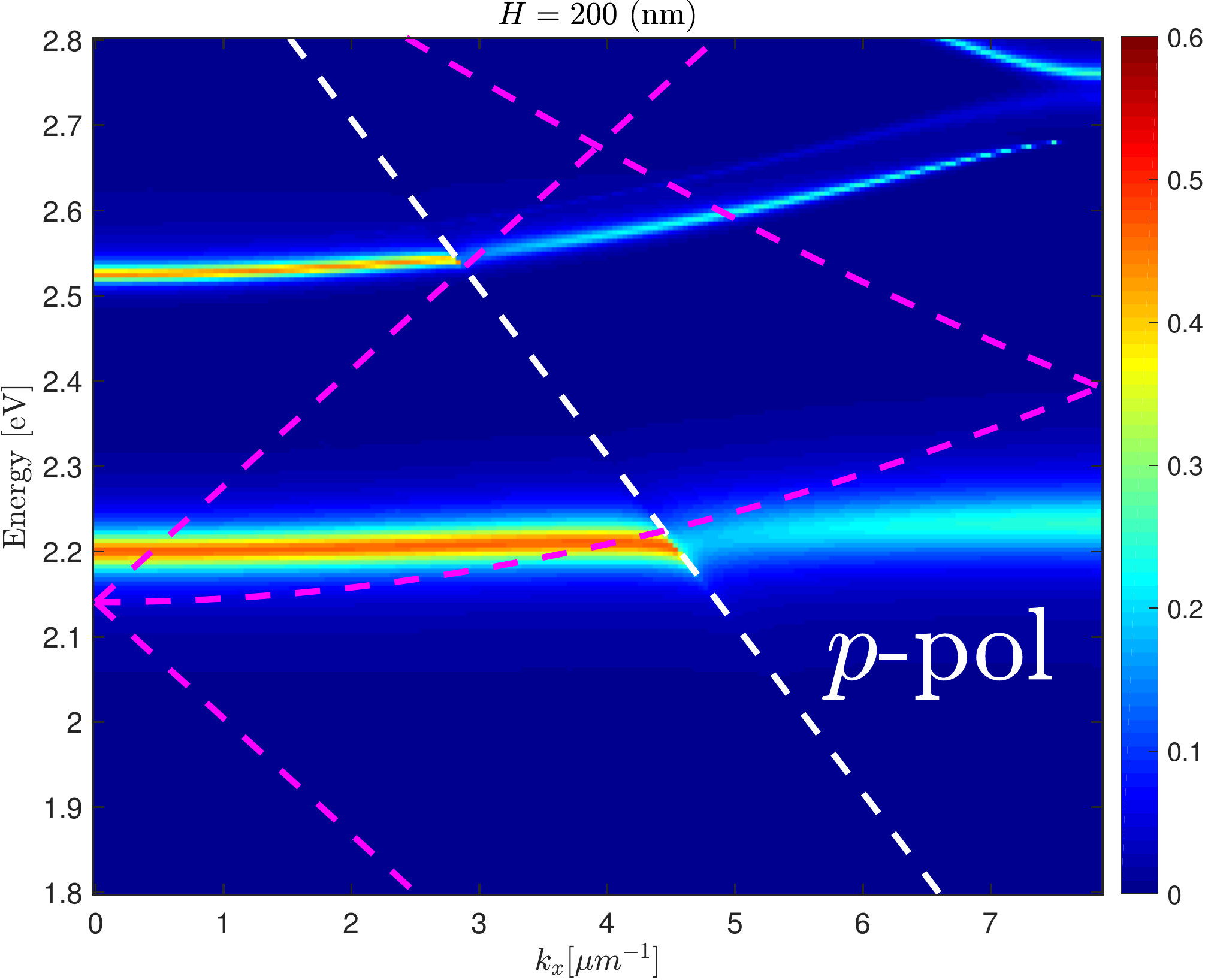}
    \end{subfigure}
    \caption{Absorption spectra in $s$ and $p$ polarizations of the plasmonic lattice of silver nanodisks in bulk silica, the center of $0.4\mu m$ and $0.2 \mu m$ membrane silica waveguides.}
    \label{fig:A_inf_0.4_0.2}
\end{figure}

Finally, we observe typical waveguides of $400 nm$ and $200 nm$ thickness (see Fig. \ref{fig:A_inf_0.4_0.2}). The main feature, which is observed immediately is that lattice resonance became almost dispersionless. It is explained by the fact, that the action of Rayleigh anomaly is strongly suppressed by the existence of two interfaces in close proximity of a lattice. That is why plasmonic particles are coupled to each other only via near field and waveguide modes and as a result, there is no conventional lattice plasmon resonance and only localized resonance, as well as resonances hybridized with photonic guided modes, are observed. An interesting deflection of LSPR on an air Rayleigh anomaly is observed in $p$ polarization. This resonance can be described according to the paper \cite{akimov2011optical}.

\clearpage
\newpage

\chapter{Comparison with experimental papers}

Theoretical calculations are very interesting by themselves. However, the final aim of any calculation is to explain experimental results or predict them. For this reason, here we compare our calculations with two experimental papers.

\section{Homogeneous environment}

\begin{figure}[h]
    \centering
    \begin{subfigure}[b]{0.41\linewidth}
    \centering
    \includegraphics[width=\textwidth]{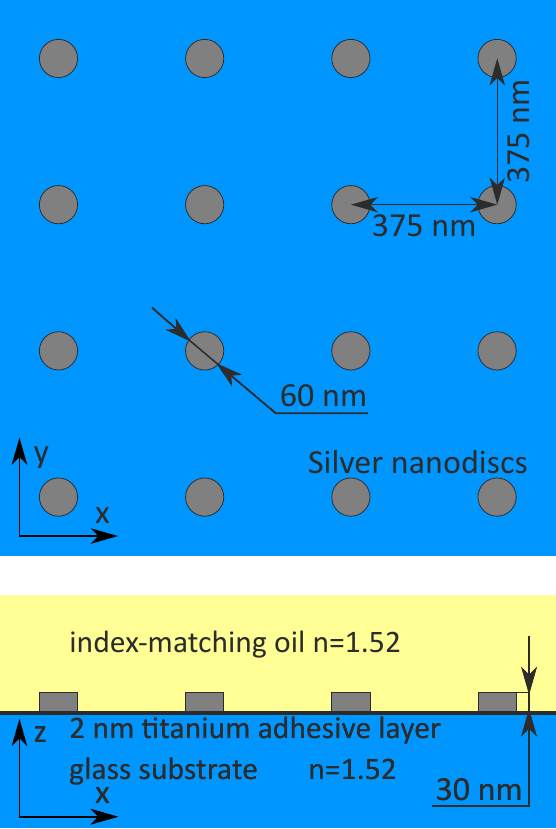}
    \end{subfigure}
    \hfill
    \begin{subfigure}[b]{0.58\linewidth}
    \centering
    \includegraphics[width=\textwidth]{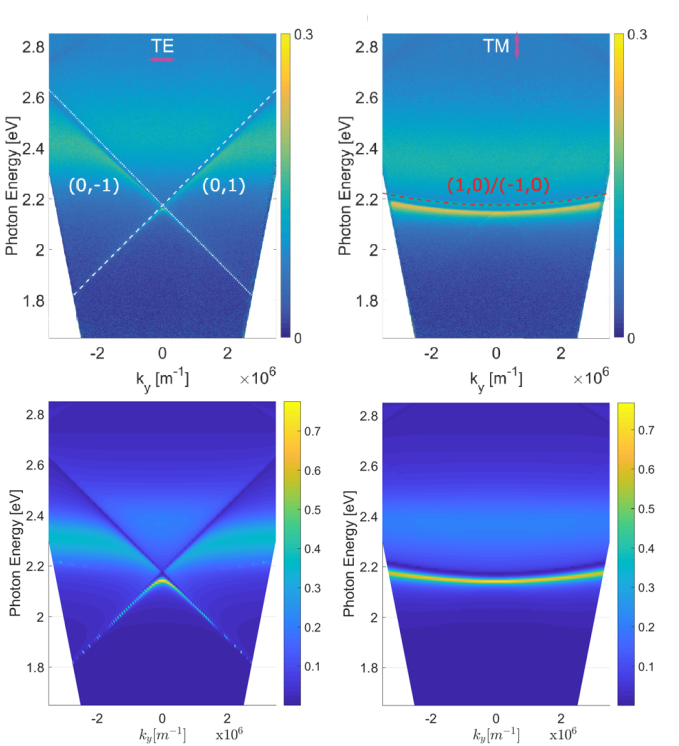}
    \end{subfigure}
    \caption{(Left) Schematic of the square lattice of silver nanodisks in an optically homogeneous environment. (Right) Experimental (top) and theoretical (bottom) extinction spectra of the lattice in $s$ and $p$ polarizations.}
    \label{fig:square_lattice}
\end{figure}

\begin{figure}[h]
    \centering
    \begin{subfigure}[b]{0.41\linewidth}
    \centering
    \includegraphics[width=\textwidth]{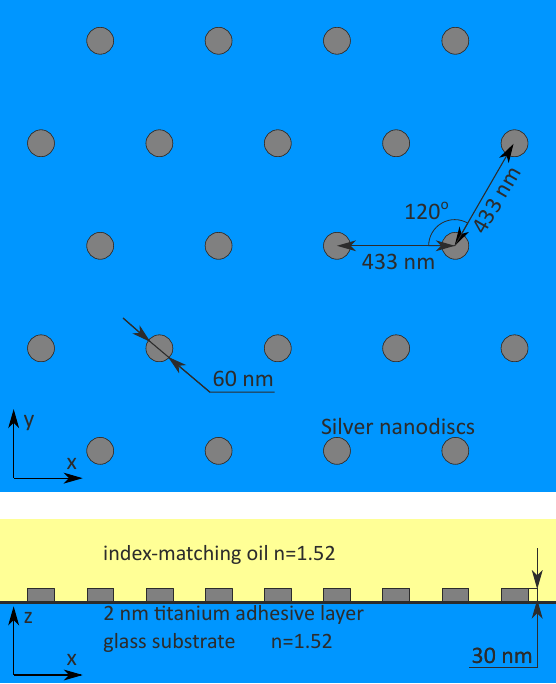}
    \end{subfigure}
    \hfill
    \begin{subfigure}[b]{0.58\linewidth}
    \centering
    \includegraphics[width=\textwidth]{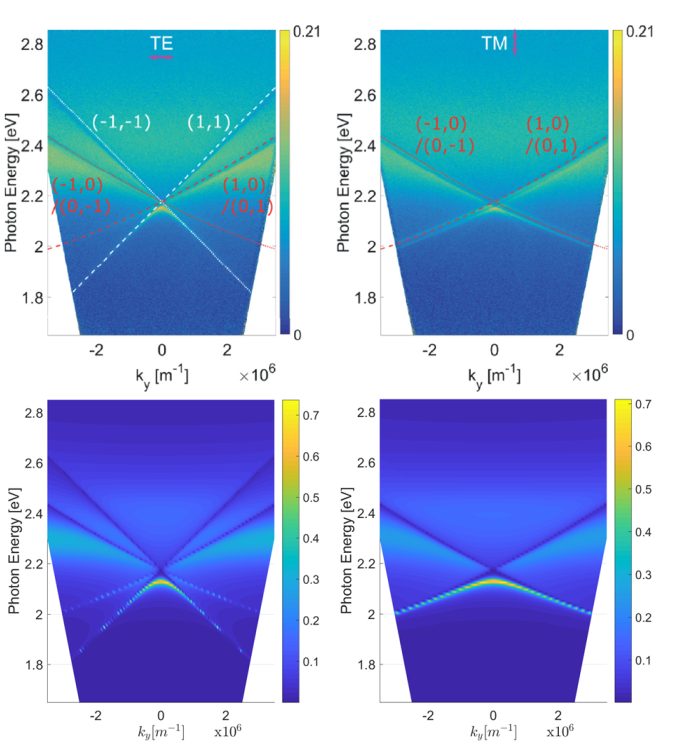}
    \end{subfigure}
    \caption{(Left) Schematic of the hexagonal lattice of silver nanodisks in an optically homogeneous environment. (Right) Experimental (top) and theoretical (bottom) extinction spectra of the lattice in $s$ and $p$ polarizations.}
    \label{fig:hexagonal_lattice}
\end{figure}

The first article, which we observe \cite{guo2017}, considers lattices with different unit cells in an optically homogeneous medium.
Experimentally such lattices are fabricated on glass substrates ($n\approx1.52$) covered by 2 nm titanium as an adhesive layer. Lattices themselves are constituted by silver nanodisks of $60$ (nm) diameter and $30$ (nm) height and are embedded into the index-matching oil (see Fig. \ref{fig:square_lattice},\ref{fig:hexagonal_lattice}).

Angle-resolved extinction spectra of both square and hexagonal lattices (see Fig. \ref{fig:square_lattice},\ref{fig:hexagonal_lattice}) show Rayleigh anomalies, which correspond to dips in the extinction spectra. Slightly below them, lattice plasmon resonances are observed. Generally speaking, both the structure and its spectra are very similar to the case considered in the previous chapter and does not require any additional analysis.

In our calculations, we did not take into account the adhesive layer. Although, titanium is very dissipative material and even such a thin layer can strongly modify LSPR it can not qualitatively change spectra and their inherent features. In this way, experimental and theoretical spectra turn out to be very similar (see Fig. \ref{fig:square_lattice}).
All the resonances and peculiarities are observed at the same frequencies. The discrepancy in magnitude of extinction is probably associated with non-ideal periodicity, the difference between the particles or diffraction in an experiment.

\section{Lattice on a waveguide}
\begin{figure}[h]
    \centering
    \begin{subfigure}{0.282\linewidth}
    \centering
    \includegraphics[width=\textwidth]{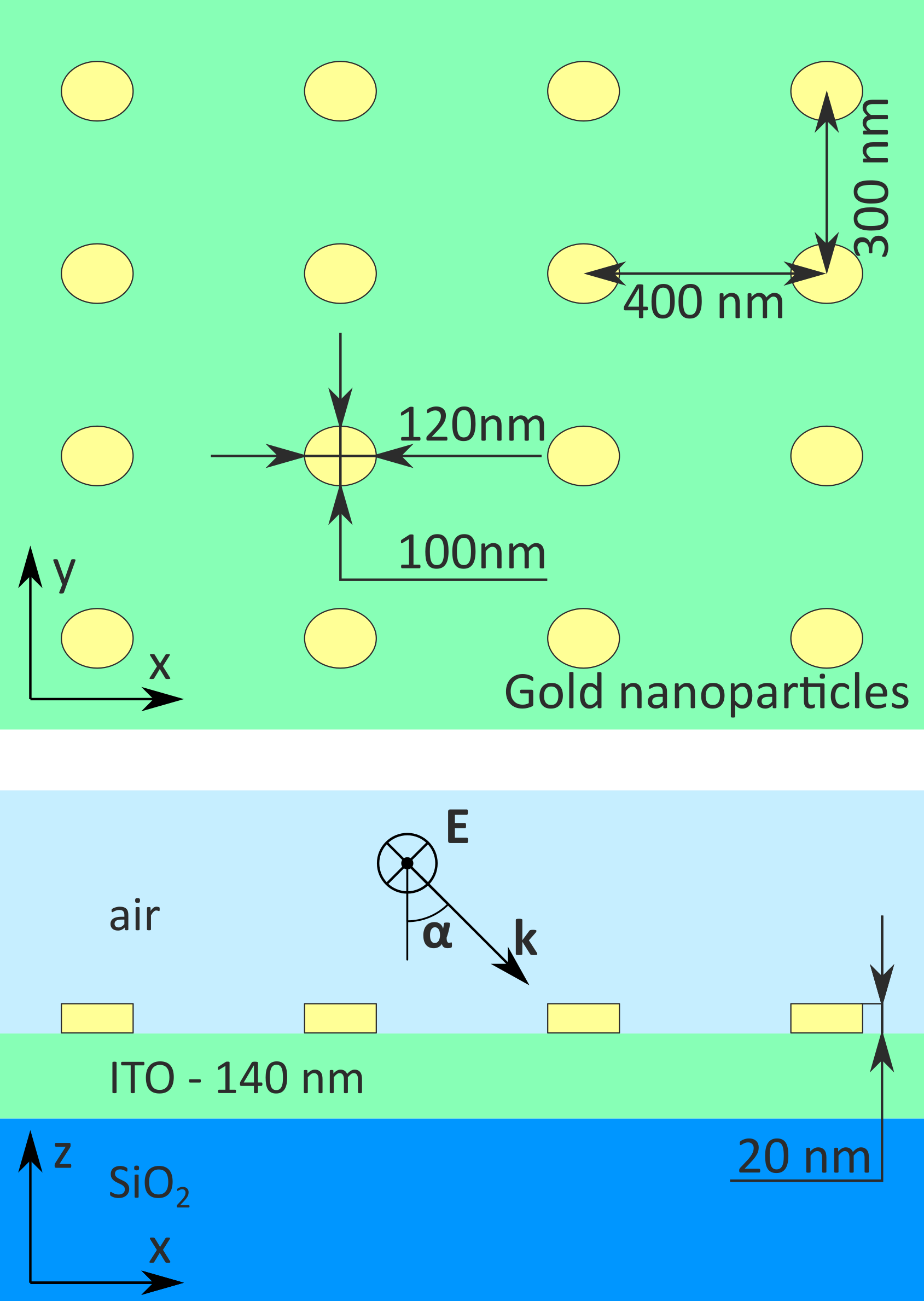}
    \end{subfigure}
    \begin{subfigure}{0.312\linewidth}
    \centering
    \includegraphics[width=\textwidth]{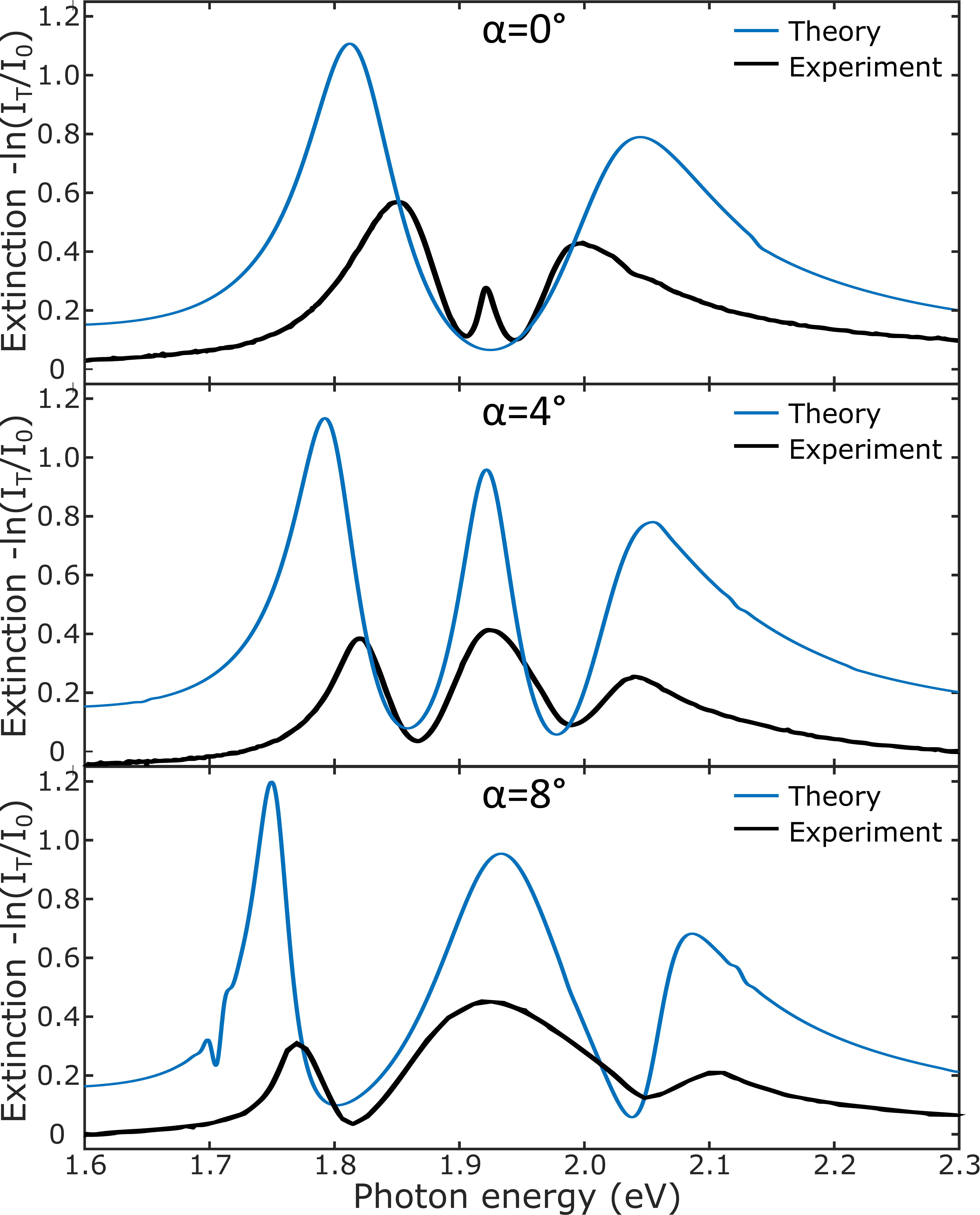}
    \end{subfigure}
    \centering
    \begin{subfigure}{0.6\linewidth}
    \centering
    \includegraphics[width=\textwidth]{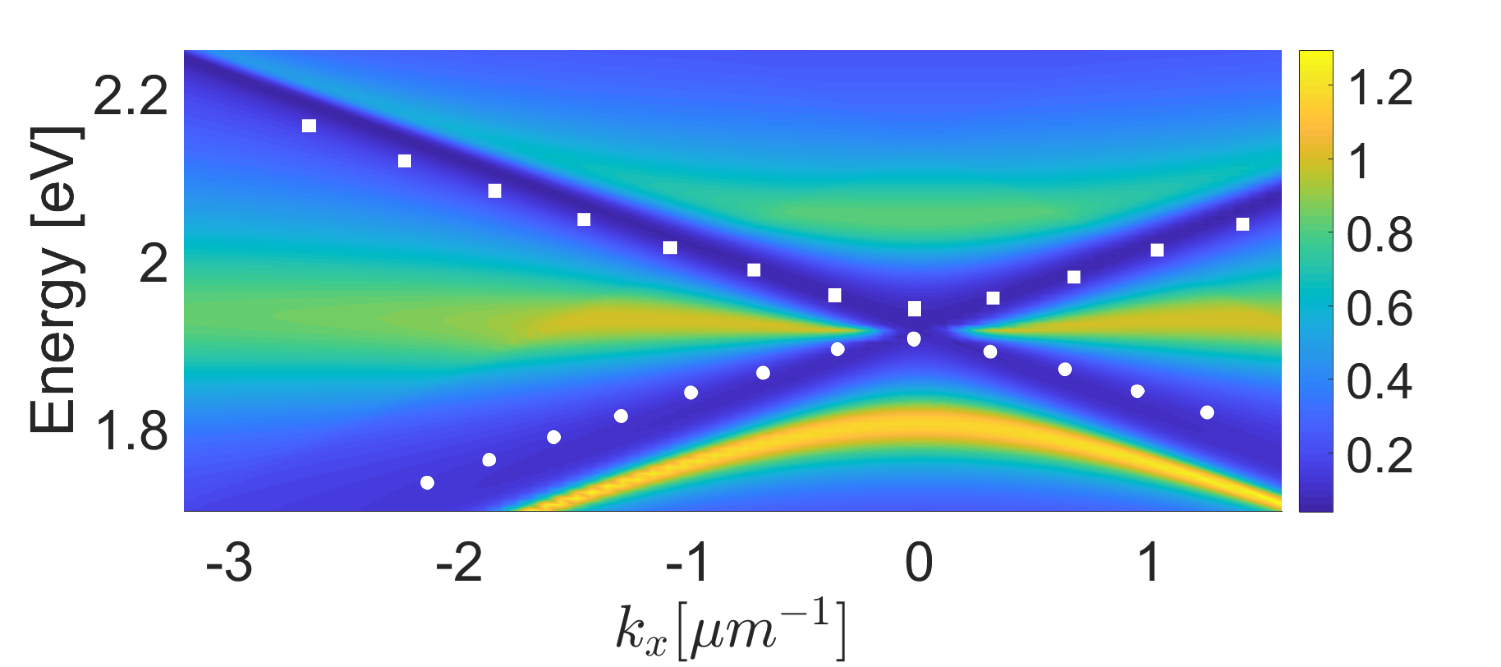}
    \end{subfigure}
    \caption{(Top-left) Schematic of the rectangular lattice of golden nanoparticles on the ITO waveguide. (Top-right) Comparison of experimental and theoretical extinction spectra of the lattice in $s$ polarization for different angles of incidence. (Bottom) Comparison of theoretically calculated extinction spectrum in $s$ polarization with dips extracted from the experiment (white squares and circles).}
    \label{fig:Giessen}
\end{figure}

As it was discussed previously, although close proximity of an interface destroys conventional lattice plasmon resonances, a similar coupling can be provided not by the photons propagating along the lattice in a homogeneous medium, but by a photonic guided mode. Such an effect for golden nanoparticles on an ITO waveguide was experimentally studied in a paper \cite{Linden2001} (see Fig. \ref{fig:Giessen} top-left panel). Two branches of Fano-Wood resonances are observed in an extinction spectrum of such a structure for different angles of incidence (see Fig. \ref{fig:Giessen} top-right panel). Experimental and theoretical peaks and dips match but also differ in amplitude, which most likely is a result of non-ideality of an experiment. However, the dispersion of dips extracted from the experiment also coincides with theoretical estimations (see Fig. \ref{fig:Giessen} bottom panel). It is also worth to mention, that energy splitting and so-called bound states in a continuum are observed in this case.

\clearpage
\newpage

\chapter{Convergence and accuracy}

\section{Convergence}

\begin{figure}[h]
\centering
\includegraphics[width=0.6\columnwidth]{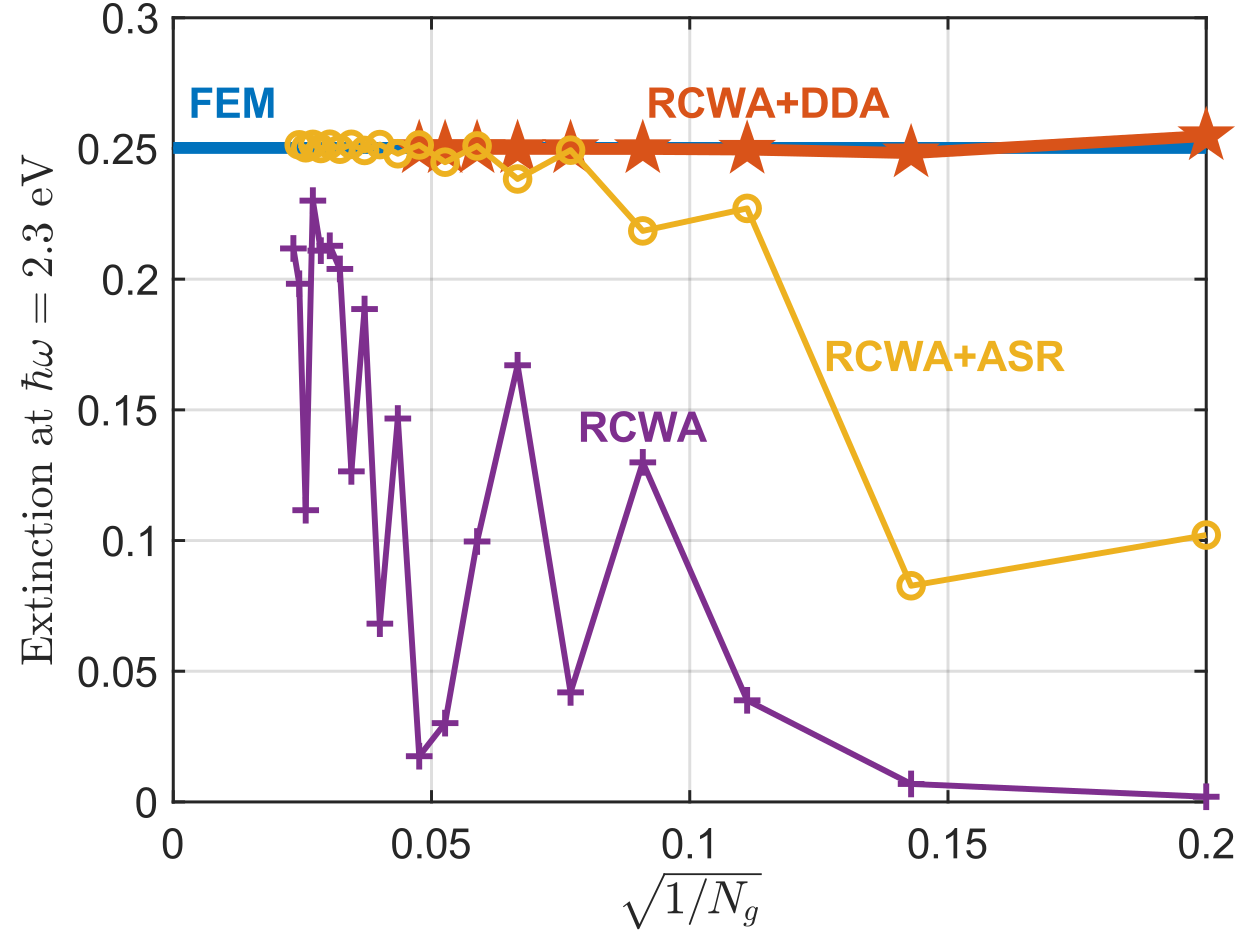}
\caption{(Color online) Extinction of the lattice in bulk silica for a normally incident light at $\hbar\omega=2.3$\,eV calculated by different methods as a function of $\sqrt{1/N_g}$, where $N_g$ is the total number of Fourier harmonics.}
\label{fig:convergence}
\end{figure}

In order to compare convergence rate with conventional RCWA method and RCWA enhanced with adaptive spatial resolution (ASR) \cite{weiss2009matched}, we have considered extinction emerging in the lattice in a homogeneous silica under the normal-incident light of $2.3$\,eV energy. It is seen in Fig.\,\ref{fig:convergence} that DDA enhanced method converges almost immediately, which is actually determined by the convergence rate of the dynamical interaction constant $\hat{C}(\mathbf{k}_\parallel)$ (see appendix \ref{App sum} for details). Concurrently, original RCWA method converges very slowly and does not provide reliable results even for $1681$ harmonics, which is the maximum available value in our calculations. RCWA+ASR calculations conducted by prof. Thomas Weiss converges much faster, however does not reach the performance of dipole approximation.

\section{Accuracy}

In order to verify the considered method, we compare our results with FEM calculations conducted in COMSOL Multiphysics. Since the calculation of extinction coefficients for each and every frequency and in-plain wavevector component takes of the order of minute it is possible to conduct computations varying only one parameter in a reasonable time. We have considered spectra corresponding to a normal incidence of light on the same structures. As it can be seen from Figs.\,\ref{sample} (j--l), our results almost perfectly match with FEM calculations. 

However, the accuracy of results might strongly depend on the structure. We know, that our main approximation is dipole one. The smaller is the characteristic size of the particle the more valid is the method. In this way, the ratio of the particle size and other spatial parameters, such as period, wavelength and distance to the closest interface, act as small parameters.

Here, we conduct a brief study of the accuracy dependence on these parameters in order to understand roughly limits of its applicability. We leave out of account study of the dependence on wavelength for several reasons. Plasmonic structures, which we consider, have strong dispersion and support bright resonance, which makes it hard to distinguish the impact of real physical effects and the drawbacks of approximation. For example, low precision can remain unnoticed, when the particle itself has an extremely small cross section. Also, the issue of the wavelength dependence refers not to lattice effects, but to a single particle and its own peculiarities. Therefore, it is better to leave this question for the study of the properties of specific particles, but not the method.

\subsection{Dense lattice}

If we consider near field of a particle then we can qualitatively assume that the wavelength is very large and we deal with the electrostatic problem. For such type of problems, it is well-known that quadrupolar and other high multipole fields decay in space much faster then dipolar one. Therefore, we can utilize dipole approximation for relatively large periods, which are smaller than a wavelength. However, when we approach the close proximity of the particle high multipole fields become comparable to a dipole contribution.

\begin{figure}[h]
    \centering
    \begin{subfigure}{0.495\linewidth}
    \centering
    \includegraphics[width=\linewidth]{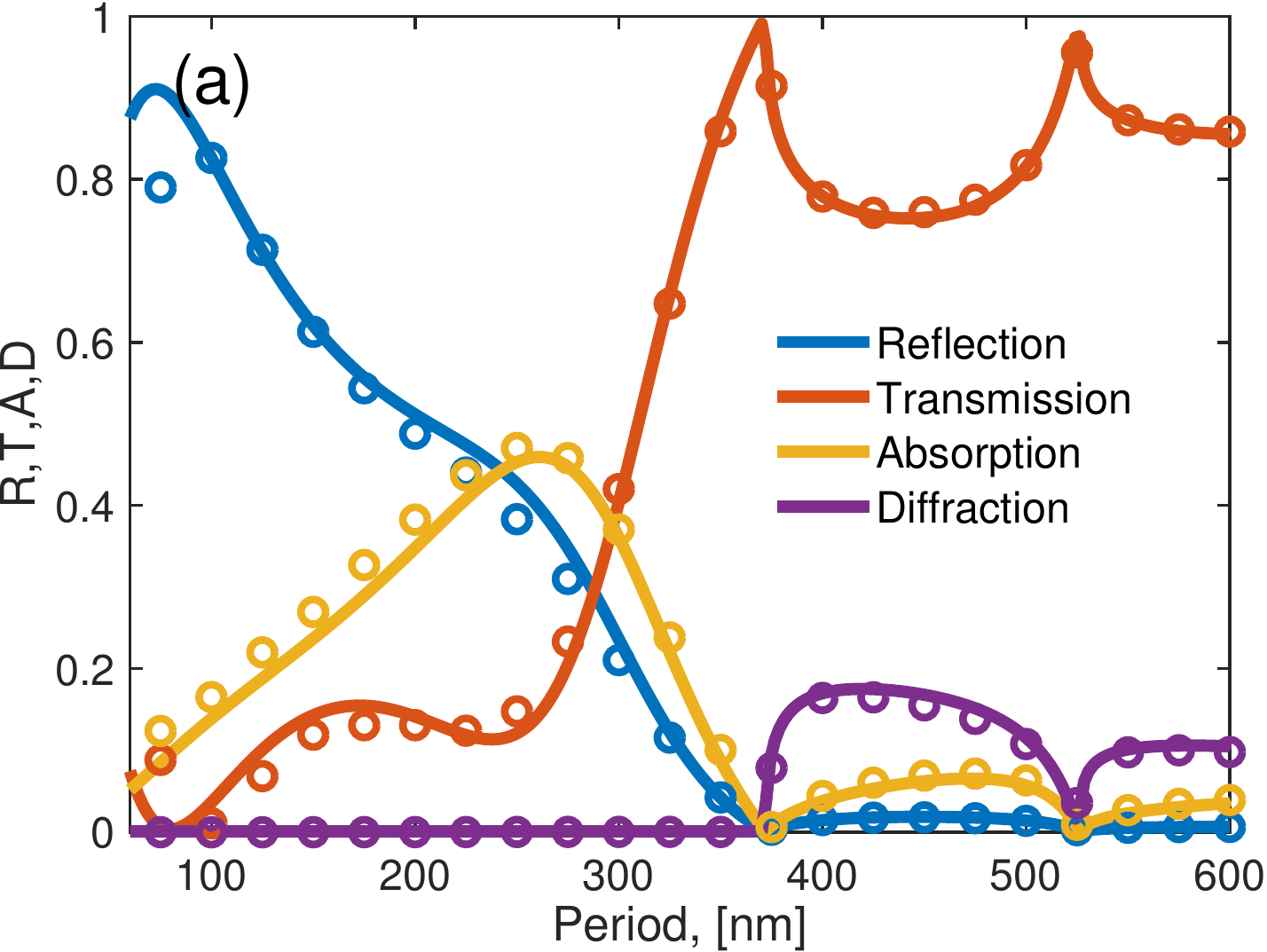}
    \end{subfigure}
    \hfill
    \begin{subfigure}{0.495\linewidth}
    \centering
    \includegraphics[width=\linewidth]{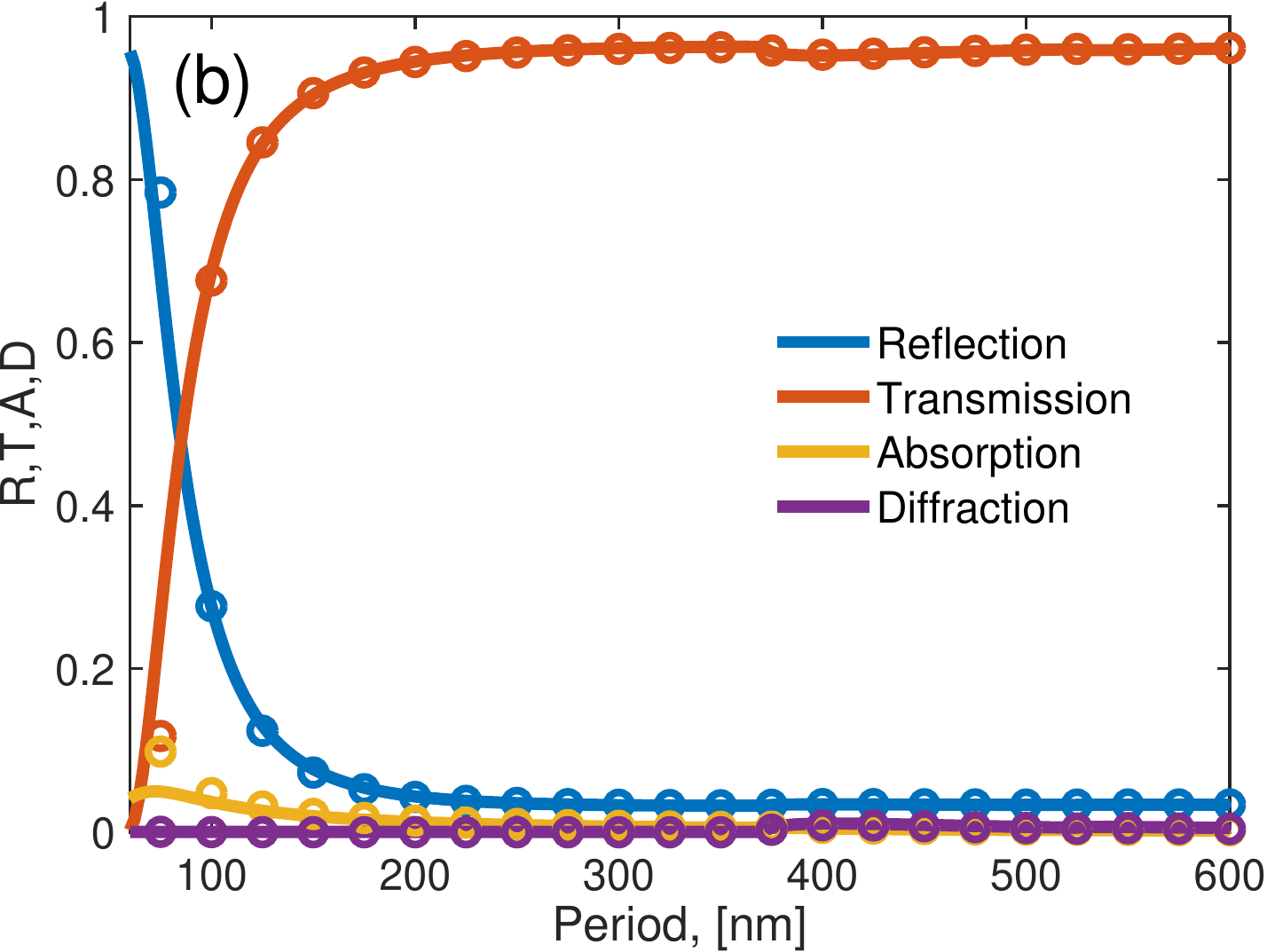}
    \end{subfigure}
    \caption{Dependence of reflection, transmission, absorption and diffraction of a plasmonic lattice in a bulk silica (panel (a)) and on the air/silica interface (panel (b)) on the period. Light of $\hbar\omega=2.3$(eV) energy is incident normally. Solid lines correspond to RCWA+DDA method, whereas circles depict FEM results.}
    \label{fig:period_sweep}
\end{figure}

In order to estimate the minimum distance on which dipolar approximation is still valid, we have considered lattice of silver nanodisks in bulk silica and on the air/silica interface for normally incident light of $\hbar\omega=2.3$(eV) energy. Their dependences of reflection, transmission, absorption, and diffraction on the period are depicted in Fig. \ref{fig:period_sweep}. In both cases, a very good match is observed for almost the whole range of periods. Significant, but still not crucial deviations arise only for a period less than $100$ (nm). Considering the fact that the diameter of silver nanodisks is $60$ (nm) this case corresponds to a very dense lattice.

\subsection{Lattice near an interface}
When we consider plasmonic lattice near an interface there are two possible approaches to conduct calculations. The first one is to include an interface in the local layer. In this approach, we account for reflection from an interface in the calculation of effective polarizability and local scattering matrix. Polarizability of a single particle is calculated near the interface as well. The main advantage of this method is the accuracy since we account for the shape of the particle and its influence on the near-field in close proximity of an interface. The main drawback is the necessity to recalculate polarizability of a particle for each distance from an interface.

Another approach is to consider plasmonic lattice in a homogeneous medium and couple it to an external interface via a combination of scattering matrices in the spirit of RCWA. This method is more convenient since we can easily change the distance between the lattice and the interface. It is accurate for large distances, since in this case interface indeed <<sees>> particles as dipoles. However, when we bring them closer, the interface should know something about the shape of the particle even though we work in the dipole approximation in the end. Local field of the particle is complicated and can not be described in dipole approximation. Another problem is the rapid increase in the required number of harmonics with approaching an interface.

From this perspective, we understand that for large distances both methods are acceptable, but the second one is much more convenient. At the same time for small distances, the first approach is not only faster than the second one, but the latter is simply incorrect. However, these speculations do not give us any estimations on the transitional distance at which we have to switch from one approach to another.

\begin{figure}[h]
    \centering
    \includegraphics[width=0.55\linewidth]{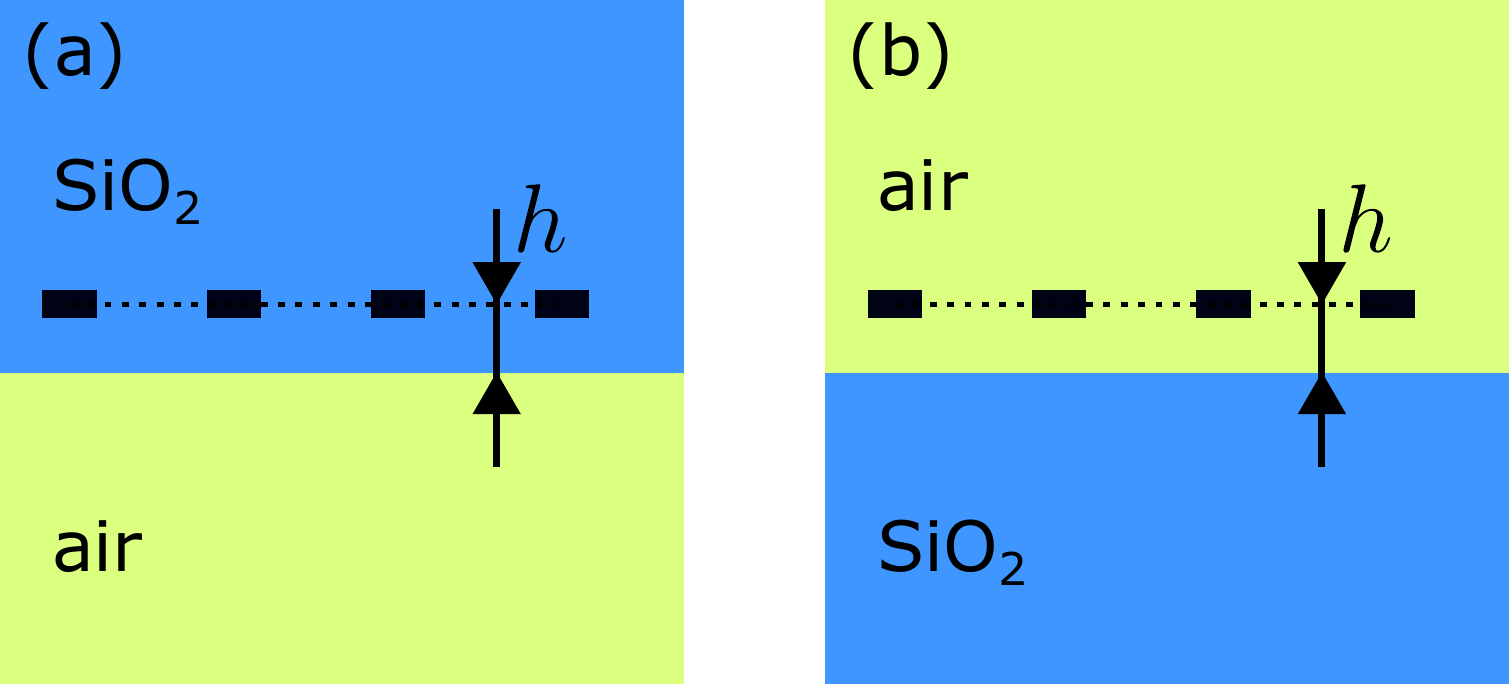}
    \caption{Schematics of the considered structures. Lattices of silver nanodisks are placed in close proximity of and interface inside silica layer (panel (a)) and air (panel (b)) correspondingly. Both structures are illuminated by a normally incident light from above.}
    \label{fig:lattice_near_interface}
\end{figure}

\begin{figure}[h]
    \centering
    \begin{subfigure}{0.495\linewidth}
    \centering
    \includegraphics[width=\linewidth]{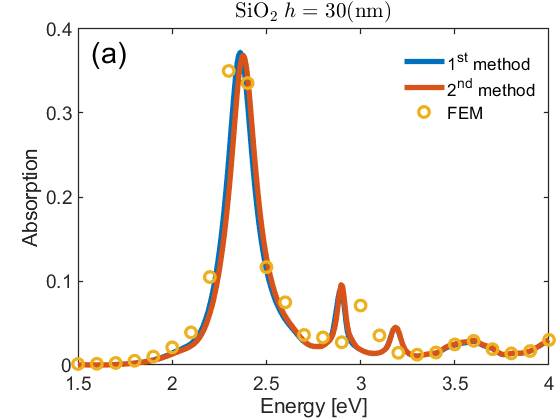}
    \end{subfigure}
    \hfill
    \begin{subfigure}{0.495\linewidth}
    \centering
    \includegraphics[width=\linewidth]{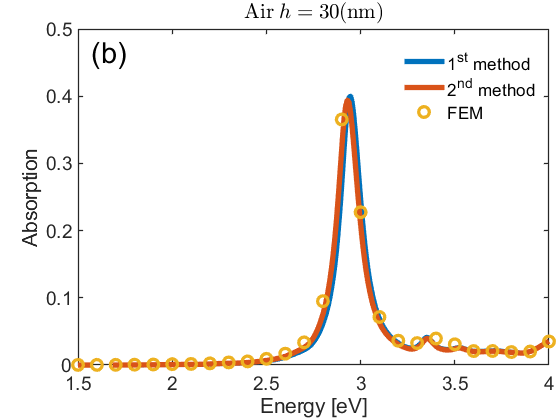}
    \end{subfigure}
    \centering
    \begin{subfigure}{0.495\linewidth}
    \centering
    \includegraphics[width=\linewidth]{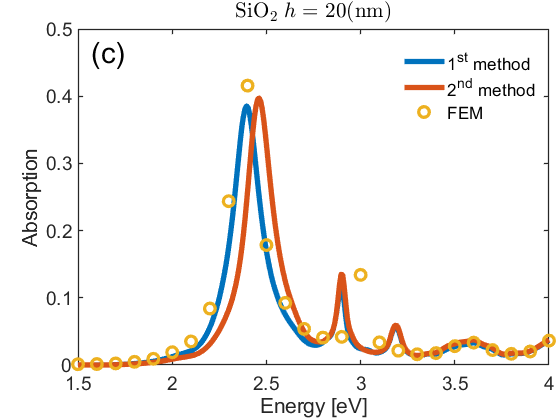}
    \end{subfigure}
    \hfill
    \begin{subfigure}{0.495\linewidth}
    \centering
    \includegraphics[width=\linewidth]{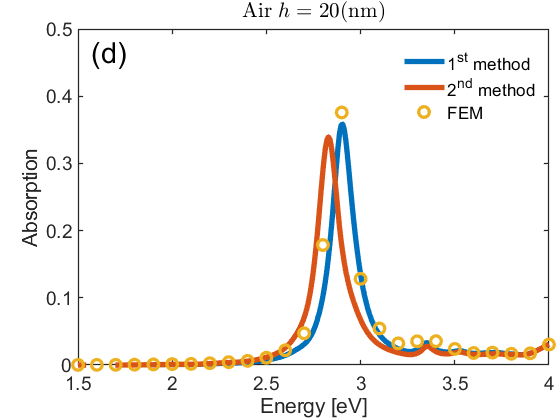}
    \end{subfigure}
    \centering
    \begin{subfigure}{0.495\linewidth}
    \centering
    \includegraphics[width=\linewidth]{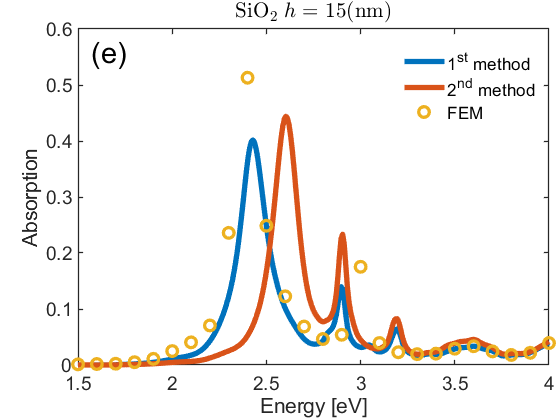}
    \end{subfigure}
    \hfill
    \begin{subfigure}{0.495\linewidth}
    \centering
    \includegraphics[width=\linewidth]{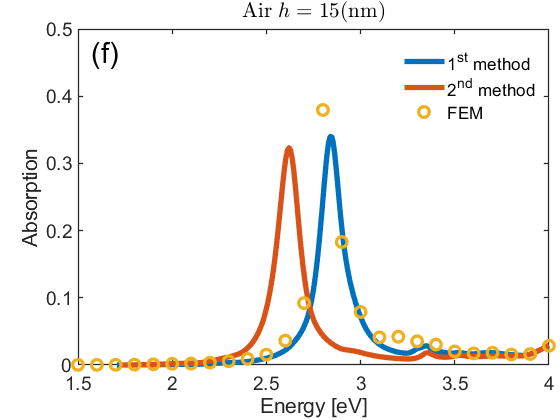}
    \end{subfigure}
    \caption{Absorption spectra of silver nanodisk $200\times200$(nm) square lattice near the air/silica interface under the normal incidence of light. Silver disks of $30$ nm radius and $20$ nm height are described by Johnson-Christy optical constants, whereas silica has $\varepsilon_{\mathrm{SiO}_2}=2.1$. Left Circle markers are related to direct calculations in COMSOL Multiphysics, while solid lines are related to DDA enhanced RCWA methods. Left column corresponds to the lattice embedded into the silica and right column to the lattice hanging in the air. In both cases, structures are illuminated from the same side of the interface from which they are located. In the first approach, the interface is included in the local layer and in the second one is not.}
    \label{fig:lattice_interface}
\end{figure}

In order to reveal this issue, we compare the results of two approaches for different distances. We have chosen dense square lattice of $200$(nm) period in order to avoid lattice resonances and effects and put them into two environments depicted in Fig. \ref{fig:lattice_near_interface}. In Fig. \ref{fig:lattice_interface} we compare absorption spectra under the normal incidence for these cases. Left column corresponds to the lattice embedded into the silica and right column to the lattice hanging in the air. In both cases, structures are illuminated from the same side of the interface from which they are located.

As it is seen from panels (a) and (b) of Fig. \ref{fig:lattice_interface} both methods give the same results for relatively large $h$. Frankly speaking, in this case, the polarizability of a particle in a homogeneous space and near the interface are almost the same. Therefore, results coincide not because of the match of the contribution of the interface calculated via different approach, but because the fact that in both cases this contribution is negligible. At the same time, FEM calculations generally match our DDA+RCWA approaches. The most significant deviation is observed for the side resonances. Most probably there are two main reasons for that. Firstly, this resonance is associated with highly-confined edge plasmons and the required dense mesh was not provided in 3D FEM calculations. Secondly, the electric field of this resonance is very heterogeneous, which makes dipole approximation less valid. Indeed, hot spots on the edges make a huge contribution in polarizability, but when we account for their interaction we should probably take the distance between the edges ($80$ nm), but not between the centers ($200$ nm). In other words, for such dense lattice and complex resonances accounting for higher multipole harmonics is required for better accuracy.

For slightly smaller distance (panels (c,d)) results of two approaches no longer match. Moreover, as we predicted the 1st method, which includes an interface into the local layer provide more precise data. When the lattice is further closer to the boundary 2nd method totally breaks down. However, the first one works much worse as well - amplitude of the main resonance no longer matches FEM results. I do not have a clear explanation for this phenomenon, but the trends observed in Fig. \ref{fig:lattice_interface} allow us to make several assumptions. We see that that the bias of a lattice even on several nanometers leads to significant deviations in the spectrum. However, when we bring a particle closer to the interface fields inside it becomes less homogeneous. From this perspective, it might be more correct to put the ideal dipole not in the geometric center of the particle, but in a slightly deviated point, which will change the spectra. In other words, we return to the discussion of the validity of the dipole approximation. Most probably accuracy is lowering near an interface because of the increase of the gradients in the field and corresponding higher order multipole moments.

At the same time, we should remember, that the period of the lattice is extremely small and is comparable with the size of the particles. Therefore, we expect that more sparse structures should provide much better results. Summing up, the obtained results confirm that we can calmly exclude interface from the local layer if it is removed from the lattice by at least several dozens of nanometers.

\clearpage
\newpage

\chapter{Conclusions}
We have proposed the method for the implementation of the DDA method in RCWA. Although our approach is very general and allows considering lattices of nanoparticles in a complex environment, in this thesis, we have implemented it for lattices in a homogeneous ambiance and on a boundary between two homogeneous media, which are the most practical cases. We have demonstrated its operational feasibility by calculation of spectra for solitary plasmonic lattice and a lattice coupled with an optical waveguide. An occurrence of Fano-Wood anomalies, lattice plasmon resonances and strong coupling between them is observed via the proposed approach. The accuracy of our approach was verified by comparison with FEM calculation, whereas the convergence rate was shown to be much faster than in conventional and adaptive-spatial-resolution enhanced RCWA calculations. Also, limits of applicability of the method, determined by the dipole approximation, were estimated. Comparison of calculations with experimental results shows a good match, which can be considered as an additional confirmation.

Fast speed of calculation (typically several dozens of milliseconds for one computation at a fixed $\omega$ and $\mathbf{k}_\parallel$) and high accuracy of results makes this approach convenient and perspective for both interpretation of experimental results and fundamental analysis of phenomena occurring in plasmonic lattices.
Although we formulate our method for a single particle in the unit cell, it can be easily generalized for the case of multiple particles, which is to be done in further research.
 
\appendix

\chapter{Polarizability tensor calculation} \label{App alpha}

\section{Total/scattering field formulation}
\subsection{General formulation}
In order to find polarizability, $\hat{\alpha}$, of an isolated nanoparticle, it is very convenient to switch to total/scattered field formulation. Let us consider a certain background structure, which does not include the particle itself and is defined by a background permittivity $\varepsilon^{\mathrm{bg}}(\mathbf{r})$. We suppose, that in the absence of a particle structure is rather simple and solution of the Maxwell equations $\mathbf{E}^{\mathrm{bg}}(\mathbf{r})$ can be obtained analytically. To give an example $\mathbf{E}^{\mathrm{bg}}(\mathbf{r})$ can represent a plane wave in a homogeneous medium or Fresnel reflection from an interface between two homogeneous media. Since $\mathbf{E}^{\mathrm{bg}}(\mathbf{r})$ is a solution of the Maxwell equations, it fulfills one of them in the following way:

\begin{equation}
    \mathrm{rot} \mathbf{H} =\frac{4\pi}{c}\mathbf{j}+\frac{\partial \mathbf{D}}{c\partial t},\label{eq:A1}
\end{equation}

\begin{equation}
    \mathrm{rot} \mathbf{H}_{\mathrm{bg}} = -\frac{i\omega}{c}\varepsilon_{\mathrm{bg}} \mathbf{E}_{\mathrm{bg}}.\label{eq:A2}
\end{equation}

Now, we will consider the total structure, including the particle itself, defined by $\varepsilon^{\mathrm{tot}}(\mathbf{r})$ permittivity. In a similar way, we define total field $\mathbf{E}^{\mathrm{tot}}(\mathbf{r})$ as a solution of the following equation:

\begin{equation}
    \mathrm{rot} \mathbf{H}_{\mathrm{tot}} = -\frac{i\omega}{c}\varepsilon_{\mathrm{tot}} \mathbf{E}_{\mathrm{tot}}.\label{eq:A3}
\end{equation}

In contradistinction to the background field, total field typically can be determined only numerically. For a practical calculations it is very convenient to express total field as a sum of background and scattered field $\mathbf{E}_{\mathrm{tot}}=\mathbf{E}_{\mathrm{bg}}+\mathbf{E}_{\mathrm{sc}}$. In this way, we can reformulate the problem to a calculation of scattered field. Subtraction of Eqn. \ref{eq:A2} from Eqn. \ref{eq:A3} gives us a rule for its calculation:

\begin{equation}
    \mathrm{rot}[ \mathbf{H}_{\mathrm{tot}}-\mathbf{H}_{\mathrm{bg}}] = -\frac{i\omega}{c}[\varepsilon_{\mathrm{tot}} \mathbf{E}_{\mathrm{tot}}-\varepsilon_{\mathrm{bg}} \mathbf{E}_{\mathrm{bg}}],\label{eq:A4}
\end{equation}

\begin{equation}
    \mathrm{rot} \mathbf{H}_{\mathrm{sc}} =
-\frac{i\omega}{c}\Delta\varepsilon \mathbf{E}_{\mathrm{bg}}-\frac{i\omega}{c}\varepsilon_{\mathrm{tot}} \mathbf{E}_{\mathrm{sc}},\label{eq:A5}
\end{equation}
where $\Delta\varepsilon(\mathbf{r}) = \varepsilon^{\mathrm{tot}}(\mathbf{r}) - \varepsilon^{\mathrm{bg}}(\mathbf{r})$ is a perturbation in permittivities introduced by a particle. As we can see from Eqns. (\ref{eq:A1}, \ref{eq:A5}), scattered field can be found as a field radiated by a current $\mathbf{j}^{\mathrm{bg}}(\mathbf{r}) = -i \omega \Delta \varepsilon(\mathbf{r}) \mathbf{E}^{\mathrm{bg}}(\mathbf{r})/(4\pi)$ induced by a background electric field \cite{craig1983,Bai2013}. It should be especially noted, that this current is distributed over the volume of a particle and that radiation is considered in the environment defined by  $\varepsilon^{\mathrm{tot}}(\mathbf{r})$. Therefore, if a particle supports a certain eigenmode (e.g. LSPR) then a Purcell effect will be observed, which leads to a resonance in scattered field. In this way, a connection between emission in resonant mode and scattering of resonator becomes obvious. Thus, in general case the problem of scattering can be considered as an emission problem, which is easier for an implementation by any near-field calculation method.

We have found out a convenient way to determine electromagnetic fields in the process of scattering, but according to the DDA, our final aim is the determination of a dipole moment, which can be inserted in a background environment in order to substitute a real particle. Eqn. \ref{eq:A5} can be easily reorganized for these purposes:

\begin{equation}
    \mathrm{rot} \mathbf{H}_{\mathrm{sc}} =
-\frac{i\omega \Delta\varepsilon \mathbf{E}_{\mathrm{tot}}}{c}-\frac{i\omega}{c}\varepsilon_{\mathrm{bg}} \mathbf{E}_{\mathrm{sc}},\label{eq:A6}
\end{equation}
which means that a current $\mathbf{j} =
-i\omega \Delta\varepsilon \mathbf{E}_{\mathrm{tot}}/(4\pi)$ is excited in a particle \cite{chen2017spectral}.

All the calculations above are very general and can be applied in a very wide range of problems. However, small particles are considered in this paper and therefore, we are only interested in a dipole moment of the induced polarization density. It can be easily found by integrating over a volume of a particle:

\begin{equation}
        \mathbf{P}=\int \frac{\Delta \varepsilon}{4\pi}\mathbf{E}^{\mathrm{tot}}(\mathbf{r}) d^3\mathbf{r}=\int \frac{\Delta \varepsilon}{4\pi}(\mathbf{E}^{\mathrm{bg}}(\mathbf{r})+\mathbf{E}^{\mathrm{sc}}_{\mathbf{j}^{\mathrm{bg}}}(\mathbf{r}))d^3\mathbf{r},
    \label{eq:6}
\end{equation}

Since, in this paper we consider small particles in media with constant permittivity, $\varepsilon^{\mathrm{bg}}(\mathbf{r})$, over their volume (boundaries between materials does not intersect particles), we assume $\mathbf{E}^{\mathrm{bg}}(\mathbf{r})$ (not $\mathbf{E}^{\mathrm{sc}}(\mathbf{r})$) to be constant in space on the dimensions of a particle $\mathbf{E}^{\mathrm{bg}}(\mathbf{r})=\mathbf{E}^{\mathrm{bg}}$. This means that the primary induced current $\mathbf{j}^{\mathrm{bg}}(\mathbf{r})$ is just a vector with $3$ components, not dependent on coordinates and makes it possible to introduce $\hat{\alpha}$ as a simple $3\times3$ tensor, connecting additional dipole moment of a particle, $\mathbf{P}$, with a background field $\mathbf{E}^{\mathrm{bg}}$:
\begin{equation}
    \mathbf{P}=\hat{\alpha}\mathbf{E}^{\mathrm{bg}}= \frac{\Delta\varepsilon V}{4\pi}(\mathbf{E}^{\mathrm{bg}}+<\mathbf{E}_{\mathbf{j}^{\mathrm{bg}}}^{\mathrm{sc}}>),
    \label{eq:A8}
\end{equation}
where $V$ is the volume of a particle and $<\mathbf{E}_{\mathbf{j}^{\mathrm{bg}}}^{\mathrm{sc}}>$ is a scattered field (determined as a field generated by the current density $\mathbf{j}^{\mathrm{bg}}$) averaged over the volume of a particle.
Such an approach clearly shows that dipole approximation can be applied even for a particle in a very complicated environment and provides an easy-to-apply procedure of $\hat{\alpha}$ tensor calculation for the cases considered in this paper.

Thus, the only thing that should be done to calculate polarizability tensor, $\Hat{\alpha}$, is the calculation of scattered field, which is a field radiated by an effective current density $\mathbf{j}^{\mathrm{bg}}$. {Since, the connection between dipole moment and electric field is linear a unitary background field (e.g. $\mathbf{E}^{\mathrm{bg}}=[1, 0, 0]^{T}$) is taken in numerical simulations}.
In general case, $3$ independent calculations are required to determine the response of a particle on each polarization of a background field. However, in the presence of additional symmetry, there can be only 2 or even 1 independent components, which simplifies calculations additionally.
We conduct {the described} calculations in COMSOL Multiphysics, whereas they might be potentially realized via any near-field calculation methods.

\subsection{Axially symmetrical case}

In case when the structure has additional axial symmetry, the calculation can be significantly simplified. Indeed, for such structures electric field can be expanded as a sum over angular harmonics:
\begin{equation}
    \mathbf{E}(r,\phi,z) = \sum_m \begin{pmatrix}E^m_r(r,z)\\E^m_\phi(r,z)\\E^m_z(r,z)
    \end{pmatrix} e^{i m \phi}.
\end{equation}
Each harmonic is absolutely independent on the others and can be determined in a 2D calculation, which is extremely faster than common 3D ones. Moreover, for the calculation of polarizability tensor we need only $m=0,\pm1$ harmonics, which strongly simplifies the general problem.

Indeed, the presence of axial symmetry means that the $z$ axis is one of the main axes and the two latter are degenerate and lay in the $x-y$ plane. For the sake of simplicity, we can choose $x$ and $y$ axes as the main ones.

As it is seen from the Eqn. \ref{eq:A8} $z$ component of the tensor corresponds to the averaged value of $E_z$ field over the volume of a particle. Obviously, only $m=0$ harmonic provides a non-zero value of the integral. In such a way, the computation of $z$ component of polarizability tensor almost does not differ from the general 3D case. We just excite $z$-directed current in a simulation of $m=0$ harmonic and then compute the corresponding averaged $E_z$ field.

However, the computation of in-plain components of the tensor is slightly more difficult. Only $m=\pm1$ harmonics make non-zero contribution in $x$ and $y$ components of averaged fields. However, each harmonic corresponds to a certain circular polarization of electromagnetic fields and therefore we can not just excite pure $x$-polarized current. Without loss of generality we can consider $m=1$ harmonic and excite corresponding circularly polarized current:
\begin{equation}
    \mathbf{j}^{\mathrm{bg}} = \begin{pmatrix}1\\i\\0   \end{pmatrix} e^{i\phi},
\end{equation}
which radiates corresponding electric field $\mathbf{E}^{\mathrm{sc}}$. Since the polarizability tensor is diagonal then we can consider $x$ and $y$ components independently. In other words, we can say that $x$ component of current radiates electric field, which has only $x$ component of averaged field and the same with $y$ components. $x$-polarized component of the $\mathbf{j}^{\mathrm{bg}}$ current, which can be easily found as:
\begin{equation} j_x^{\mathrm{bg}}=j^{\mathrm{bg},1}_re^{i\phi} \cos \phi-j^{\mathrm{bg},1}_\phi e^{i\phi}\sin\phi = 1. \end{equation}
Similarly averaged $x$ component of corresponding field can be determined as follows:
\begin{equation} <E_x^{\mathrm{sc}}>=\frac{1}{2}<E_r^{\mathrm{sc},1}>-i\frac{1}{2}<E_\phi^{\mathrm{sc},1}>, \end{equation}
where averaging  means the following:
\begin{equation} <E_r^{\mathrm{sc},1}>=\int E_r^{\mathrm{sc},1}(r,z) r dr dz d \phi/\int  r dr dz d \phi =2\pi \int E_r^{\mathrm{sc},1}(r,z) r dr dz /V. \end{equation}

\begin{figure}[h]
\centering
\includegraphics[width=1\columnwidth]{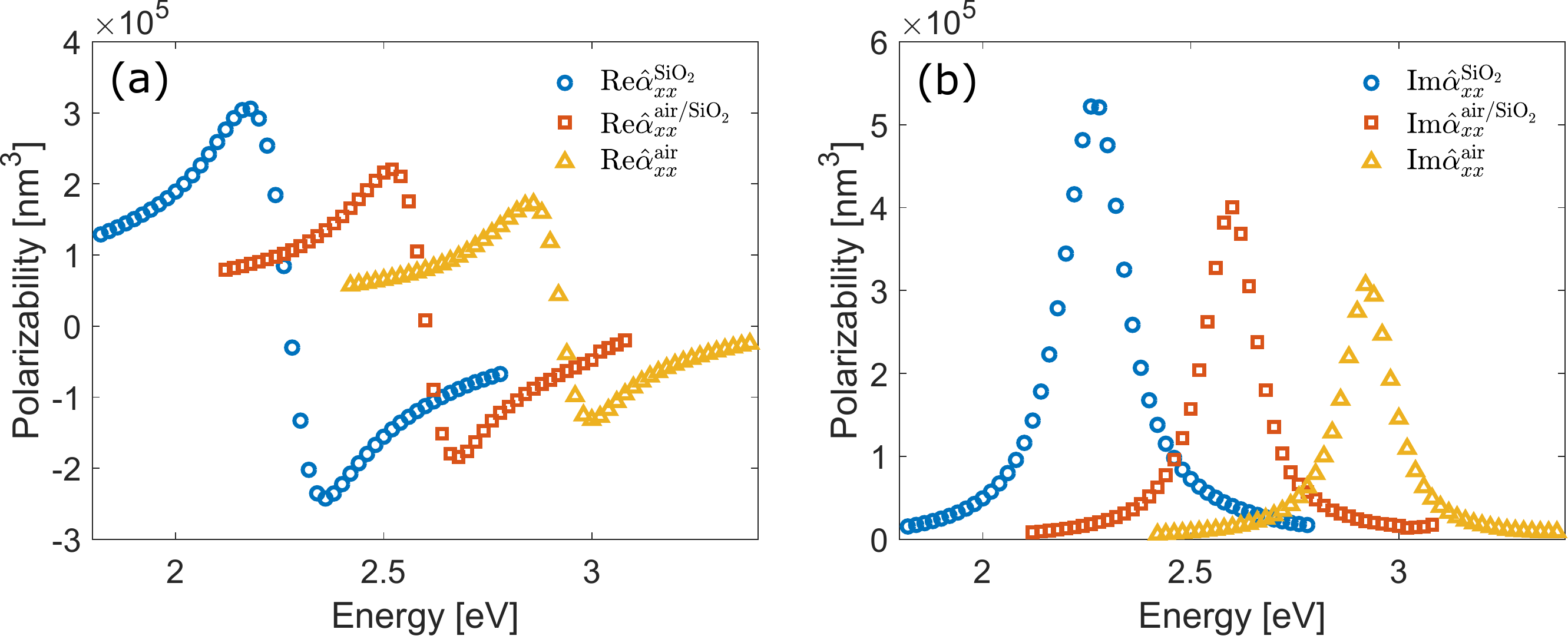}
\caption{(Color online)  Energy dependence of {real (a) and imaginary (b) parts of} in-plain polarizability of {the same} silver nanodisks embedded in {bulk} silica{, air} and laying {on the air/silica interface. All the data is calculated in COMSOL Multiphysics.}}
\label{fig:Homogeneous_alpha}
\end{figure}

\section{Example of calculations}

In Fig.\,\ref{fig:Homogeneous_alpha} we show energy dependence of in-plain component {$\hat{\alpha}_{xx}$} of a polarizability tensor of the silver nanodisk in {bulk} silica{, air} an{d} in the air on the air/silica interface. It is seen, that wide resonances of LSPRs correlate with the obtained spectra  (see Fig. \ref{sample} (d--i)). {Since silica has permittivity higher than air, corresponding plasmonic resonance in a disc is strongly red-shifted. Resonance of a particle laying on a boundary is obviously located between them.} Almost perfect matching of extinction spectra {with FEM calculations} (see Fig. \ref{sample} (j--l)) indirectly proves that our approach provides precise results and substantiates our choice.

In this way, there are several other important advantages of the method applied for the calculation of $\hat{\alpha}$ tensor. First of all, it is very universal so that allows considering particles of any, even very complex shape, in any environment. Secondly, it provides high accuracy, which as a result leads to high-precision spectra.

\section{Resonant approximation for polarizability of indivisual particle}
\subsection{General formulation}

The proposed method of calculation of polarizability tensor is accurate and fast enough. However, the most interesting objects for consideration are resonant nanoparticles including plasmonic ones and they allow to simplify the computational problem additionally.

Indeed, theory of radiation in low-$Q$ resonators \cite{sauvan2013,Bai2013,lalanne2018} states that in our case, scattering field can be expressed as a sum of the fields of quasinormal modes with definite frequency-dependent amplitudes $\mathbf{E}_{\mathrm{sc}}(\mathbf{r})=\sum_l A_l (\omega)\tilde{\mathbf{E}}_l$. In most cases, when resonances' spectra do not intersect or they have orthogonal polarizations, then the amplitude of each mode can be approximately determined independently on the others:
\begin{equation}
    A_l(\omega) = \frac{ -4\pi i \int  \mathbf{j}^{\mathrm{bg}} \tilde{\mathbf{E}}_l d^3 \mathbf{r}}{\omega-\tilde{\omega}_l},
\end{equation}where the field of the mode is normalized as follows:
\begin{equation}
\int \left[ \tilde{\mathbf{E}}_l \left.\frac{\partial \omega \varepsilon (\omega)}{\partial \omega}\right|_{\omega=\omega_l} \tilde{\mathbf{E}}_l-\tilde{\mathbf{H}}_l \left.\frac{\partial \omega \varepsilon (\omega)}{\partial \omega}\right|_{\omega=\omega_l} \tilde{\mathbf{H}_l} \right] d^3 \mathbf{r}=1.
\end{equation}

There are several methods to normalize quasinormal modes, which exponentially diverge at infinity \cite{lalanne2018}. However, in this work, we apply a convenient method described in the paper \cite{Bai2013}. The main idea is that if we excite a certain mode with a current $\mathbf{j}^{\mathrm{probe}}$ on a complex frequency $\omega$ very close to the eigenfrequency $\tilde{\omega}$, then the total field $\mathbf{E}^{\mathrm{probe}}$ is very close to the field of this mode:
\begin{equation}
    \mathbf{E}^{\mathrm{probe}} =  A(\omega)\tilde{\mathbf{E}} = \frac{ -4\pi i \int  \mathbf{j}^{\mathrm{probe}} \tilde{\mathbf{E}} d^3 \mathbf{r}}{\omega-\tilde{\omega}}\tilde{\mathbf{E}}.
\end{equation}
Therefore
\begin{equation}
    \int  \mathbf{j}^{\mathrm{probe}} \mathbf{E}^{\mathrm{probe}} d^3 \mathbf{r}= \frac{ -4\pi i( \int  \mathbf{j}^{\mathrm{probe}} \tilde{\mathbf{E}} d^3 \mathbf{r})^2}{\omega-\tilde{\omega}},
\end{equation}
and finally we obtain:
\begin{equation}
    \tilde{\mathbf{E}} = \mathbf{E}\sqrt{\frac{\omega-\tilde{\omega}}{-4\pi i\int  \mathbf{j}^{\mathrm{probe}} \mathbf{E}^{\mathrm{probe}} d^3 \mathbf{r} }}.
\end{equation}

The pole itself is found by an iterative search via the Pade approximation according to the paper. Another important issue, which raises during the search of the eigenmodes, is the extension of permittivity to the plane of complex frequencies. One way to tackle this problem is to fit the data for real frequencies by some analytical model (for instance, Drude-Lorentz one) and apply it for the calculation on a complex frequency. Although this approach is very powerful it has several drawbacks as well. Sometimes it is hard to fit the data onto a wide spectral range, in some cases the choice of the fitting model is not evident and not unique. That is why in our calculations we apply more straightforward approach - we just employ Taylor decomposition. Since the typical $Q$-factor of plasmonic resonance is approximately $10$, we estimate first and second derivatives via the divided differences on a spectral range of approximately $\delta\omega\approx\mathrm{Re}\tilde{\omega}/10$ width. Finally, Taylor series together with Cauchy–Riemann equations give us an estimation on permittivity at the complex frequency.

\subsection{Polarizability tensor calculation}

Accounting for the fact that background current, as well as background field, are constant over the particle in dipole approximation, we simplify the expression for the amplitude of the mode:
\begin{equation}
    A_l(\omega) = \frac{ -\omega \Delta \varepsilon (\omega) V \mathbf{E}^{\mathrm{bg}} <\tilde{\mathbf{E}}_l> }{\omega-\tilde{\omega}_l}.\label{eq:A9}
\end{equation}
Therefore, dipole moment of a particle can be expressed as follows:
\begin{multline}
    \mathbf{P}= \frac{\Delta\varepsilon V}{4\pi}(\mathbf{E}^{\mathrm{bg}}+\sum_l A_l <\tilde{\mathbf{E}}_l>)=\\=\frac{\Delta\varepsilon V}{4\pi}(\mathbf{E}^{\mathrm{bg}}+\sum_l \frac{ -\omega \Delta \varepsilon (\omega) V  \mathbf{E}^{\mathrm{bg}} <\tilde{\mathbf{E}}_l> }{\omega-\tilde{\omega}_l} <\tilde{\mathbf{E}}_l>).
    \label{eq:A10}
\end{multline}
Finally, polarizability tensor can be found as follows:
\begin{equation}
    \hat{\alpha}_{ij}=\frac{\Delta\varepsilon V}{4\pi}(\delta_{ij}+\sum_l \frac{ -\omega \Delta \varepsilon (\omega) V  }{\omega-\tilde{\omega}_l}
    <\tilde{E}_i^l><\tilde{E}_j^l>).
    \label{eq:A11}
\end{equation}

This is a general expression, which can potentially deal with an arbitrary number of modes. However, in most cases, small particles have only 3 pronounced dipole resonances. Moreover, in the presence of symmetry, they are degenerate. For instance, if all the 3 modes are symmetric, then:
\begin{equation}
    \hat{\alpha}_{ij}=\frac{\Delta\varepsilon V}{4\pi}(1+ \frac{ -\omega \Delta \varepsilon (\omega) V  }{\omega-\tilde{\omega}}
    {<\tilde{E}_{x}>}^2)\delta_{ij},
    \label{eq:A12}
\end{equation}
where $\tilde{\mathbf{E}}$ is an $x$-polarized mode. If modes are degenerate only in $x-y$ plain, then:
\begin{equation}
    \hat{\alpha}_{ij}=\frac{\Delta\varepsilon V}{4\pi}\left(\delta_{ij} -\omega \Delta \varepsilon (\omega) V  
    \left[\frac{{<\tilde{E}^{1}_x>}^2}{\omega-\tilde{\omega}_1} \mathrm{diag}(1,1,0)_{ij}+
    \frac{{<\tilde{E}^{2}_{z}>}^2}{\omega-\tilde{\omega}_2}\mathrm{diag}(0,0,1)_{ij}\right]\right),
    \label{eq:A13}
\end{equation}
where $\tilde{\mathbf{E}}^1$ and $\tilde{\mathbf{E}}^2$ are $x$ and $z$-polarized modes correspondingly.

\subsection{Axially symmetrical case}

As it was discussed for the straightforward calculations, axially symmetrical structures can be treated by much simpler simulations. Calculation of modes in symmetrical resonators is easier as well. Resonance along $z$ axis corresponds to $m=0$ axial harmonic whereas in-plain resonances along $x$ and $y$ axis are superpositions of circularly-polarized modes with $m=\pm1$.

Circularly polarized modes have the following fields:
\begin{equation}
    \tilde{\mathbf{E}}^\pm (r,\phi,z)
    = \left(
\begin{array}{c}
\tilde{E}_r \\
\pm \tilde{E}_\phi\\
\tilde{E}_z
\end{array}\right)
e^{\pm i \phi},
\qquad
\tilde{\mathbf{H}}^\pm (r,\phi,z)
= \left(
\begin{array}{c}
\pm \tilde{H}_r \\
\tilde{H}_\phi\\
\pm \tilde{H}_z
\end{array}\right)
e^{\pm i \phi}.
\end{equation}

This means that if the current 
\begin{equation}
    \mathbf{j}^{\mathrm{probe},\pm} (r,\phi,z)
    = \left(
\begin{array}{c}
j^{\mathrm{probe}}_r \\
0\\
0\end{array}\right)
e^{\pm i \phi}
\end{equation} excites fields of corresponding angular dependences, then current 
\begin{equation}
    \mathbf{j}^{\mathrm{probe}} (r,\phi,z)
    = \left(
\begin{array}{c}
j^{\mathrm{probe}}_r \\
0\\
0\end{array}\right)
\cos \phi
\end{equation} will excite field
\begin{equation}
    \mathbf{E}^{\mathrm{probe}} (r,\phi,z)
    = \left(
\begin{array}{c}
E_r^{\mathrm{probe}} \cos\phi\\
 E_\phi^{\mathrm{probe}} i \sin\phi\\
 E_z^{\mathrm{probe}} \cos\phi
\end{array}\right),
\end{equation}which corresponds to the $x$ -polarized mode.

Thus, the integral required for the calculation of the normalized field of the mode can be found as follows:
\begin{equation}
    \int \mathbf{j}^{\mathrm{probe}} \mathbf{E}^{\mathrm{probe}} d^3\mathbf{r} = \int j^{\mathrm{probe}}_r E_r^{\mathrm{probe}} r dr dz \cos^2\phi d\phi= \pi\int j^{\mathrm{probe}}_r E_r^{\mathrm{probe}} r dr dz 
\end{equation}

The field averaged over the volume is found correspondingly:

\begin{equation}
    <E_x> = \int (E_r \cos^2 \phi - i E_\phi \sin^2\phi ) r dr dz d\phi/V=\pi\int (E_r  - i E_\phi ) r dr dz /V.
\end{equation}

\section{Analytical approach}

In some simple cases, polarizability of a particle can be easily found analytically. One of the most well-known examples is the calculation polarizability of an ellipsoid in a homogeneous medium, which is described in many books and papers (e.g. \cite{landau2013electrodynamics,jensen1999,rodriguez2012}). When the silver disk is embedded in homogeneous ambiance it can be approximated by an oblate spheroid with semiaxes of $b = 30\cdot(4/3)^{1/3} (nm)$ and $a = 10\cdot(4/3)^{1/3} (nm)$ in order to provide the same volume and axes ratio. In-plain polarizability of such ellipsoid can be found analytically in quasistatic approximation \cite{landau2013electrodynamics} as follows:
\begin{equation}
    \hat{\alpha}_{xx}^{\mathrm{static}} = \varepsilon_m \frac{V}{4\pi} \frac{\Delta\varepsilon}{\varepsilon_m +\Delta\varepsilon n_x},
\end{equation}
where $n_x$ is the depolarization coefficient, which is determined by an asymmetry of an ellipsoid.

However, this approximation is valid only for very small particles and we apply so-called modified long-wavelength approximations (MLWA) \cite{rodriguez2012,jensen1999}, which accounts for the first dynamical corrections to obtain more precise results:
\begin{equation}
    \hat{\alpha}_{xx} = \frac{\hat{\alpha}^{\mathrm{static}}_{xx}}{1-\frac{2}{3}i\frac{k^3}{\varepsilon_m}\hat{\alpha}^{\mathrm{static}}_{xx}-\frac{k^2}{b\varepsilon_m}\hat{\alpha}^{\mathrm{static}}_{xx}},
\end{equation}
where $k=\sqrt{\varepsilon_m}k_0$ is the wavevector in a medium.



\section{Rotating $\hat{\alpha}$ tensor}

In the case, when a particle does not have rotational symmetry (for instance, nanorod) an issue of particle rotation arise. Conventionally, polarizability, $\hat{\alpha}$, is calculated in a basis of main axes $x'-y'$ and therefore should be recalculated for a fixed $x-y$ basis. 

First, we consider only in-plain rotation. A particle is rotated on an angle $\varphi$ clockwise when we are looking in the direction of propagation of $z$-axis. In this way:
\begin{equation}
    \mathbf{P}_{xy}=
    \begin{pmatrix}
    \cos \varphi & -\sin \varphi \\
    \sin \varphi & \cos \varphi
    \end{pmatrix}
    \mathbf{P}_{x'y'}, \quad     \mathbf{E}_{xy}=
    \begin{pmatrix}
    \cos \varphi & -\sin \varphi \\
    \sin \varphi & \cos \varphi
    \end{pmatrix}
    \mathbf{E}_{x'y'},
\end{equation}

Therefore:
\begin{multline}
    \hat{\alpha}_{xy}=
    \begin{pmatrix}
    \cos \varphi & -\sin \varphi & 0 \\
    \sin \varphi & \cos \varphi & 0 \\
    0 & 0 & 1
    \end{pmatrix} \hat{\alpha}_{x'y'}
    {\begin{pmatrix}
    \cos \varphi & -\sin \varphi & 0 \\
    \sin \varphi & \cos \varphi & 0 \\
    0 & 0 & 1
    \end{pmatrix}}^{-1}=\\
    =\begin{pmatrix}
    \cos \varphi & -\sin \varphi & 0 \\
    \sin \varphi & \cos \varphi & 0 \\
    0 & 0 & 1
    \end{pmatrix} \hat{\alpha}_{x'y'} \begin{pmatrix}
    \cos \varphi & \sin \varphi & 0 \\
    -\sin \varphi & \cos \varphi & 0 \\
    0 & 0 & 1
    \end{pmatrix},
\end{multline}



\section{Comparison of different approaches}

Optical properties of plasmonic nanodisks are mainly determined by its in-plain component and particularly localized plasmon resonance, which shapes the spectrum. Here, we compare $\hat{\alpha}_{xx}$ calculated via different approaches and estimate their accuracy in such a way.

\begin{figure}[h]
    \centering
    \begin{subfigure}{0.495\linewidth}
    \centering
    \includegraphics[width=\linewidth]{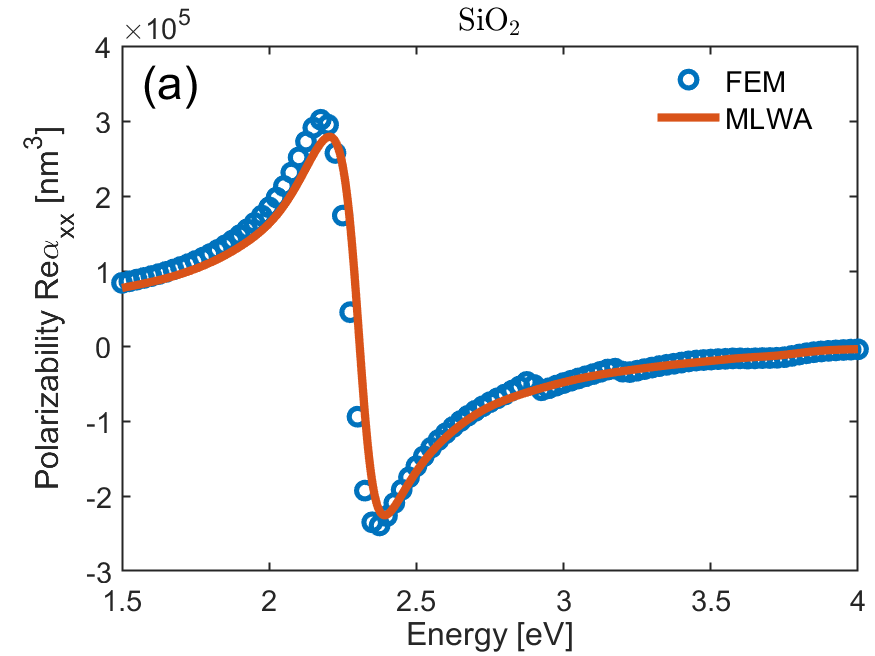}
    \end{subfigure}
    \hfill
    \begin{subfigure}{0.495\linewidth}
    \centering
    \includegraphics[width=\linewidth]{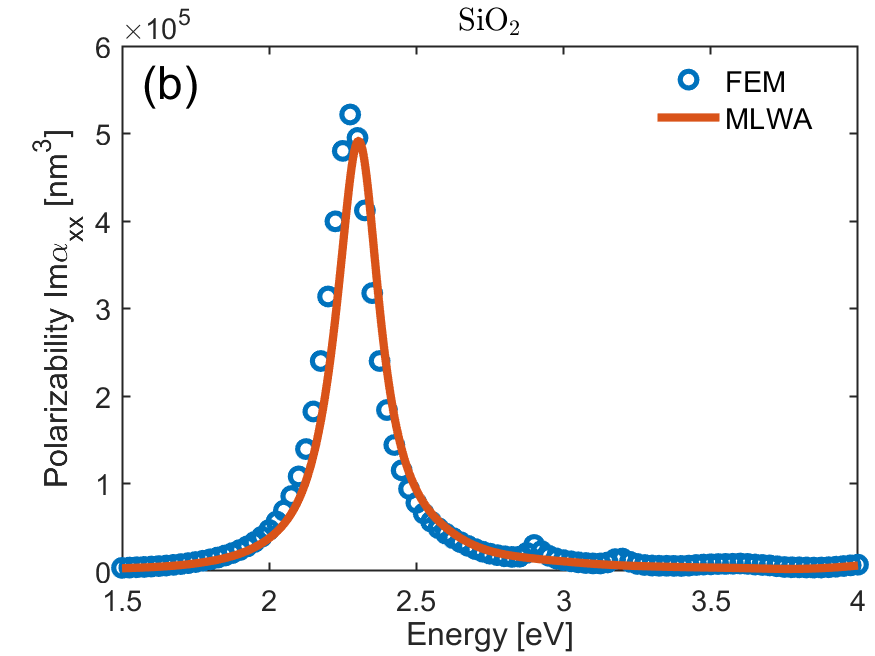}
    \end{subfigure}
    \centering
    \begin{subfigure}{0.495\linewidth}
    \centering
    \includegraphics[width=\linewidth]{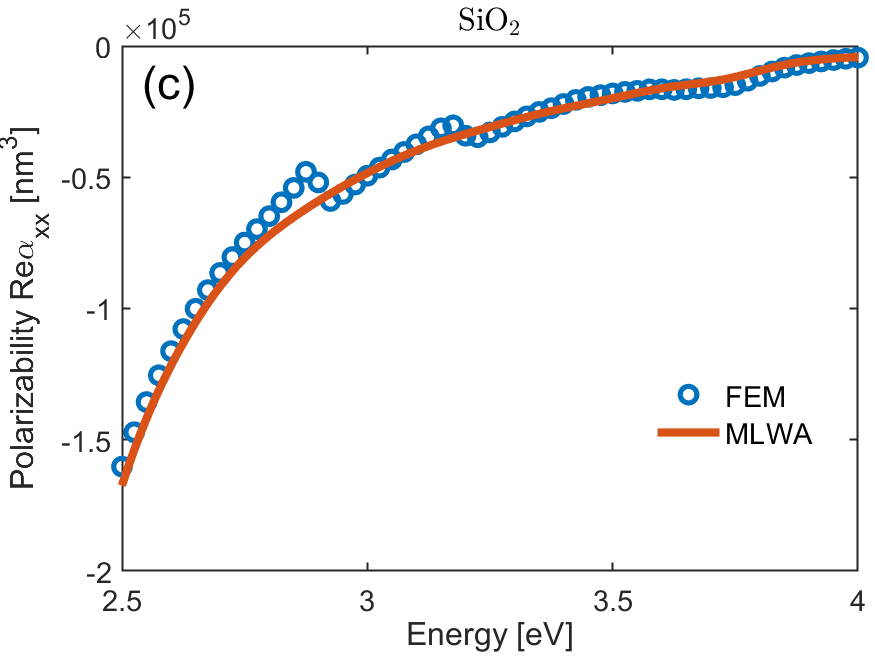}
    \end{subfigure}
    \hfill
    \begin{subfigure}{0.495\linewidth}
    \centering
    \includegraphics[width=\linewidth]{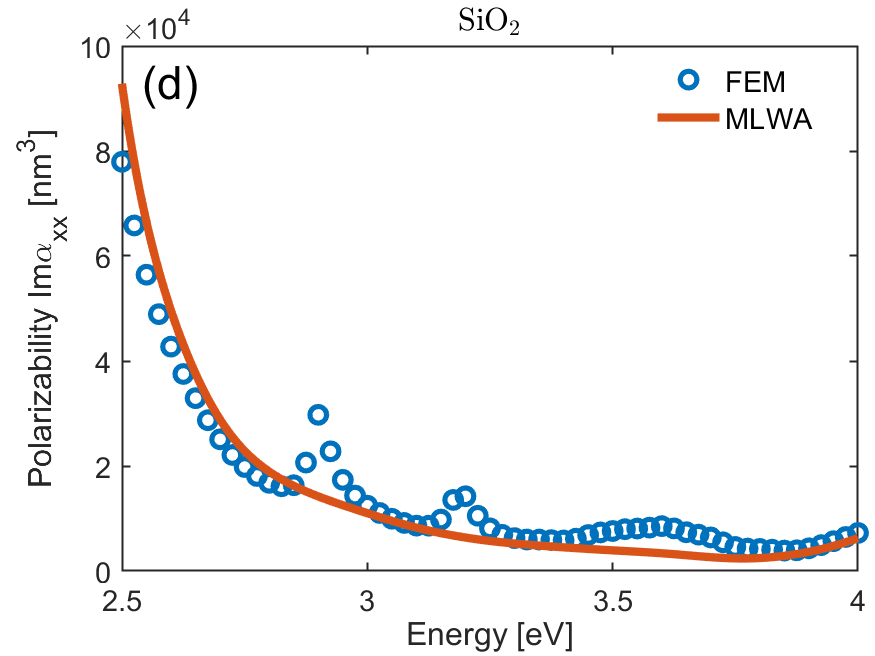}
    \end{subfigure}
    \caption{Energy dependence of in-plain component of polarizability tensor of silver nandisk in silica. Silver disk of $30$ nm radius and $20$ nm height is described by Johnson-Christy optical constants, whereas silica has $\varepsilon_{\mathrm{SiO}_2}=2.1$. Panels (a,c) and (b,d) depiсt real and imaginary parts of $\hat{\alpha}_{xx}$ correspondingly. Circle markers are related to direct calculations in COMSOL Multiphysics, while solid lines are related to MLWA analytical model.}
    \label{fig:silica_MLWA}
\end{figure}

We start with the consideration of the same silver disks in homogeneous silica and air, which is the most simple case and might be tackled with any approach. As can be seen from Fig. \ref{fig:silica_MLWA}, MLWA model generally describes the localized plasmon resonance. Although it is possible to slightly adjust $a$ and $b$ semiaxes and make the match almost perfect, it does not make sense. Indeed, if we need to conduct FEM calculations in order to find parameters for an analytical model, then this model does not provide any benefits not in time consumption nor in accuracy.

\begin{figure}[h]
    \centering
    \begin{subfigure}{0.495\linewidth}
    \centering
    \includegraphics[width=\linewidth]{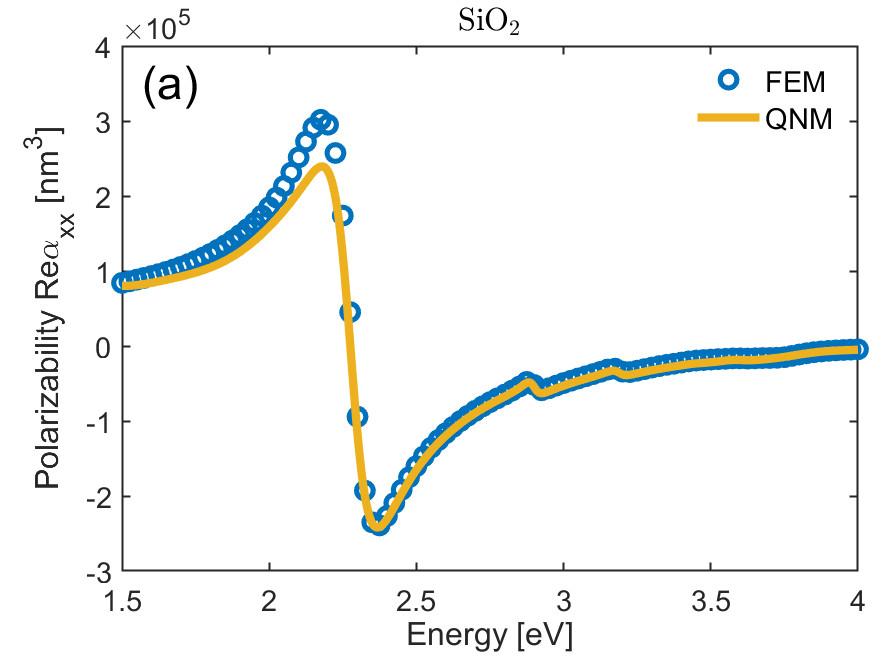}
    \end{subfigure}
    \hfill
    \begin{subfigure}{0.495\linewidth}
    \centering
    \includegraphics[width=\linewidth]{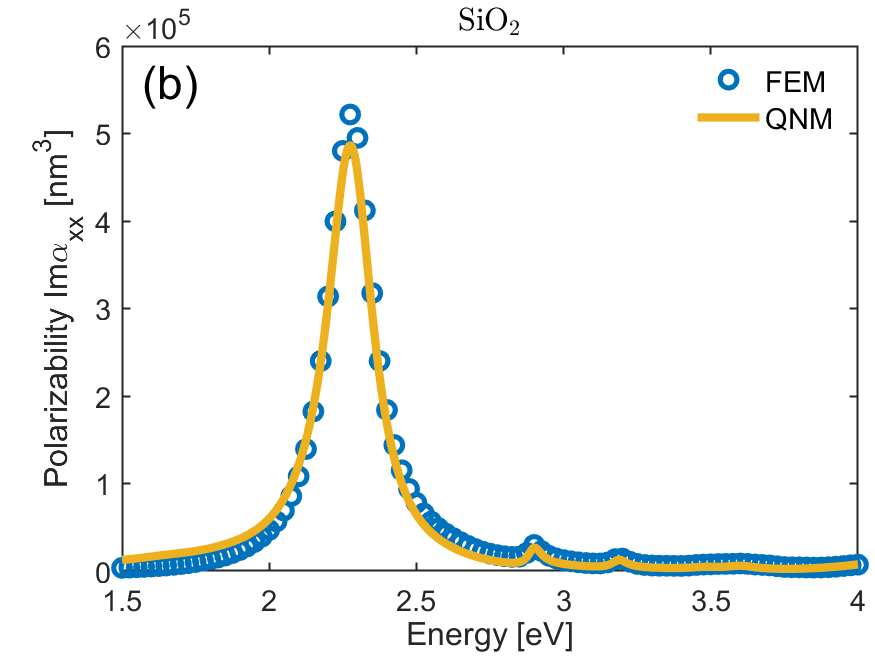}
    \end{subfigure}
    \centering
    \begin{subfigure}{0.495\linewidth}
    \centering
    \includegraphics[width=\linewidth]{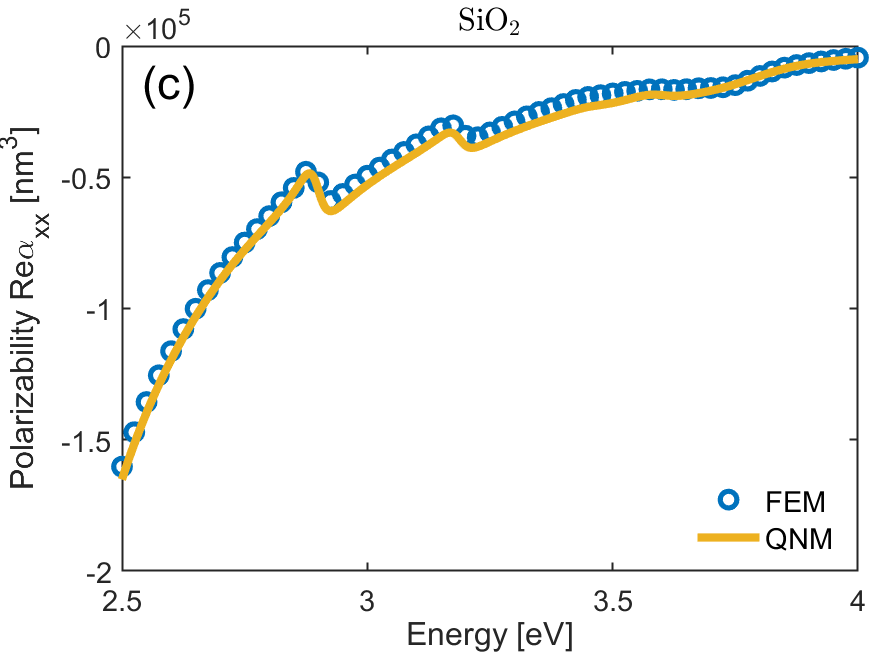}
    \end{subfigure}
    \hfill
    \begin{subfigure}{0.495\linewidth}
    \centering
    \includegraphics[width=\linewidth]{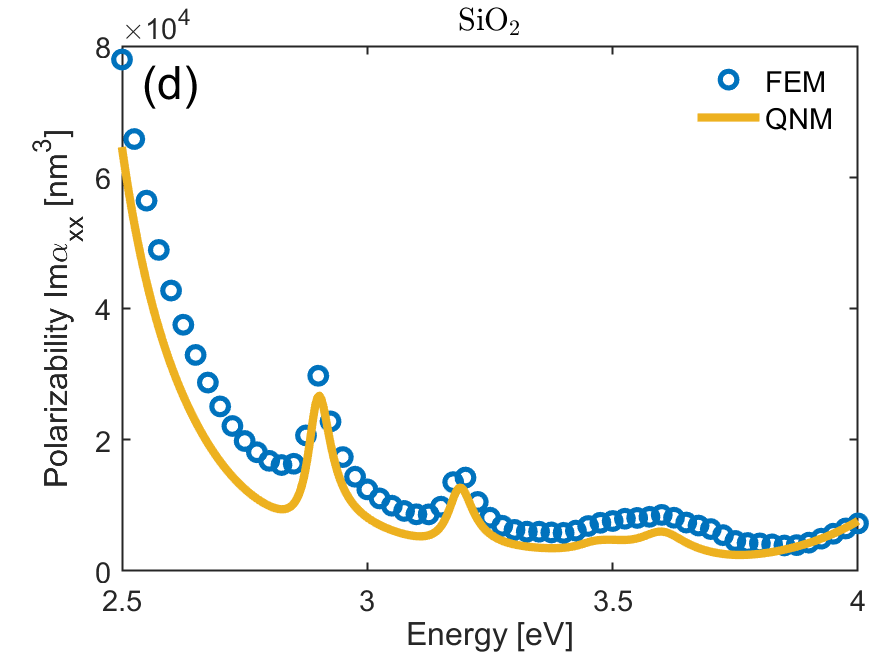}
    \end{subfigure}
    \caption{Energy dependence of in-plain component of polarizability tensor of silver nandisk in silica. Silver disk of $30$ nm radius and $20$ nm height is described by Johnson-Christy optical constants, whereas silica has $\varepsilon_{\mathrm{SiO}_2}=2.1$. Panels (a,c) and (b,d) depiсt real and imaginary parts of $\hat{\alpha}_{xx}$ correspondingly. Circle markers are related to direct calculations in COMSOL Multiphysics, while solid lines are related to quasinormal modes approach.}
    \label{fig:Silica_QNM}
\end{figure}

\begin{figure}[h]
    \centering
    \includegraphics[width=\linewidth]{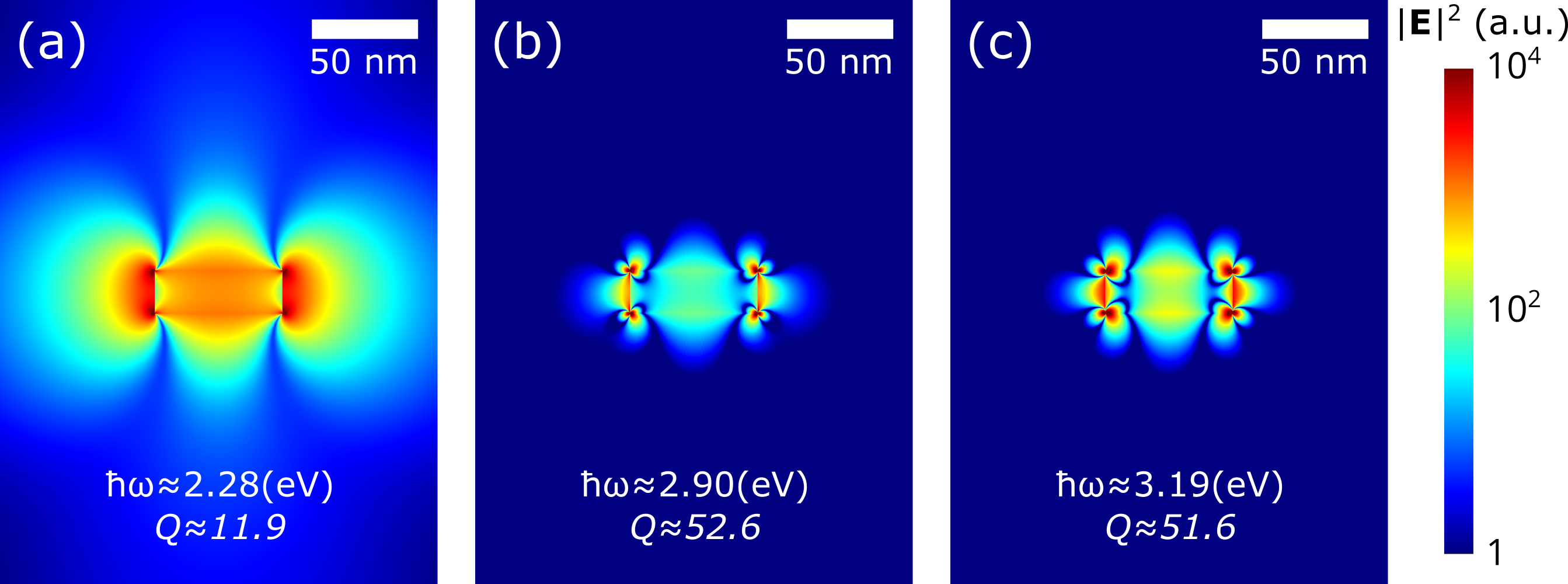}
    \caption{Distribution of electric field in eigenmodes of a silver nanodisc in silica.}
    \label{fig:Silica_QNM_fields}
\end{figure}

As can be seen from Fig. \ref{fig:silica_MLWA} (c,d) there are several side resonances, which are too complex to be described by any analytical model. At the same time, the theory of quasinormal modes is developed specifically to work with them and therefore provides us such an opportunity. Each peculiarity in the polarizability is associated with corresponding plasmonic resonance. We have found several resonances and calculated their contribution to polarizability (see Fig. \ref{fig:Silica_QNM}). It is seen that the curve calculated in resonant approximation matches the precise data in a good way and follows all the peculiarities. The only significant deviation is observed at $2.2$(eV). The reason for this phenomenon remains unveiled, but probably it is related to a significant change of permittivity on a width of the resonance.
The main resonance ($\hbar \omega \approx 2.28-0.096i$(eV), $Q\approx11.9$) is a classical LSPR, whose field is almost homogeneous inside the metal (see Fig. \ref{fig:Silica_QNM_fields} (a)). Second and third resonances are slightly red-shifted and are much narrower ($\hbar \omega \approx 2.90-0.028i$(eV), $Q\approx52.6$ and $\hbar \omega \approx 3.19-0.031i$(eV), $Q\approx51.6$ correspondingly). Distribution of their electric fields (see Fig. \ref{fig:Silica_QNM_fields} (b,c)) shows that most probably they are formed as a hybridization of LSPR with so-called edge plasmons. Although this result is clearly observed theoretically and helps us to show the power of resonant approximation, in practice it will be almost impossible to observe this phenomenon. The edge rounding by a few nanometers fillet (which will obviously occur in the experiment) almost destroys these modes. The low and wide peak at $3.5-3.7$(eV) constitutes from many closely-pitched resonances. It does not make sense to resolve all the modes and consider the peak in resonant approximation, therefore we just found several modes to confirm their existence and nature of the peak.

Silver disk in the air has very similar optical properties. The wonderful fact is that although resonances are strongly red-shifted and have completely different quality factors, distributions of electric fields are almost unchanged. They are so similar that it is impossible to notice any difference by eye.

\begin{figure}[h]
    \centering
    \begin{subfigure}{0.495\linewidth}
    \centering
    \includegraphics[width=\linewidth]{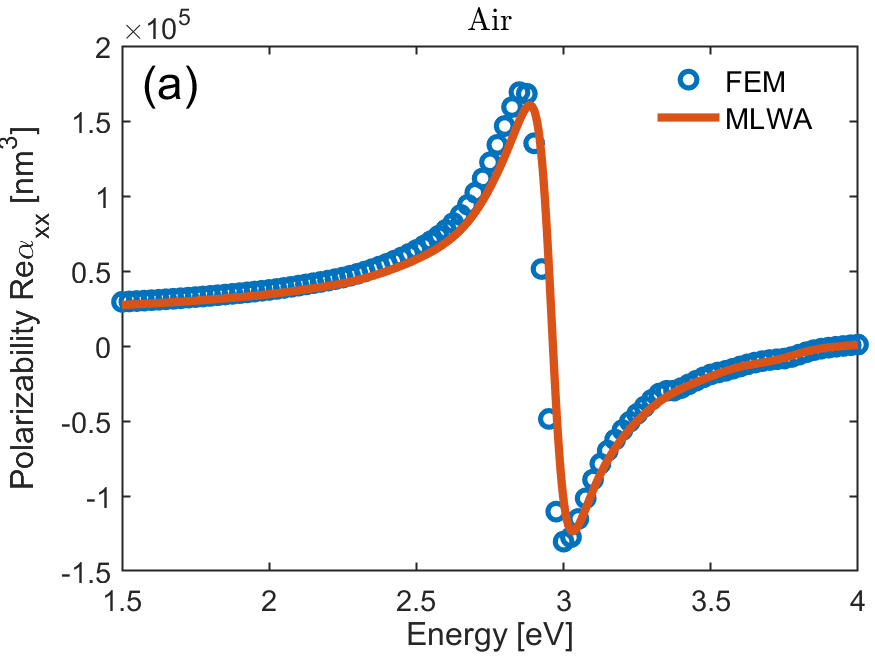}
    \end{subfigure}
    \hfill
    \begin{subfigure}{0.495\linewidth}
    \centering
    \includegraphics[width=\linewidth]{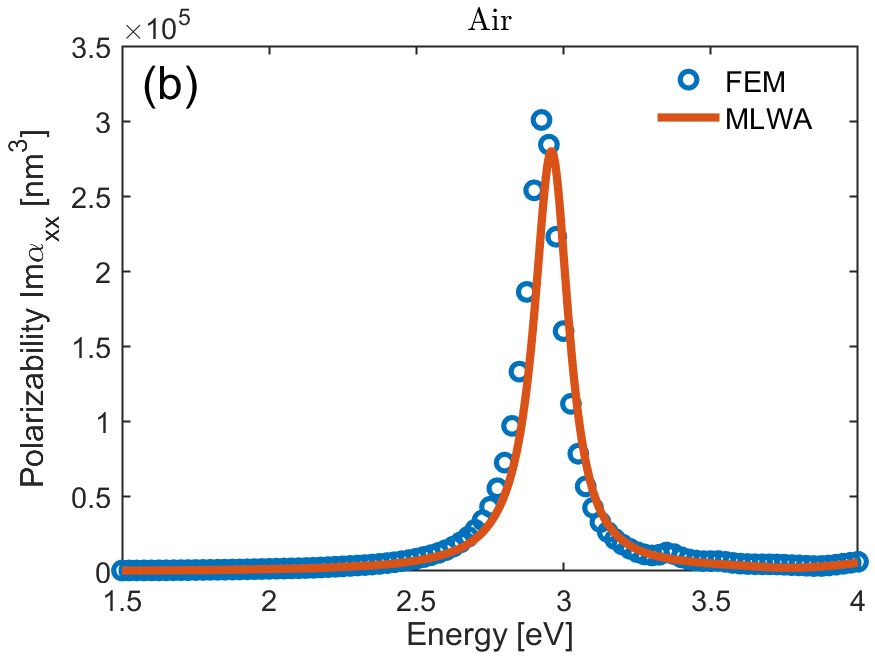}
    \end{subfigure}
    \centering
    \begin{subfigure}{0.495\linewidth}
    \centering
    \includegraphics[width=\linewidth]{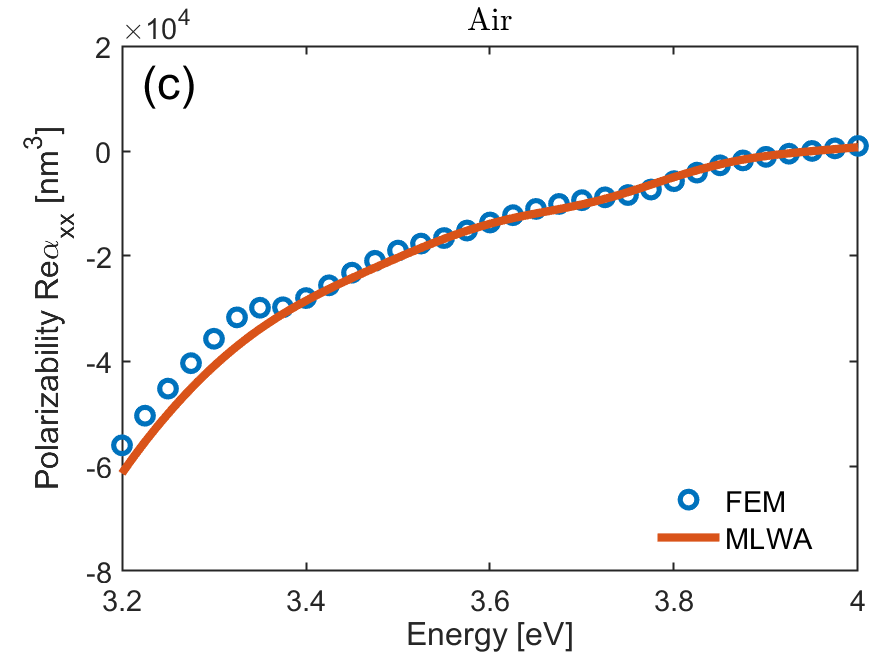}
    \end{subfigure}
    \hfill
    \begin{subfigure}{0.495\linewidth}
    \centering
    \includegraphics[width=\linewidth]{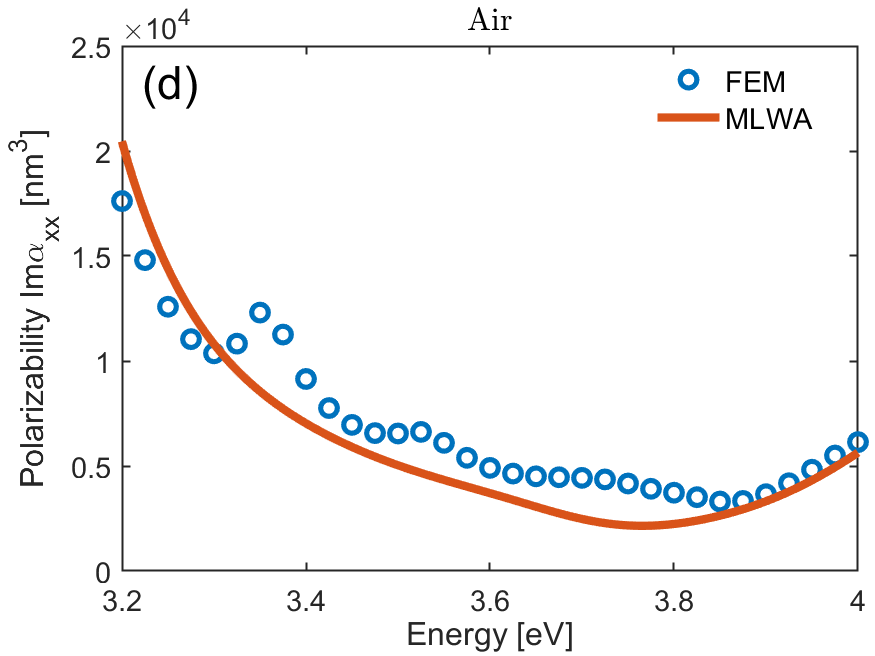}
    \end{subfigure}
    \caption{Energy dependence of in-plain component of polarizability tensor of silver nandisk in air. Silver disk of $30$ nm radius and $20$ nm height is described by Johnson-Christy optical constants. Panels (a,c) and (b,d) depiсt real and imaginary parts of $\hat{\alpha}_{xx}$ correspondingly. Circle markers are related to direct calculations in COMSOL Multiphysics, while solid lines are related to MLWA analytical model.}
\end{figure}

\begin{figure}[h]
    \centering
    \begin{subfigure}{0.495\linewidth}
    \centering
    \includegraphics[width=\linewidth]{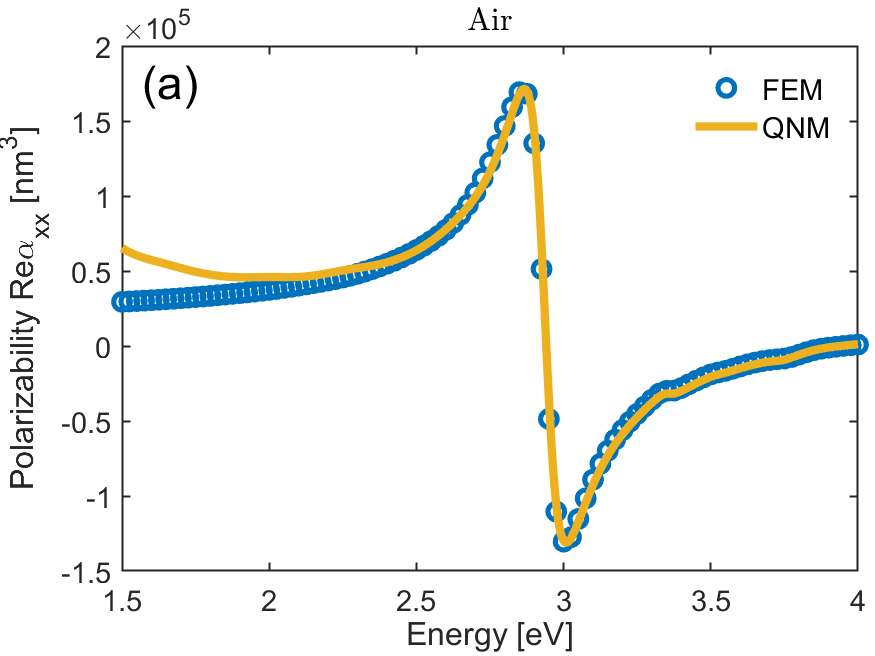}
    \end{subfigure}
    \hfill
    \begin{subfigure}{0.495\linewidth}
    \centering
    \includegraphics[width=\linewidth]{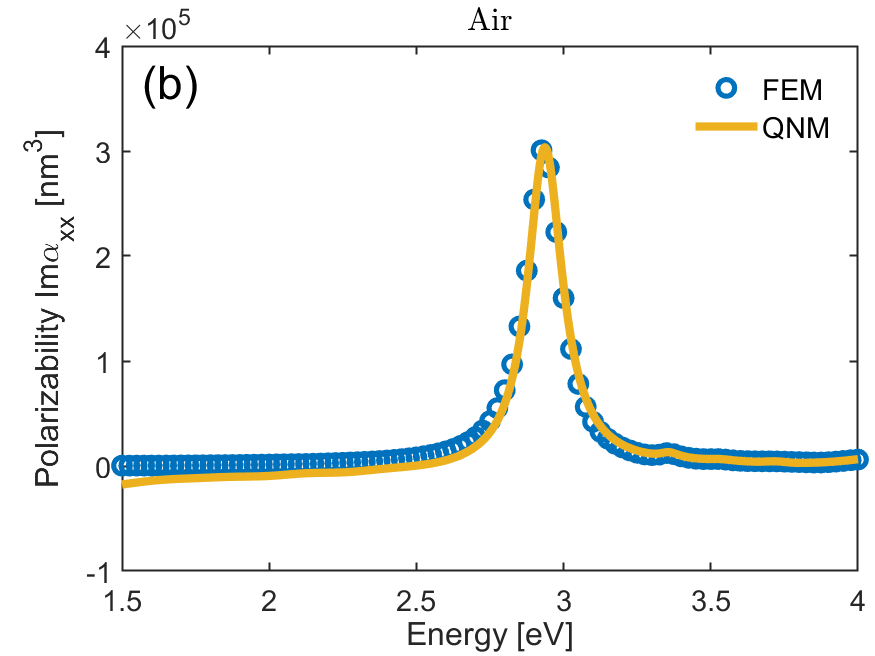}
    \end{subfigure}
    \centering
    \begin{subfigure}{0.495\linewidth}
    \centering
    \includegraphics[width=\linewidth]{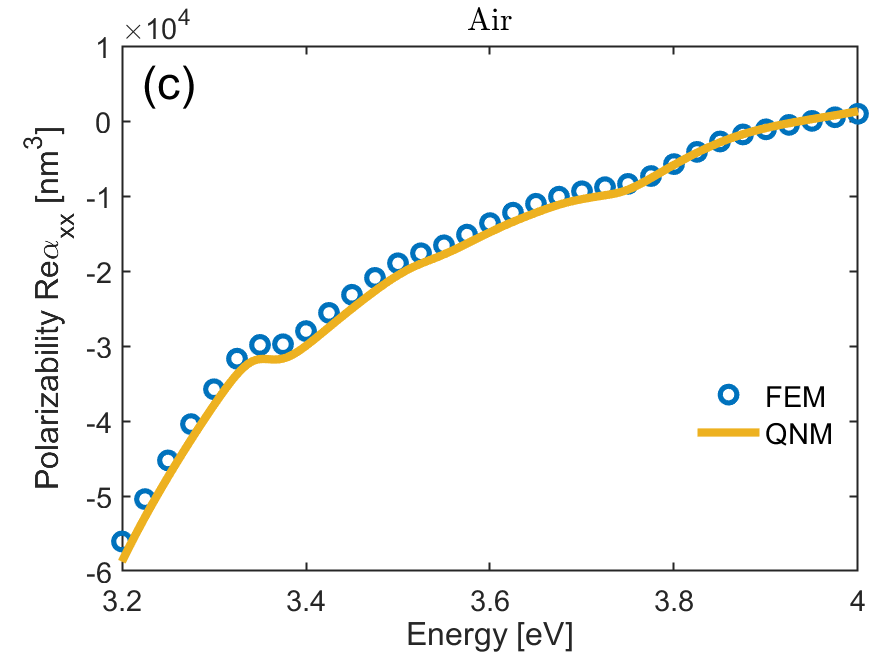}
    \end{subfigure}
    \hfill
    \begin{subfigure}{0.495\linewidth}
    \centering
    \includegraphics[width=\linewidth]{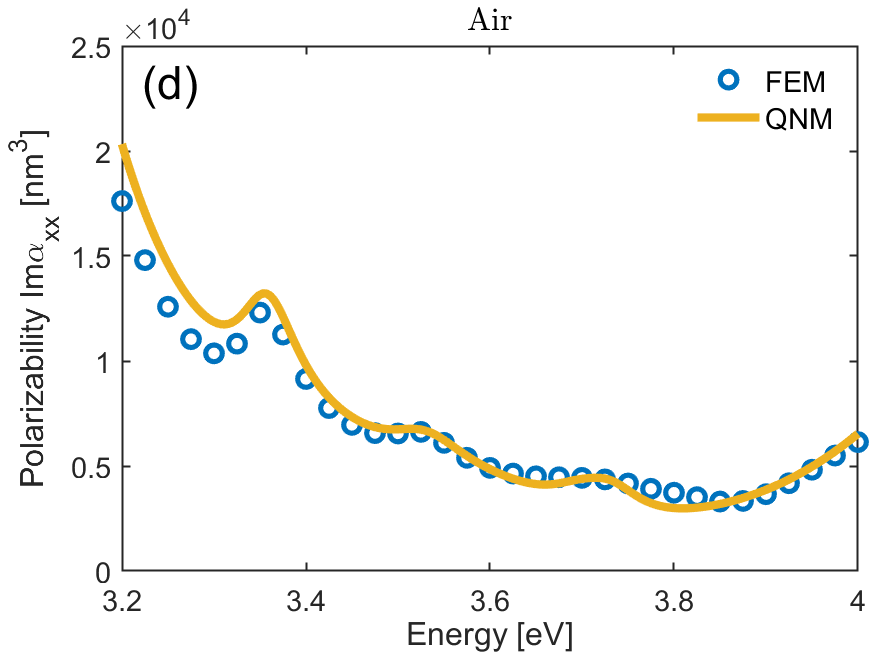}
    \end{subfigure}
    \caption{Energy dependence of in-plain component of polarizability tensor of silver nandisk in air. Silver disk of $30$ nm radius and $20$ nm height is described by Johnson-Christy optical constants. Panels (a,c) and (b,d) depiсt real and imaginary parts of $\hat{\alpha}_{xx}$ correspondingly. Circle markers are related to direct calculations in COMSOL Multiphysics, while solid lines are related to quasinormal modes approach.}
\end{figure}

\begin{figure}[h]
    \centering
    \includegraphics[width=\linewidth]{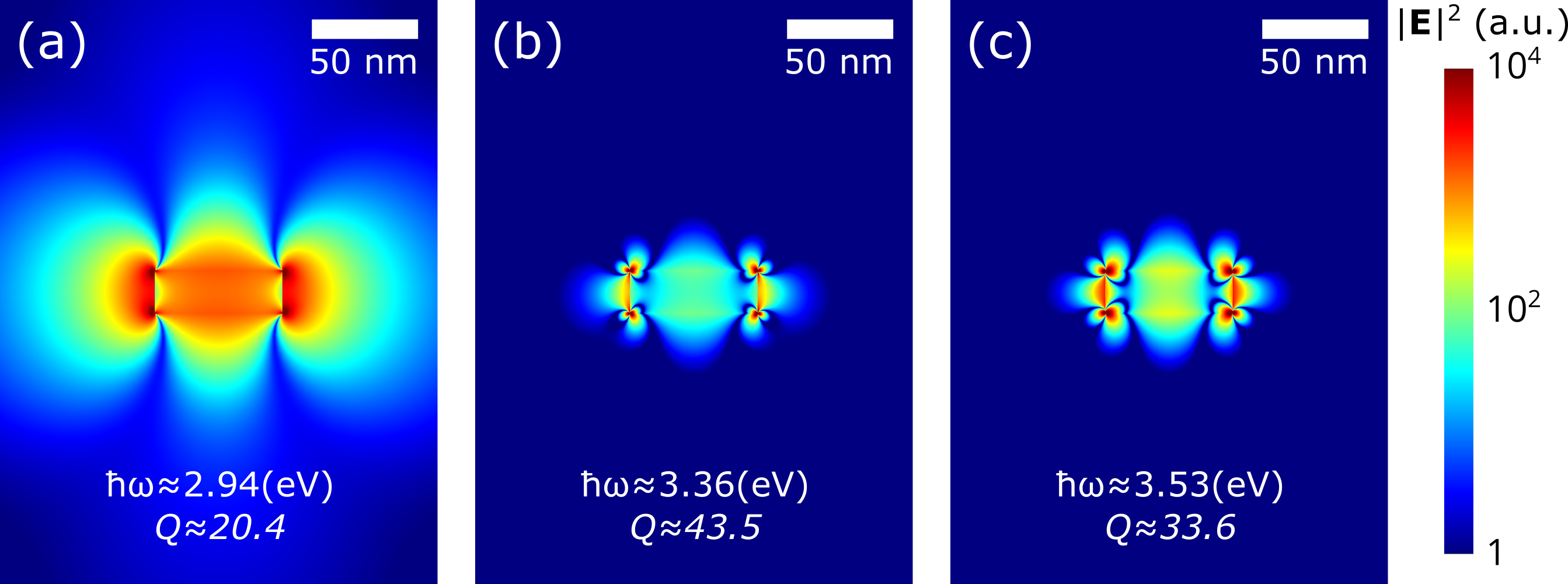}
    \caption{Distribution of electric field in eigenmodes of a silver nanodisc in air.}
    \label{fig:Air_QNM_fields}
\end{figure}

\clearpage

\section{Crossing the interface}

As it was discussed above, two most commonly occurring cases are a particle in a homogeneous ambiance or placed onto an interface. The first one is the most simple for theoretical consideration and provides several unique phenomena, whereas the second one is the most natural for experimental realization. However, from the theoretical point of view, it is very interesting to observe the transition between these states and to study the dependence of the polarizability on the distance from the interface, at least on a certain example.

\begin{figure}[h]
    \centering
    \includegraphics[width=0.3\linewidth]{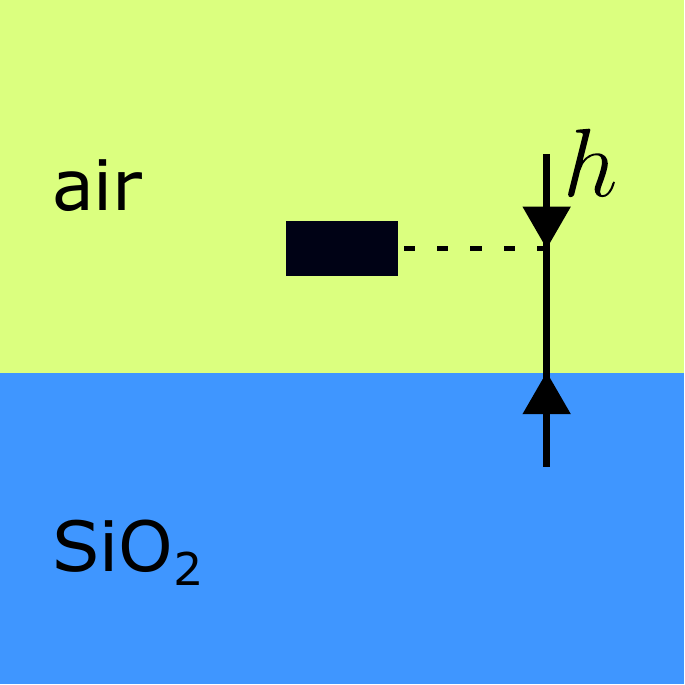}
    \caption{Schematic of a plasmonic disk near an air/silica interface.}
    \label{fig:disk_near_interface}
\end{figure}

\begin{figure}[h]
    \centering
    \begin{subfigure}{0.495\linewidth}
    \centering
    \includegraphics[width=\linewidth]{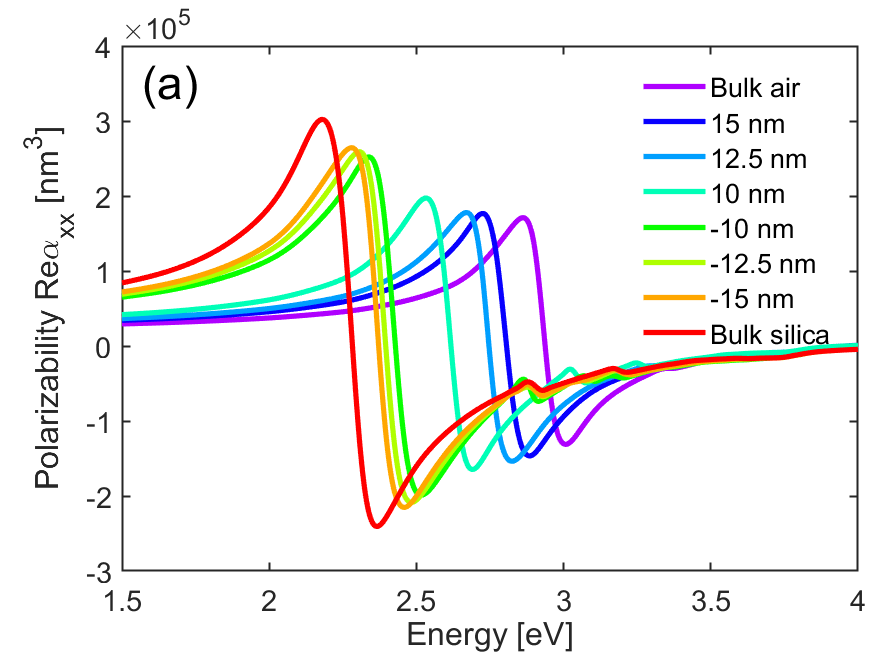}
    \end{subfigure}
    \hfill
    \begin{subfigure}{0.495\linewidth}
    \centering
    \includegraphics[width=\linewidth]{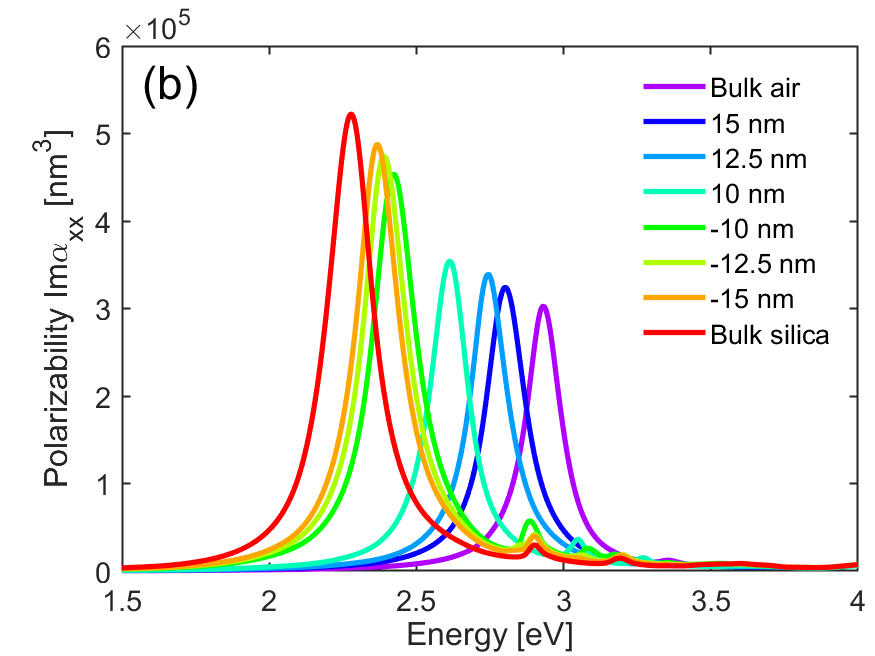}
    \end{subfigure}
    \centering
    \begin{subfigure}{0.495\linewidth}
    \centering
    \includegraphics[width=\linewidth]{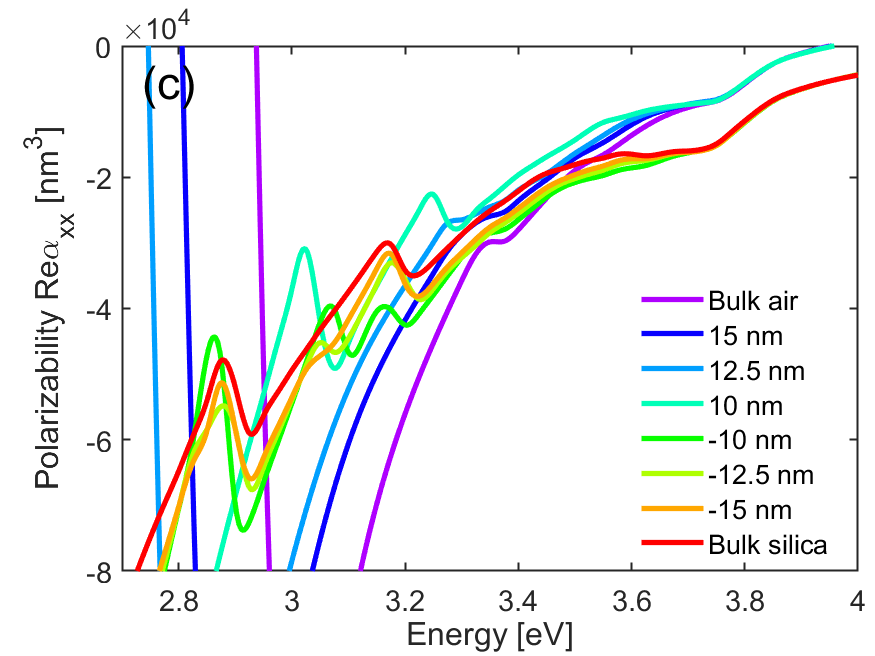}
    \end{subfigure}
    \hfill
    \begin{subfigure}{0.495\linewidth}
    \centering
    \includegraphics[width=\linewidth]{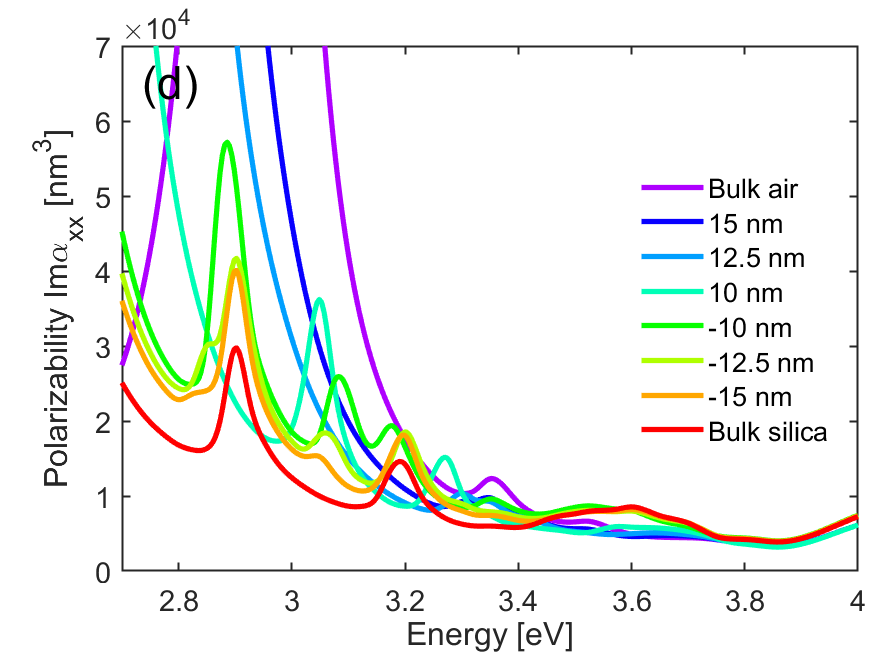}
    \end{subfigure}
    \caption{Energy dependence of in-plain component of polarizability tensor of silver nandisk near the air/silica interface. Silver disk of $30$ nm radius and $20$ nm height is described by Johnson-Christy optical constants. Distance $h$ specified in the legend shows the spacing between the center of the disk and the interface. Panels (a,c) and (b,d) depiсt real and imaginary parts of $\hat{\alpha}_{xx}$ correspondingly. Solid lines are obtained by an interpolation of calculations in COMSOL Multiphysics.}
    \label{fig:Interface}
\end{figure}

We have already considered silver nanodisks in silica and air environments, so here, we place the particle on different distances from their interface (see Fig. \ref{fig:disk_near_interface}). Figure \ref{fig:Interface} shows how the polarizability tensor transforms while crossing the interface. The distance specified in the legend shows the spacing between the center of the disk and the interface. In this way lines related to $\pm10$(nm) distance correspond to the particles touching an interface from different sides.

The most important fact is that the particle moved away from the border by only 5 nanometers behaves almost as in a bulk medium. This means that polarizability is mainly determined by the local environment and therefore it is not possible to predict the polarizability of a particle on an interface, knowing only bulk polarizabilities, since it strongly depends on the shape of a certain particle.

The behavior of the side resonances depicted in Fig. \ref{fig:Interface} (c-d) is too complex to be analyzed from this graph.

\clearpage
\newpage

\chapter{Sum calculation} \label{App sum}

{\section{Green's function filtering}}\label{App:B1}
First of all, since in this paper we consider only examples of environments, which have translational symmetry in the $x-y$ plane, and all the particles are located in the same plane, for a fixed $z$ {coordinate}, Green's function depends only on the difference of coordinates. Therefore, sum \ref{eq:4} transforms to a slightly simpler form:
\begin{equation}
    \hat{C}(\mathbf{k}_\parallel)=
    \sum_{j\neq 0}
    \hat{G}(\mathbf{t}_j,z=z_p) e^{-i\mathbf{k}_\parallel\mathbf{t}_j},
    \label{eq:B1}
\end{equation}
where $\mathbf{t}_j$ is the $j$-th translational vector of a lattice in real space.

Typically this sum converges very slowly. For instance, in a homogeneous medium dyadic Green's function $\hat{G}_{\alpha\beta}(\mathbf{r})=k_0^2(\delta_{\alpha\beta}+\frac{1}{ k^2}\partial_{\alpha}\partial_{\beta})\frac{e^{i k r}}{r}$ (where $k=\sqrt{\varepsilon}_m k_0$ is a wavevector in a medium) decays as $e^{ikr}/r$, which makes the sum calculation not a trivial problem. Various methods have been developed for an efficient calculation of a lattice sum for homogeneous ambience \cite{ewald1921,stevanovic2006,papanicolaou1999,wette1965,Belov2005}. Many of them, starting from the classical Ewald's method are based on the following idea: if a lattice sum of a function converges slowly in real space, then this function can be represented as a sum of two auxiliary functions. The first one accounts for high gradients and decays very fast at infinity, whereas the second one, on the contrary, should be very smooth. In this way, the sum of the first function converges fast in real space and the sum of the second one can be calculated efficiently in reciprocal space (via the Poisson formula \cite{collin1960}).

\begin{figure}[h]
    \centering
    \includegraphics[width=1\linewidth]{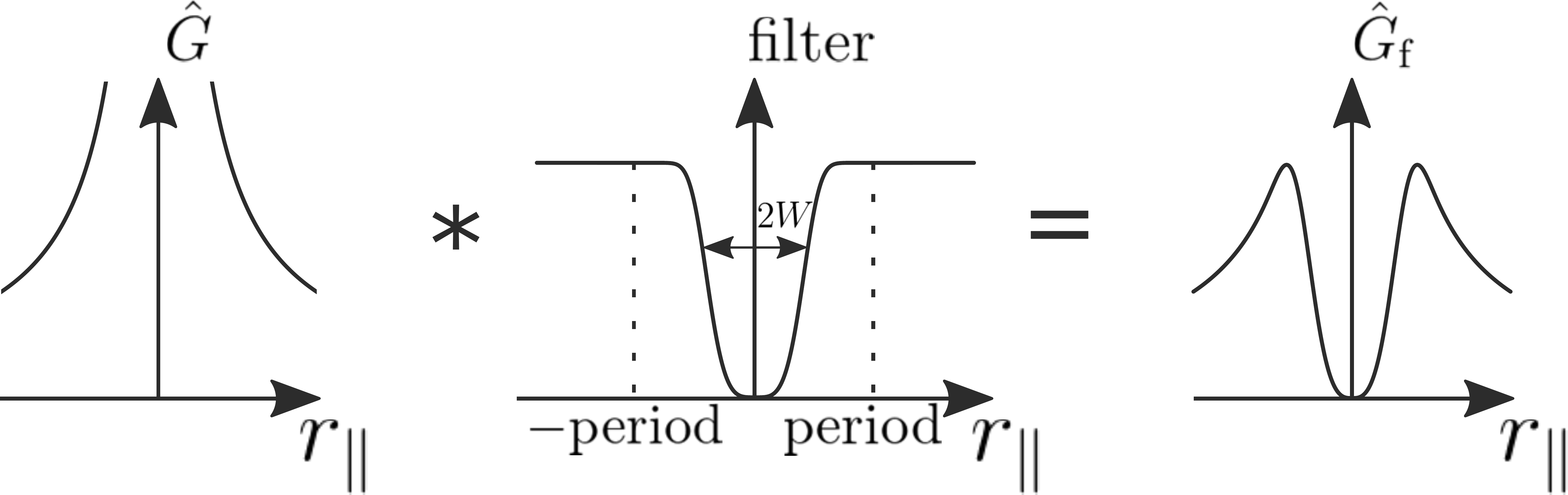}
    \caption{Schematic of filtering of the dyadic Green's function. {$W$ is the width of the filter, which should be less than a period.}}
    \label{fig:filter}
\end{figure}

Following this idea, we implement a very similar method, which helps us to obtain fast convergence for lattices both in homogeneous ambiance and on an interface between two media and has very clear physical sense. Green's function has high gradients only in the proximity of the zero point, but in our case, this point is not included in the sum, since self-action of a particle is already accounted for by the polarizability tensor, $\Hat{\alpha}$.
This fact gives us an opportunity to split the Green's function in such a way that the contribution of the first, high-gradient function is negligible. 
In order to do that we multiply the original Green's function by a special filter, which is equal to zero at the zero point and {tends} to unity for distances larger or equal to the period of a structure $\hat{G}_{\mathrm{f}}(\mathbf{r})=f(\mathbf{r})\hat{G}(\mathbf{r})$ (see Fig. \ref{fig:filter}). In this way the summation can be conducted over all the nodes of a lattice, having exactly the same result:
\begin{equation}
    \hat{C}(\mathbf{k}_\parallel)=
    \sum_{j\neq 0}
    \hat{G}(\mathbf{t}_j) e^{-i\mathbf{k}_\parallel     \mathbf{t}_j}=
    \sum_{j}
    \hat{G}_{\mathrm{f}}(\mathbf{t}_j) e^{-i\mathbf{k}_\parallel     \mathbf{t}_j}
    =
    \frac{4\pi^2}{s}
    \sum_j
    \hat{M}_{\mathrm{f}}(\mathbf{k}_\parallel+\mathbf{g}_j),
    \label{eq:B2}
\end{equation}
where $\hat{M}_{\mathrm{f}}(\mathbf{k}_\parallel) = \frac{1}{4\pi^2}\int \hat{G}_{\mathrm{f}}(\mathbf{r})e^{-i\mathbf{k}_\parallel\mathbf{r}}d^2\mathbf{r}$, is the filtered dyadic Green's function in reciprocal space, $s$ is an area of a unit cell in real space and $\mathbf{g}_j$ is a $j$-th vector of reciprocal lattice. If the filtered Green's function, $\hat{G}_{\mathrm{f}}$, is a smooth enough that it has $n$ derivatives and all of them are absolutely integrable, then its Fourier image $\hat{M}_{\mathrm{f}}(\mathbf{k}_\parallel) = o(k_\parallel^{-n})$ for $k_\parallel\rightarrow \infty$, which provides fast asymptotic convergence. 

This simple approach not only provides fast convergence but also has a clear physical meaning. Indeed, if we apply a filter, which is smooth, then each summand in reciprocal space has a sense of a contribution of corresponding diffraction order in the interaction between particles. Therefore, in most cases, it is possible to account for the same number of diffraction orders in sum calculation and in the scattering matrix $\mathbb{S}$, which is very convenient. 

However, calculation of Fourier image of a filtered Green's function $\hat{M}_{\mathrm{f}}(\mathbf{k}_\parallel)$ is not always an easy task. When the homogeneous space is considered, then it is possible to compose such a filtering function, that Fourier transform can be calculated analytically. But, when a particle near an interface is considered, then expressions become too complex. Since in real space filtering is just a multiplication then in reciprocal space {we deal with a} convolution $\hat{M}_{\mathrm{f}}(\mathbf{k}_\parallel) = (F*\hat{M})(\mathbf{k}_\parallel)$, where $F(\mathbf{k}_\parallel)=\frac{1}{4\pi^2}\int f(\mathbf{r}_\parallel)e^{-i\mathbf{k}_\parallel\mathbf{r}}d^2\mathbf{r}$ and $\hat{M}(\mathbf{k}_\parallel)=\frac{1}{4\pi^2}\int \hat{G}(\mathbf{r}_\parallel)e^{-i\mathbf{k}_\parallel\mathbf{r}_\parallel}d^2\mathbf{r}_\parallel$. However, since filtering function tends to unity at infinity, it is convenient to introduce complementary filtering function, $\bar{f}(\mathbf{r}_\parallel) = 1-f(\mathbf{r}_\parallel)$, for further derivations. In this way we introduce $\hat{M}_{\bar{\mathrm{f}}}(\mathbf{k}_\parallel) = (\bar{F}*\hat{M})(\mathbf{k}_\parallel)$, which helps to find the originally filtered Green's function as
$\hat{M}_{\mathrm{f}}(\mathbf{k}_\parallel) =\hat{M}(\mathbf{k}_\parallel)-\hat{M}_{\bar{\mathrm{f}}}(\mathbf{k}_\parallel)$.

\section{Green's function near an interface}

Here, we consider Green's function for an emitter located in the proximity of an interface between two media since it might be easily reduced to a case of a homogeneous medium. It is known, that this function can be expressed as a sum of two parts, $\hat{M} = \hat{M}^{{0}} + \hat{M}^r$ \cite{novotny2012}. The first term is just Green's function of the homogeneous medium and the second one accounts for the field reflected from the boundary, which is naturally calculated in reciprocal space because it requires just multiplication of amplitudes of plane waves by Fresnel coefficients.

Green's function for a homogeneous ambience in reciprocal space $\hat{M}^{{0}}(\mathbf{k}_\parallel)$ can be expressed as a sum of $s$- and $p$- polarized contributions:

\begin{multline}
\hat{M}^{{0,}\pm} =\hat{M}^{{0,}\pm}_s +\hat{M}^{{0,}\pm}_p=
\frac{i k_0^2}{2\pi k_z k_\parallel^2}\left(
\begin{array}{ccc}
     k_y^2& -k_x k_y & 0 \\
     -k_x k_y& k_x^2& 0\\
     0 & 0 & 0
\end{array}
\right)+\\
+
\frac{ik_0^2}{2\pi k^2 k_\parallel^2}\left(
\begin{array}{ccc}
     k_x^2 k_z& k_x k_y k_z & \mp k_x k_\parallel^2 \\
     k_x k_y k_z& k_y^2 k_z & \mp k_y k_\parallel^2\\
     \mp k_x k_\parallel^2 & \mp k_y k_\parallel^2 & k_\parallel^4/k_z
\end{array}
\right),
\label{eq:B3}
 \end{multline}
where $k_{z}=\sqrt{k^2-\mathbf{k}_\parallel^2}$ is the $z$ component of the wavevector (positive imaginary part should be chosen for $k_\parallel>k$). Sign $\pm$ corresponds to upper/lower half-spaces, however{,} when we consider plane $z=z_p${,} the choice of the sign does not matter and all the alternating-sign components finally do not make any contribution.

Splitting of the tensor into two terms corresponding to different polarizations is very convenient for an accounting of reflection from an interface. We just have to multiply each term by corresponding reflection coefficient, choose appropriate signs {keeping in mind} the fact that the reflected wave propagates in the direction opposite to an incident one and take into account an additional phase, which is gained during the propagation to the interface and back.

Without loss of generality, we consider a lattice of particles placed above an interface, whose centers are located at a distance of $h$ from a boundary {(see Fig. \ref{fig:2cases} (b))}. The reflected part of Green's function has the following form:

\begin{multline}
    \hat{M}^r = \hat{M}^r_s + \hat{M}^r_p =
e^{2ik_{z}h}\Bigl[r_s(k_\parallel)\frac{i k_0^2}{2\pi k_z k_\parallel^2}\left(
\begin{array}{ccc}
     k_y^2& -k_x k_y & 0 \\
     -k_x k_y& k_x^2& 0\\
     0 & 0 & 0
\end{array}
\right)-\\
-
r_p(k_\parallel)
\frac{ik_0^2}{2\pi k^2 k_\parallel^2}\left(
\begin{array}{ccc}
     k_x^2 k_z& k_x k_y k_z &  k_x k_\parallel^2 \\
     k_x k_y k_z& k_y^2 k_z & k_y k_\parallel^2\\
     - k_x k_\parallel^2 & - k_y k_\parallel^2 & - k_\parallel^4/k_z
\end{array}\right)\Bigr],
\label{eq:B4}
\end{multline}
where $r_s$ and $r_p$ are Fresnel reflection coefficients for corresponding polarizations, and $e^{2ik_zh}$ is a phase delay.

\section{Convenient representation of $\hat{M}$ matrices}

Although, formal expressions and representations  for Green's functions such as \ref{eq:B3},\ref{eq:B4} are very well known and can be used in any purposes, it is very useful to study them in a rigorous way. Understanding the structure of these tensors as well as their possible representations helps to deal with them efficiently both analytically and numerically.

We start from the definition of Green's function for a homogeneous ambiance  $\hat{M}^0(\mathbf{k}_\parallel)$:
\begin{multline}
    \hat{M}^0_{\alpha\beta}(\mathbf{k}_\parallel)=\frac{1}{4\pi^2}\int\left[ k_0^2(\delta_{\alpha\beta}+\frac{1}{ k^2}\partial_{\alpha}\partial_{\beta})\frac{e^{i k r_\parallel}}{r_\parallel}\right]e^{-i\mathbf{k}_\parallel\mathbf{r}_\parallel}d^2\mathbf{r}_\parallel=\\=\frac{k_0^2}{4\pi^2}(\delta_{\alpha\beta}-\frac{k_{\alpha}k_{\beta}}{ k^2})\int\frac{e^{i k r_\parallel}}{r_\parallel}e^{-i\mathbf{k}_\parallel\mathbf{r}_\parallel}d^2\mathbf{r}_\parallel.
    \label{eq:B5} 
\end{multline}
The latter integral can be easily calculated:
\begin{equation}
\int\frac{e^{i k r_\parallel}}{r_\parallel}e^{-i\mathbf{k}_\parallel\mathbf{r}_\parallel}d^2\mathbf{r}_\parallel = \int_0^\infty \int_0^{2\pi}\frac{e^{i k r_\parallel}}{r_\parallel}e^{-i{k}_\parallel {r}_\parallel \cos\phi} r_\parallel d{r}_\parallel d\phi
= 2\pi \int_0^\infty e^{i k r_\parallel}J_0({k}_\parallel {r}_\parallel )  d{r}_\parallel=\frac{2\pi i}{k_z},
\label{eq:B6}
\end{equation}
where $k_z>0$ for $k_\parallel<k$ and $\mathrm{Im} k_z>0$ for $k_\parallel>k$. It is important to understand, that $k_z$ in Eqn. \ref{eq:B5} emerges from the differentiation and therefore has different signs depending on the half-space, which is considered. It is the only place in the whole text in which we do not state the sign of $k_z$.

We know, that electromagnetic wave, which propagates in a homogeneous medium are perpendicular, which means that $\mathbf{k}\mathbf{\epsilon}=0$, where $\epsilon$ is a vector of polarization (not necessarily unit). Therefore, for a fixed direction of wave propagation subspace of its polarization is two-dimensional, which means that it makes sense to switch $xyz$-basis to a basis of some accessible polarizations. In this way, the following expression acts as the unity operator on an electric field (and $\hat{M}$ as a consequence):
\begin{equation}
    \hat{1}=\sum_{p=1}^2|\epsilon_p><\epsilon_p|,\label{eq:B7}
\end{equation}
where $p$ denotes the polarization. Application of this operator to $\hat{M}$ matrix gives its elegant representation \cite{wijnands1997}:

\begin{multline}
\hat{M}_{\alpha \beta}^{0,\pm}
=\sum_\gamma \sum_p(\epsilon_p^\pm[\mathbf{k}_\parallel])_\alpha(\epsilon_p^{*\pm}[\mathbf{k}_\parallel])_\gamma \hat{M}_{\gamma \beta}^{0,\pm}=\\=\frac{i k_0^2 }{2\pi k_z} \sum_\gamma \sum_p(\epsilon_p^\pm[\mathbf{k}_\parallel])_\alpha(\epsilon_p^{*\pm}[\mathbf{k}_\parallel])_\gamma(\delta_{\gamma\beta}-\frac{k_{\gamma}k_{\beta}}{ k^2})
=\frac{i k_0^2 }{2\pi k_z}  \sum_p(\epsilon_p^\pm[\mathbf{k}_\parallel])_\alpha(\epsilon_p^{*\pm}[\mathbf{k}_\parallel])_\beta,
\label{eq:B8}
\end{multline}
where $(\epsilon_p^\pm[\mathbf{k}_\parallel])_\alpha$ is a polarization vector defined by polarization $p$, in-plain component of a wavevector $\mathbf{k}_\parallel$ and sign defining the direction of propagation, $(\epsilon_p^{*\pm}[\mathbf{k}_\parallel])_\alpha$ is a reciprocal vector, which is orthogonal to vectors of another polarization and gives unity under scalar product with vector of the same polarization.

It should be noted, that any basis can be implemented in this formula. Most commonly used is the basis of $s$- and $p$-polarized waves.
\begin{equation}
    \epsilon_{s}^{\pm}=\epsilon_{s}^{*\pm}=\frac{1}{k_\parallel}
\left (
\begin{array}{c}
     -k_y  \\
     k_x   \\
     0
\end{array}
\right )
\qquad
\epsilon_{p}^{\pm}=\epsilon_{p}^{*\pm}=\frac{1}{kk_\parallel}
\left (
\begin{array}{c}
     \mp k_xk_z  \\
     \mp k_yk_z   \\
     k_{\parallel}^2
\end{array}
\right )\label{eq:B9_2}
\end{equation}
Explicit representation of \ref{eq:B8} sum in this basis results in well-known expression \ref{eq:B3}. However, in some cases another basis might be more convenient. For instance, $xy$ one is commonly used in RCWA:
\begin{equation}
    \epsilon_{x}^{\pm}=
\left (
\begin{array}{c}
     1  \\
     0   \\
     \mp \frac{k_x}{k_z}
\end{array}
\right )
\qquad
\epsilon_{y}^{\pm}=
\left (
\begin{array}{c}
     0  \\
     1   \\
     \mp \frac{k_y}{k_z}
\end{array}
\right ),\label{eq:B10_2}
\end{equation}
\begin{equation}
    \epsilon_{x}^{*\pm}=
\frac{1}{k^2}
\left (
\begin{array}{c}
     k^2-k_x^2  \\
     -k_x k_y   \\
     \mp k_x k_z
\end{array}
\right )
\qquad
\epsilon_{y}^{*\pm}=
\frac{1}{k^2}
\left (
\begin{array}{c}
     -k_x k_y  \\
     k^2-k_y^2   \\
     \mp k_y k_z
\end{array}
\right ),\label{eq:B11_2}
\end{equation}

Basis of left- and right- hand polarized waves is also convenient for some applications:
\begin{equation}
\epsilon_{L}^{\pm}=\epsilon_{R}^{*\pm}=\frac{\epsilon_{p}^{\pm}-i\epsilon_{s}^{\pm}}{\sqrt{2}}=
\frac{1}{\sqrt{2}k k_\parallel}\left (
\begin{array}{c}
     \mp k_x k_z +i k k_y  \\
     \mp k_y k_z -i k k_x   \\
     k_\parallel^2
\end{array}
\right ),\label{eq:B10.25_2}
\end{equation}
\begin{equation}
    \epsilon_{R}^{\pm}=\epsilon_{L}^{*\pm}=\frac{\epsilon_{p}^{\pm}+i\epsilon_{s}^{\pm}}{\sqrt{2}}=
\frac{1}{\sqrt{2}k k_\parallel}\left (
\begin{array}{c}
     \mp k_x k_z -i k k_y  \\
     \mp k_y k_z +i k k_x   \\
     k_\parallel^2
\end{array}
\right )
.\label{eq:B10.5_2}
\end{equation}
Now, we are ready to consider Green's function of a dipole located near an interface. Its reflected part, $\hat{M}^r$, can be easily constructed in an analogous way. According to the previous section we consider a particle, which is located in $z>z_p$ half-space. According to the choice of the $z$-axis direction depicted in several figures, this corresponds to a lattice beneath an interface, although it does not matter. In order to find reflected part, we should take, $\hat{M}^{0-}$, tensor, multiply it by a phase factor, reflection coefficient and substitute polarization vector by a polarization vector of wave propagating in an opposite direction. This results in the following formula:

\begin{equation}
\hat{M}_{\alpha \beta}^{0,\pm}
=\frac{i k_0^2 }{2\pi k_z} e^{2 i k_z h} \sum_p(\epsilon_p^+[\mathbf{k}_\parallel])_\alpha \sum_{p^{'}} r_{p p^{'}}(\epsilon_{p^{'}}^{*-}[\mathbf{k}_\parallel])_\beta,\label{eq:B12_2}
\end{equation}
where $r_{p p^{'}}$ is a reflection coefficient from $p'$ polarization to $p$ one. It should be noted, that each  multiplier in Eqn. \ref{eq:B12_2} has very clear meaning and therefore it can be easily reconstructed "mnemonically" in similar situations. The rule is that reciprocal polarization vector accounts for emission and therefore corresponds to the direction of initially radiated wave ($-$ in our case), whereas common polarization vector corresponds to the wave in its eventual state ($+$ in our case). Since Eqn. \ref{eq:B12_2} is derived in rather general form it can be potentially implemented for any basis and reflection from a photonic crystal, which can even mix $s-p$ polarizations.

\section{Analytical calculation of convolution for a homogeneous environment}

We have already described the process of the filtering of Green's function. However, when we deal with a case of homogeneous space, this general idea might be simplified. Indeed, operator $k_0^2(\delta_{\alpha \beta} +\frac{1}{k^2} \partial_\alpha \partial_\beta) $ acts locally in real space. Therefore, if some function has a plateau on which it is equal to $0$, then obviously, this plateau will remain after an application of the operator. This gives us an opportunity to filter not a dyadic Green's function itself, which has lots of components, but so-called scalar Green's function of the vector potential, $e^{i k r}/r$, which is an argument of the discussed operator.

This approach is much easier since the simple form of the internal part of Green's function allows to find such a filtering function that has an analytical Fourier image. Together with the technique presented in the previous section, we understand that this leads to the fact that calculation of filtered dyadic Green's function in reciprocal space is reduced to a substitution of the value of the integral \ref{eq:B6} with a specialized $m^0_\mathrm{f}(k_\parallel)$, which is found below.

Let $\bar{f}(r_\parallel)$ be a filter as it was defined above. However, we should still remember, that now we filter another function.
\begin{multline}
    m^0_\mathrm{\bar{f}}(k_\parallel) = \int \bar{f}(r_\parallel)\frac{e^{i k r_\parallel}}{r_\parallel}e^{-i\mathbf{k}_\parallel\mathbf{r}_\parallel}d^2\mathbf{r}_\parallel =\\= \int_0^\infty \int_0^{2\pi}\bar{f}(r_\parallel) \frac{e^{i k r_\parallel}}{r_\parallel}e^{-i{k}_\parallel {r}_\parallel \cos\phi} r_\parallel d{r}_\parallel d\phi
= 2\pi \int_0^\infty \bar{f}(r_\parallel) e^{i k r_\parallel}J_0({k}_\parallel {r}_\parallel )  d{r}_\parallel
\end{multline}

We have found a class of functions of the following shape that allows analytical calculation of the latter integral:
\begin{equation}
    \bar{f}^\mathrm{s}_{N}(x)=e^{- Nx}\sum_{n=0}^{N-1}\frac{N^n x^n}{n!}=\frac{\Gamma(N,N x)}{\Gamma( N)},
\end{equation}
where $\Gamma$ is a famous gamma function. Width of this function can be roughly estimated as unity:
\begin{equation}
    \int_0^\infty \bar{f}^\mathrm{s}_{N}(x) dx = 1.
\end{equation}
\begin{figure}[h]
    \centering
    \includegraphics[width=0.6\columnwidth]{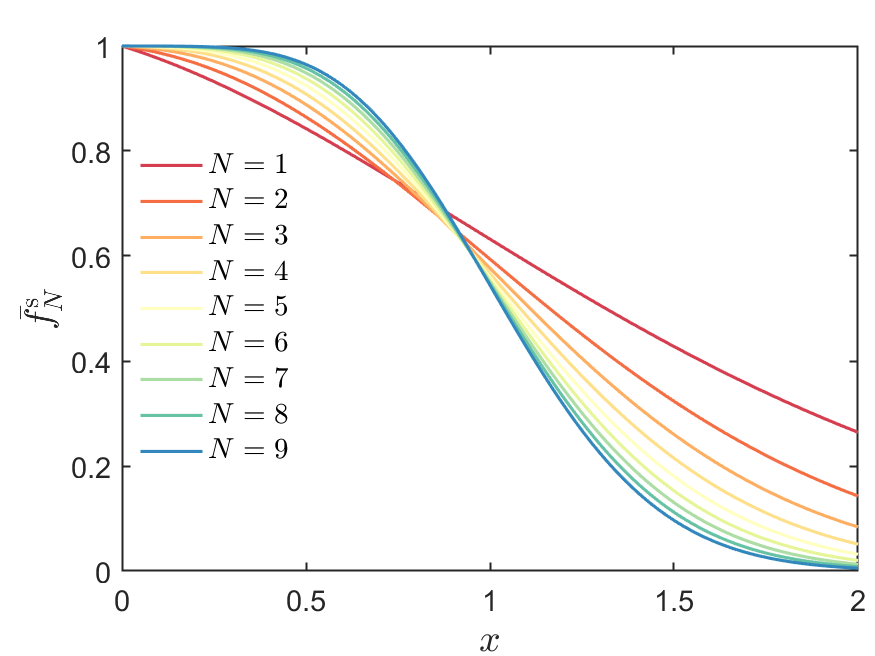}
    \caption{Graph of the filtering function $\bar{f}^\mathrm{s}_N(x)$.}
    \label{fig:filter_family}
\end{figure}

Therefore, filtering function of width $W$ can be found as $\bar{f}(r_\parallel) = \bar{f}^\mathrm{s}_{N}(r_\parallel/W)$. Finally, we return to $m_{\bar{f}}^0(k_\parallel)$:

\begin{multline}
    m^0_\mathrm{\bar{f}}(k_\parallel) = 2\pi \int_0^\infty \bar{f}(r_\parallel) e^{i k r_\parallel}J_0({k}_\parallel {r}_\parallel )  d{r}_\parallel=\\=2\pi \int_0^\infty     e^{-N r_\parallel/W}\sum_{n=0}^{N-1}\frac{N^n(r_\parallel/W)^n}{n!} e^{i k r_\parallel}J_0({k}_\parallel {r}_\parallel )  d{r}_\parallel=\\
    =2\pi \sum_{n=0}^{N-1}\frac{N^n}{n! W^n}\int_0^\infty  r_\parallel^n   e^{(i k -N/W)r_\parallel } J_0({k}_\parallel {r}_\parallel )  d{r}_\parallel=\\
    =2\pi \sum_{n=0}^{N-1}\frac{N^n}{n! W^n}\frac{\partial^n}{\partial a^n}\left. \frac{1}{\sqrt{a^2+k_\parallel^2}}\right|_{a=i k-N/W}\\
    =\frac{2\pi W}{N}\sum_{n=0}^{N-1}\frac{1}{n!}\frac{\partial^n}{\partial a^n}\left. \frac{1}{\sqrt{a^2+(k_\parallel W/N)^2}}\right|_{a=i k W/N-1}
\end{multline}
The wonderful fact is that the expression $H_{N}(x,y)=\sum_{n=0}^{N-1}\frac{1}{n!}\frac{\partial^n}{\partial x^n} \frac{1}{\sqrt{x^2+y^2}}$ can be found as a solution of recursive equation (i.e. iteratively):

\begin{align}
    H_{0}(x,y) &= 0\\
    H_1(x,y) &= \frac{1}{\sqrt{x^2+y^2}},\\    
    H_2(x,y) &= \frac{x^2-x+y^2}{\sqrt{x^2+y^2}^3},\\
    H_n(x,y)&=-\frac{ (x^2+y^2-3x - n(x^2-2x+y^2))H_{n-1}+(3x-2+n(1-2x))H_{n-2}+(2-n)H_{n-3}}{(n-1) (x^2+y^2) },
\end{align}
which brings us to the final expression:

\begin{equation}
    m^0_\mathrm{\bar{f}}(k_\parallel)
    = \frac{2\pi W}{N} H_N(i k W/N-1,k_\parallel W/N).
\end{equation}

As it was announced above, filtered Green's function might be easily found as:

\begin{equation}
\hat{M}_{\alpha \beta}^{0,\pm}
=\frac{ k_0^2 }{4\pi^2 } m^0_\mathrm{\bar{f}}(k_\parallel) \sum_p(\epsilon_p^\pm[\mathbf{k}_\parallel])_\alpha(\epsilon_p^{*\pm}[\mathbf{k}_\parallel])_\beta.
\end{equation}

The interesting fact is that even though the filter is a smooth function and provides a narrow space spectrum of the filtered vector potential, its space spectrum of an electric field is significantly wider because of specific peculiarities of the operator connecting vector potential with an electric field. This phenomenon results in necessity to take into account much more diffraction orders than it makes sense physically (typically several hundred instead of a dozen), however it is still much less than is needed to describe local field of plasmonic resonance (if we apply RCWA directly) and require just summation over them, which has a complexity of $O(N)$ instead of standard matrix inversion $O(N^3)$ in RCWA. At the same time, the opportunity to filter only one function instead of at least 6 and feasibility to do it analytically makes such an approach extremely efficient. The main advantage is that we need just to calculate analytically several terms of the sum, which is done extremely fast, much faster than any other required operations.

\section{Fast calculation of convolution}

2D convolution, which is required for filtering of $\hat{M}$ is a rather expensive operation. In order to tackle this problem, we boost it in several ways.

It can be easily noticed, that components of {of all the $\hat{M}$ tensor summands} have some trivial angular {dependence}. {Here, we consider axial symmetric filtering functions $\bar{f}(r_{\parallel})$, which means that $\bar{F}(k_\parallel)$ also depends only on the absolute value of the wavevector in-plain component.} This makes it possible to reduce the {convolution} to a much simpler form.

Indeed, let us consider a convolution for a certain component of the tensor $m_{\bar{\mathrm{f}}}(\mathbf{k}_\parallel) = (\bar{F}*m)(\mathbf{k}_\parallel)$. From the explicit expressions for tensors' components it can be easily noticed, that any of them can be represented in the following way:

\begin{equation}
    m(k_\parallel,\alpha) = m_k(k_\parallel) m_\alpha(\alpha),
\end{equation}
where angular part takes one of the following forms:

\begin{equation}
    m_\alpha(\alpha) = \left\{
    \begin{array}{c}
         1(\alpha) \\
         \sin \alpha \\
         \cos \alpha \\
         \sin^2 \alpha \\
         \cos^2 \alpha \\
         \sin \alpha \cos \alpha
    \end{array}
    \right\}
\end{equation}

In this way convolution can be calculated as follows:
\begin{multline}
    m_{\bar{\mathrm{f}}}(k_\parallel,\alpha) = \int m_k(k'_\parallel) m_\alpha(\alpha')\bar{ F}(\sqrt{k^2_\parallel+k^{'2}_\parallel-2k'_\parallel k_\parallel\cos(\alpha'-\alpha)})  k'_\parallel dk'_\parallel d\alpha'=\\
    =\int m_k(k'_\parallel) m_\alpha(\alpha'+\alpha) \bar{F}(\sqrt{k^2_\parallel+k^{'2}_\parallel-2k'_\parallel k_\parallel\cos(\alpha')})  k'_\parallel dk'_\parallel d\alpha'
\end{multline}

For each and every element of a set $m_\alpha$ a decomposition of $m_\alpha(\alpha'+\alpha)$ can be conducted. To give an example, a case of $m_\alpha(\alpha) = \sin^2 (\alpha)$ is considered below. Since $\sin^2(\alpha+\alpha') = 0.5-0.5\cos (2\alpha) \cos(2\alpha') + 0.5 \sin (2\alpha) \sin(2\alpha')$ we obtain:

\begin{equation}
m_{\bar{\mathrm{f}}}(k_\parallel,\alpha) =
0.5 \mathfrak{m}(k_\parallel)-0.5 \cos (2\alpha) \mathfrak{m}_{\cos{2\alpha}}(k_\parallel)+0.5 \sin (2\alpha) \mathfrak{m}_{\sin{2\alpha}}(k_\parallel),
\end{equation}
where 
\begin{equation}
\mathfrak{m}(k_\parallel) = \int m_k(k'_\parallel) \bar{F}(\sqrt{k^2_\parallel+k'^2_\parallel-2k'_\parallel k_\parallel\cos(\alpha')})  
k'_\parallel dk'_\parallel d\alpha',
\label{eq:B9}
\end{equation}

\begin{equation}
\mathfrak{m}_{\sin{2\alpha}}(k_\parallel) = \int m_k(k'_\parallel) \sin(2\alpha')\bar{F}(\sqrt{k^2_\parallel+k'^2_\parallel-2k'_\parallel k_\parallel\cos(\alpha')})  k'_\parallel dk'_\parallel d\alpha',
\label{eq:B10}
\end{equation}

\begin{equation}
\mathfrak{m}_{\cos{2\alpha}}(k_\parallel) = \int m_k(k'_\parallel) \cos(2\alpha')\bar{F}(\sqrt{k^2_\parallel+k'^2_\parallel-2k'_\parallel k_\parallel\cos(\alpha')})  k'_\parallel dk'_\parallel d\alpha',
\label{eq:B11}
\end{equation}

In this way, for a definite frequency, $k_\parallel$-dependence might be calculated on a grid and then interpolated. The fact that angular dependence is determined analytically and all pre-calculations are conducted for a 1D grid as well as the possibility not to calculate functions many times in close{ly-spaced} points increase the speed of computations drastically. The strongest speedup is observed when the angle dependence of a spectrum is considered.
Moreover, for some filters, angular part of integrals {(\ref{eq:B9}-\ref{eq:B11}) and similar ones} can be calculated analytically, which additionally speeds the calculations up.

\begin{figure}[h]
    \centering
    \includegraphics[width=0.6\columnwidth]{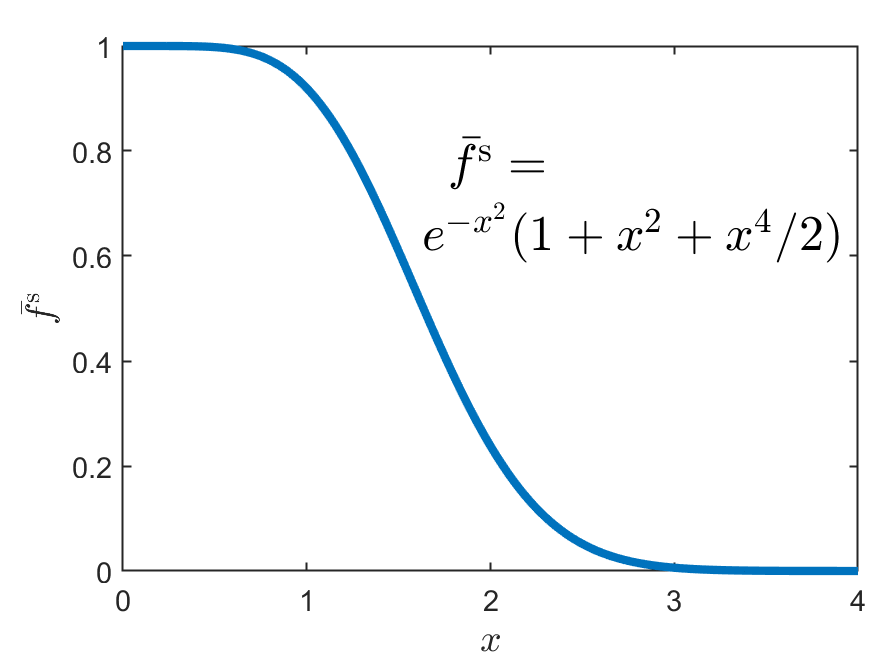}
    \caption{Graph of the filtering function $\bar{f}^{\mathrm{s}}(x)$.}
    \label{fig:filter_shape}
\end{figure}

{Hereinafter, we observe a specific filter of the following shape $\bar{f}^{\mathrm{s}}(x) =  e^{-x^2}(1+x^2+x^4/2)$ (see Fig. \ref{fig:filter_shape}), which we use in our calculations. This filter is convenient for a practical utilization because of a high rate of decay and the possibility to operate with it analytically, which is demonstrated below.} Its width can be roughly estimated as $\int_0^{\infty}\bar{f}^{{s}}(x)dx=15\sqrt{\pi}/16\approx1.66$. Therefore, in order to obtain the filter of the provided width, $W$ {(see Fig. \ref{fig:filter})}, we should use $\bar{f}(r_{{\parallel}}) = \bar{f}^{\mathrm{s}}(1.66r_\parallel/W)$. Since original Green's function diverges as $r_\parallel^{-3}$ at the zero point and filter $f(r_\parallel)$ decays as $r_\parallel^6$, then the filtered Green's function has to have at least 2 derivatives and therefore $\hat{M}_{\mathrm{f}}(\mathbf{k}_\parallel)=o(1/k_\parallel^2)$ for $k_\parallel\rightarrow\infty$.

{Fourier image of the filter $\bar{f}^{\mathrm{s}}(x)$ can be easily calculated analytically:}
\begin{equation}
    {\bar{F}^{\mathrm{s}}(y)=\frac{1}{4\pi^2}\int_0^{\infty} dx [x \bar{f}^{\mathrm{s}}(x) \int_0^{2\pi}d\varphi  e^{-x y \cos \varphi}] =  e^{-y^2/4}(96-24y^2+y^4)/(128\pi).}
\end{equation}
{At the same time, Fourier image of the original filter $\bar{f}(r_\parallel)$ is subsequently derived as $\bar{F}(\mathbf{k}_\parallel) =(W/1.66)^2 \bar{F}^{\mathrm{s}}(\mathbf{k}_\parallel W/1.66)$}

{The following expressions are valid for the particular filter utilized in this paper:} 

\begin{multline}
\int  \bar{F}(\sqrt{k^2_\parallel+k'^2_\parallel-2k'_\parallel k_\parallel\cos(\alpha')})  d\alpha'
= \\
= (W/1.66)^2[(\kappa^4_\parallel+\kappa'^4_\parallel+6 \kappa^2_\parallel \kappa'^2_\parallel-24 (\kappa'^2_\parallel+ \kappa^2_\parallel)+96) I_0(\kappa_\parallel \kappa'_\parallel/2)-\\-4 \kappa_\parallel \kappa'_\parallel (\kappa^2_\parallel+\kappa'^2_\parallel-10) I_1(\kappa_\parallel \kappa'_\parallel/2)]e^{ -(\kappa^2_\parallel+\kappa'^2_\parallel)/4}/64,
\end{multline}

\begin{equation}
\int  \sin(2\alpha') \bar{F}(\sqrt{k^2_\parallel+k'^2_\parallel-2k'_\parallel k_\parallel\cos(\alpha')})  d\alpha'
= 0,
\end{equation}

\begin{multline}
\int  \cos(2\alpha')\bar{F}(\sqrt{k^2_\parallel+k'^2_\parallel-2k'_\parallel k_\parallel\cos(\alpha')})  d\alpha'
= \\
=(W/1.66)^2[(\kappa^4_\parallel+\kappa'^4_\parallel+6 \kappa^2_\parallel \kappa'^2_\parallel-8 (\kappa^2_\parallel+ \kappa'^2_\parallel)) I_0(\kappa_\parallel \kappa'_\parallel/2)-\\-4  (\kappa^4_\parallel+\kappa'^4_\parallel-8(\kappa^2_\parallel+\kappa'^2_\parallel)+\kappa^2_\parallel \kappa'^2_\parallel (\kappa^2_\parallel+\kappa'^2_\parallel-4))\frac{ I_1(\kappa_\parallel \kappa'_\parallel/2)}{\kappa_\parallel \kappa'_\parallel}] e^{ -(\kappa^2_\parallel+\kappa'^2_\parallel)/4} /64,
\end{multline}

\begin{equation}
{\int  \sin(\alpha') \bar{F}(\sqrt{k^2_\parallel+k'^2_\parallel-2k'_\parallel k_\parallel\cos(\alpha')})  d\alpha'= 0,}
\end{equation}

\begin{multline}
{\int  \cos(\alpha')\bar{F}(\sqrt{k^2_\parallel+k'^2_\parallel-2k'_\parallel k_\parallel\cos(\alpha')})  d\alpha' =} \\
{=(W/1.66)^2[(\kappa^4_\parallel+\kappa'^4_\parallel+6 \kappa^2_\parallel \kappa'^2_\parallel-32 (\kappa^2_\parallel+ \kappa'^2_\parallel)+192) I_1(\kappa_\parallel \kappa'_\parallel/2)-}\\{-4 \kappa_\parallel \kappa'_\parallel (\kappa^2_\parallel+\kappa'^2_\parallel-10) I_2(\kappa_\parallel \kappa'_\parallel/2)] e^{ -(\kappa^2_\parallel+\kappa'^2_\parallel)/4} /64,}
\end{multline}
{where $I_0$, $I_1$ and $I_2$ are the modified Bessel functions of the first kind of corresponding orders, $\kappa = k_\parallel W/1.66$ and $\kappa' = k'_\parallel W/1.66$.}

To conclude, in order to find a filtered Green's function in reciprocal space, $\hat{M}_{\mathrm{f}}(\mathbf{k}_\parallel)$, at any point, we should calculate just several functions on a 1D grid and continue them analytically to the whole plane. Pre-calculation of this auxiliary functions on a grid and subsequent interpolation makes these computations much faster. Moreover, for specific filtering functions calculation of the angular part of an integral can be conducted analytically, which boosts calculations additionally.

Finally, this approach, allows us to calculate angle-dependent spectra in a few minutes on a regular laptop, which is more than enough for practical utilization. Moreover, calculation of convolution is even not a bottleneck in our calculations and does not limit the performance of the whole program.

\clearpage
\newpage

\chapter{Details on RCWA matrix calculations}\label{App rcwa}
\section{Combination of scattering matrices}

The elements of the scattering matrix $\mathbb{S}$ which is a combination of two scattering matrices $\mathbb{S}^1$ and $\mathbb{S}^2$ denoted as $\mathbb{S} = \mathbb{S}^1 \otimes \mathbb{S}^2$  is given by the following formula
\begin{align}
    \mathbb{S}_{11} & = \mathbb{S}^2_{11}(\hat{I}-\mathbb{S}^1_{12}\mathbb{S}^2_{21})^{-1}\mathbb{S}^1_{11}\\
    \mathbb{S}_{12} & = \mathbb{S}^2_{12} + \mathbb{S}^2_{11}(\hat{I}-\mathbb{S}^1_{12}\mathbb{S}^2_{21})^{-1}\mathbb{S}^1_{12}\mathbb{S}^2_{22}\\
    \mathbb{S}_{21} & = \mathbb{S}^1_{21} + \mathbb{S}^1_{22}(\hat{I}-\mathbb{S}^2_{21}\mathbb{S}^1_{12})^{-1}\mathbb{S}^2_{21}\mathbb{S}^1_{11}\\
    \mathbb{S}_{22} & = \mathbb{S}^1_{22}(\hat{I}-\mathbb{S}^2_{21}\mathbb{S}^1_{12})^{-1}\mathbb{S}^2_{22}
\end{align}

\section{Method of oscillating currents}
The RCWA formalism allows to calculate the emission of oscillating dipoles embedded in an arbitrary layer. The basic principle to do that is to construct the amplitude discontinuity vector which connects the vectors of amplitudes at coordinates infinitesimally above and below the dipole's plane \cite{lobanov2012emission}:
\begin{equation}
\begin{bmatrix}\vec{\mathrm{d}}\\\vec{\mathrm{u}}\end{bmatrix}_{z_p{+}0}
-\begin{bmatrix}\vec{\mathrm{d}}\\\vec{\mathrm{u}}\end{bmatrix}_{z_p{-}0}
=\begin{bmatrix}\vec{\mathrm{j}}_d\\\vec{\mathrm{j}}_u\end{bmatrix}=\mathbb{A}\begin{bmatrix}\vec{\mathrm{d}}_0\\\vec{\mathrm{u}}_0\end{bmatrix}.
\label{eq:appc:5}
\end{equation}

In this paper, we consider only examples, when the currents are located in a section of a homogeneous layer (which does not exclude the existence of other layers in near-field slightly higher and lower), therefore all the derivations below are conducted under this assumption. Application of the material matrix, $\mathbb{F}$, \cite{tikhodeev2002} to the equation \ref{eq:appc:5} gives us jumps of tangential components of electric and magnetic fields:

\begin{equation}
    \begin{bmatrix}{E}_{x}\\{E}_{y}\\{H}_{x}\\{H}_{y}\end{bmatrix}_{z_p{+}0}-
    \begin{bmatrix}{E}_{x}\\{E}_{y}\\{H}_{x}\\{H}_{y}\end{bmatrix}_{z_p{-}0} =
    \begin{bmatrix}{J}_{Ex}\\{J}_{Ey}\\{J}_{Hx}\\{J}_{Hy}\end{bmatrix}=J = \mathbb{F}\begin{bmatrix}\vec{\mathrm{j}}_d\\\vec{\mathrm{j}}_u\end{bmatrix},
\label{eq:appc:6}
\end{equation}
where $J$ is the vector of discontinuities of Fourier components of electric and magnetic fields.
The elements of this vector are found from the Fourier components of the surface current $\mathbf{i}$ \cite{lobanov2012emission}:
\begin{equation}
    \begin{bmatrix}{{J}}_{Ex}\\{{J}}_{Ey}\end{bmatrix}=\frac{4\pi}{c\varepsilon}
    \begin{pmatrix}\hat{K}_x\\\hat{K}_y\end{pmatrix}i_z,
    \hspace{20pt}
    \begin{bmatrix}{{J}}_{Hx}\\{{J}}_{Hy}\end{bmatrix}=\frac{4\pi}{c}
    \begin{pmatrix}i_y\\{-}i_x\end{pmatrix},
    \label{eq:appc:7}
\end{equation}
where $\hat{K}_x$ and  $\hat{K}_y$ are defined by formulas (\ref{eq:kxky}).

In this way, in order to find the matrix $\mathbb{A}$, we should just express harmonics of this current through the vector $[\vec{\mathrm{d}}_0, \vec{\mathrm{u}}_0]^T$. The first step is the application of the material matrix, which gives us fields:
\begin{equation}
    \begin{bmatrix}{E}_{x}^0\\{E}_{y}^0\\{H}_{x}^0\\{H}_{y}^0\end{bmatrix}=\mathbb{F}\begin{bmatrix}\vec{\mathrm{d}}_0\\\vec{\mathrm{u}}_0\end{bmatrix}.
\label{eq:appc:8}
\end{equation}
This vector of electric and magnetic fields harmonics is consequently transformed to the vector of Fourier harmonics of all the components of electric field:
\begin{equation}
    \begin{bmatrix}{E}_{x}^0\\{E}_{y}^0\\{E}_{z}^0\end{bmatrix}=\begin{bmatrix}
    \hat{I} & \hat{0} & \hat{0}  & \hat{0} \\
    \hat{0} & \hat{I} & \hat{0}  & \hat{0} \\
    \hat{0} & \hat{0} & \hat{K}_y/\varepsilon & -\hat{K}_x/\varepsilon \\
    \end{bmatrix}\begin{bmatrix}{E}_{x}^0\\{E}_{y}^0\\{H}_{x}^0\\{H}_{y}^0\end{bmatrix}.
    \label{eq:appc:9}
\end{equation}
In turn, this vector is used for calculation of electric field in real space at the position of the $i$-th particle:

\begin{equation}
    \mathbf{E}^0=
    \begin{bmatrix}\mathrm{E}_{x}^0\\\mathrm{E}_{y}^0\\\mathrm{E}_{z}^0\end{bmatrix}=\begin{bmatrix}
    \vec{\Gamma} & \vec{0} & \vec{0}\\
    \vec{0} & \vec{\Gamma} & \vec{0}\\
    \vec{0} & \vec{0} & \vec{\Gamma}
    \end{bmatrix}\begin{bmatrix}{E}_{x}^0\\{E}_{y}^0\\{E}_{z}^0\end{bmatrix},
    \label{eq:appc:10}
\end{equation}
where $1 \times N_g$ hypervector $\vec{\Gamma}$ is set by the following expression:
\begin{equation}
    \vec{\Gamma} = \mathrm{exp}(i(k_x+\vec{g}_x)x+i(k_y+\vec{g}_y)y)
\end{equation}
where $(x, y)$ is the coordinate of a nanoparticle in the unit cell in a real space.

Thus, the current of the $i$-th particle is determined as:
\begin{equation}\mathbf{I}=-i\omega \hat{\alpha}^{\mathrm{\mathrm{eff}}} \mathbf{E}^0.\label{eq:appc:11}\end{equation}
Finally, according to the Poisson formula a grid of point currents in real space corresponds to the following harmonics in Fourier space:
\begin{equation}
    \begin{bmatrix}i_{x}\\i_{y}\\i_{z}\end{bmatrix}=\frac{1}{s}
    \begin{bmatrix}
    \vec{\Gamma} & \vec{0} & \vec{0}\\
    \vec{0} & \vec{\Gamma} & \vec{0}\\
    \vec{0} & \vec{0} & \vec{\Gamma}
    \end{bmatrix}^\dagger
    \begin{bmatrix}\mathrm{I}_{x}\\\mathrm{I}_{y}\\\mathrm{I}_{z}\end{bmatrix},
\label{eq:appc:12}
\end{equation}
where $s$ is a surface of a unit cell in a real space and dagger denotes Hermitian conjugate.

Combining equations (\ref{eq:appc:5}-\ref{eq:appc:12}) we obtain the following expression for $\mathbb{A}$ tensor:
\begin{equation}
    \mathbb{A}=\mathbb{F}^{-1}\hat{A}\mathbb{F},
    \label{eq:appc:13}
\end{equation}
where
\begin{equation}
    \hat{A} = \frac{-4 \pi i  k_0}{s}\begin{bmatrix}
    \hat{0} & \hat{0}& \hat{K}_x/\varepsilon\\
    \hat{0} & \hat{0}& \hat{K}_y/\varepsilon\\
    \hat{0} & \hat{I}&\hat{0} \\
    {-}\hat{I} & \hat{0} &\hat{0} 
    \end{bmatrix}
   \begin{bmatrix}
    \vec{\Gamma} & \vec{0} & \vec{0}\\
    \vec{0} & \vec{\Gamma} & \vec{0}\\
    \vec{0} & \vec{0} & \vec{\Gamma}
    \end{bmatrix}^{\dagger}
    \hat{\alpha}^{\mathrm{eff}}
    \begin{bmatrix}
    \vec{\Gamma} & \vec{0} & \vec{0}\\
    \vec{0} & \vec{\Gamma} & \vec{0}\\
    \vec{0} & \vec{0} & \vec{\Gamma}
    \end{bmatrix}
    \begin{bmatrix}
    \hat{I} & \hat{0} & \hat{0}  & \hat{0} \\
    \hat{0} & \hat{I} & \hat{0}  & \hat{0} \\
    \hat{0} & \hat{0} & \hat{K}_y/\varepsilon & -\hat{K}_x/\varepsilon \\
    \end{bmatrix}.
    \label{eq:appc:14}
\end{equation}

\section*{Acknowledgements}
This work was supported by the Russian Foundation for Basic Research (Grant No. 18-29-20032).

\bibliographystyle{unsrt}


\end{document}